\documentclass[twocolumn,twocolappendix]{aastex701}

\usepackage[T1]{fontenc}

\begin{document}

\title{SPURS: An Ultra-deep  View Inside the Compact, Nitrogen-Enriched Nuclei of Little Red Dots}

\author[0000-0001-5940-338X]{Mengtao Tang}
\affiliation{Tsung-Dao Lee Institute, Shanghai Jiao Tong University, 1 Lisuo Road, Shanghai 201210, People’s Republic of China}
\affiliation{School of Physics and Astronomy, Shanghai Jiao Tong University, 800 Dongchuan Road, Shanghai 200240, People’s Republic of China}
\affiliation{State Key Laboratory of Dark Matter Physics, Shanghai Jiao Tong University, 1 Lisuo Road, Shanghai 201210, People’s Republic of China}
\email[show]{mengtao.tang@sjtu.edu.cn}

\author[0000-0001-6106-5172]{Daniel P. Stark}
\affiliation{Department of Astronomy, University of California, Berkeley, Berkeley, CA 94720, USA}
\email{dpstark@berkeley.edu}

\author[0000-0002-3407-1785]{Charlotte A. Mason}
\affiliation{Cosmic Dawn Center (DAWN)}
\affiliation{Niels Bohr Institute, University of Copenhagen, Jagtvej 128, 2200 Copenhagen N, Denmark}
\email{charlotte.mason@nbi.ku.dk}

\author[0000-0002-2178-5471]{Zuyi Chen}
\affiliation{Cosmic Dawn Center (DAWN)}
\affiliation{Niels Bohr Institute, University of Copenhagen, Jagtvej 128, 2200 Copenhagen N, Denmark}
\email{zuyi.chen@nbi.ku.dk}

\author[0000-0001-5860-3419]{Tucker Jones}
\affiliation{Department of Physics and Astronomy, University of California, Davis, 1 Shields Avenue, Davis, CA 95616, USA}
\email{tdjones@ucdavis.edu}

\author{Sarah Searle Grannis}
\affiliation{Cosmic Dawn Center (DAWN)}
\affiliation{Niels Bohr Institute, University of Copenhagen, Jagtvej 128, 2200 Copenhagen N, Denmark}
\email{wbh841@alumni.ku.dk}

\author[0000-0002-9132-6561]{Peter Senchyna}
\affiliation{The Observatories of the Carnegie Institution for Science, 813 Santa Barbara Street, Pasadena, CA 91101, USA}
\email{psenchyna@carnegiescience.edu}

\author[0000-0003-1432-7744]{Lily Whitler}
\affiliation{Kavli Institute for Cosmology, University of Cambridge, Madingley Road, Cambridge, CB3 0HA, UK}
\affiliation{Cavendish Laboratory, University of Cambridge, JJ Thomson Avenue, Cambridge, CB3 0US, UK}
\email{lw851@cam.ac.uk}

\author[0000-0002-2645-679X]{Keerthi Vasan G. C.}
\affiliation{The Observatories of the Carnegie Institution for Science, 813 Santa Barbara Street, Pasadena, CA 91101, USA}
\email{kvch153@gmail.com}

\author[0000-0001-5487-0392]{Viola Gelli}
\affiliation{Cosmic Dawn Center (DAWN)}
\affiliation{Niels Bohr Institute, University of Copenhagen, Jagtvej 128, 2200 Copenhagen N, Denmark}
\email{viola.gelli@nbi.ku.dk}

\begin{abstract}
We present the first ultra-deep rest-UV spectroscopy of four UV-bright Little Red Dots (LRDs), obtained from the SPURS Cycle 4 Large Program. The spectra reveal broad C~{\small IV} (FWHM $\approx2700-2800$~km~s$^{-1}$) in two LRDs, alongside narrow-line densities elevated above star-forming galaxies ($n_e\sim10^4-10^5$~cm$^{-3}$, reaching $10^6$~cm$^{-3}$ in the most extreme source) and nitrogen-enhancements in all four LRDs. We detect broad He~{\small II} emission (FWHM $\approx930$~km~s$^{-1}$) in one LRD, and two others with fast P-Cygni absorption ($\gtrsim2200$~km~s$^{-1}$). Strong interstellar absorption lines and Ly$\alpha$ damping wings reveal the UV continuum is deeply embedded in neutral gas ($N_{\rm HI}\gtrsim10^{22}$~cm$^{-2}$) in all four LRDs. Detections of fluorescent Fe~{\small II} and O~{\small I} emission and fine-structure absorption indicate this gas lies close to the UV-emitting region. In archival $z>4$ samples, we find nitrogen and strong C~{\small III}] emission are significantly more common in LRDs than in the galaxy population. The transmission of broad C~{\small IV}, tracing the broad-line region or cocoon, depends on rest-optical color within our sample, consistent with an orientation-dependent picture in which bluer, less obscured sightlines offer a more direct, polar view of the central engine and its outflows. We find several potential signatures of very massive stars, whose winds may contribute to nitrogen enhancement. We investigate other abundance patterns expected from supermassive stars but our results are inconclusive. Our results place the UV-emitting region within $\lesssim8$~pc of the LRD nucleus, consistent with an actively assembling nuclear star cluster. Dynamical interactions in this extremely dense environment, including tidal disruption of stars, may explain the high incidence of nitrogen enhancements in LRDs.
\end{abstract}



\section{Introduction} \label{sec:intro}

JWST \citep{Gardner2023,Rigby2023} has opened a new window onto the growth of supermassive black holes in the early universe. Deep spectroscopy has enabled the identification of broad-line (BL) active galactic nuclei (AGN) out to very high redshifts, revealing black hole growth already underway within the first billion years of cosmic history. Much of the early work in this area has focused on Little Red Dots (LRDs), a population of compact sources identified by the combination of red rest-frame optical colors and bluer rest-frame ultraviolet (UV) colors \citep{Furtak2023,Kocevski2023,Labbe2023,Greene2024,Matthee2024}. Following their discovery in JWST/Near Infrared Camera \citep[NIRCam;][]{Rieke2023} imaging \citep{Barro2024,Kokorev2024,Kocevski2025,Labbe2025}, LRDs were shown to exhibit broad Balmer lines ($\gtrsim1000$~km~s$^{-1}$) alongside narrow forbidden lines ($\simeq300$~km~s$^{-1}$), a combination characteristic of broad-line AGN \citep{Greene2024,Killi2024,Maiolino2024b,Matthee2024,Hviding2025}. 

Application of standard virial relations \citep{Greene2005,Reines2015} to the broad Balmer lines suggested that LRDs host black holes with large masses ($\simeq10^6-10^8$~$M_\odot$; e.g., \citealt{Greene2024,Lin2024,Maiolino2024b,Matthee2024,Kocevski2025}). Under the assumption that the rest-frame UV emission is dominated by the host galaxy, the inferred stellar masses were found to be relatively small ($\simeq10^8-10^9$~$M_\odot$), implying black hole-to-stellar mass ratios in excess of local scaling relations \citep[e.g.,][]{Maiolino2024b,Kocevski2025}. 
These findings led to the conclusion that LRDs trace a population of galaxies with overmassive black holes. While it has been shown that LRDs only represent a subset of the high-$z$ AGN JWST has discovered \citep{Hainline2025,Taylor2025,Kocevski2026}, their space density is large enough to suggest that they may represent an important channel of early black hole growth \citep[e.g.,][]{Barro2024,Kokorev2024,Akins2025,Labbe2025}.

As more data on LRDs arrived, evidence began to mount that they differed in several important ways from lower redshift broad-line AGN. Deep stacking experiments revealed that LRDs are X-ray weak \citep[e.g.,][]{Ananna2024,Yue2024,Akins2025,Maiolino2025,Brazzini2026}, suggesting either heavy gas obscuration or an intrinsically weak corona. Observations with MIRI and ALMA revealed little evidence of dust emission \citep[e.g.,][]{Perez-Gonzalez2024,Williams2024,Akins2025}, indicating the red colors are unlikely to be dominated by reddening. High ionization lines were found to be rare \citep[e.g.,][]{Tang2025,Lambrides2026,Wang2026}, perhaps indicating a softer ionizing spectrum. Deep rest-frame optical spectroscopy revealed Balmer line absorption and Balmer breaks, both suggesting very dense gas absorption by a population of hydrogen atoms in the $n=2$ level \citep[e.g.][]{Juodzbalis2024,deGraaff2025b,Inayoshi2025,Ji2025,Naidu2026}. 

These observations have motivated new theoretical frameworks for LRDs. Some have argued that LRDs are black holes embedded in a dense gas cocoon that thermalizes the accretion disk spectrum into a pseudo-blackbody (so-called ``black hole stars''; \citealt{deGraaff2025b,Naidu2026}). More recently, radiative transfer models have shown that the red optical continuum of LRDs can be explained as Paschen free-bound nebular emission from the ionized cocoon \citep{Sneppen2026b} without requiring full thermalization of the spectrum. In this and related frameworks, electron scattering broadens the Balmer lines as photons traverse the dense gas. The intrinsic line widths are therefore narrower than the observed profiles, implying correspondingly lower black hole masses. Recent work has demonstrated that these lower masses bring LRDs into closer agreement with the local black hole-to-stellar mass scaling relations \citep[e.g.,][]{Rusakov2026,Torralba2026b}. Whether these new frameworks provide the correct interpretation of the LRDs remains debated in the literature, with some groups arguing that LRDs can be more naturally explained as overmassive black holes in small galaxies \citep[e.g.,][]{Maiolino2024b,Brazzini2025,Juodzbalis2026a,Scholtz2026b}. 

Progress on this debate has relied almost entirely on rest-frame optical and longer-wavelength observations. The rest-frame UV of LRDs remains comparatively unexplored, leaving key questions open. The UV continuum and narrow emission lines in LRDs are generally attributed to the host galaxy \citep[e.g.,][]{deGraaff2025c,Rinaldi2025,Inayoshi2026,Torralba2026a}, but it is possible that emission from the AGN or gas cocoon may contribute significantly to both \citep[e.g.,][]{Leung2025,Ando2026}. Studying the host galaxy via narrow emission lines in the optical is challenging owing to the underlying LRD continuum. The UV continuum offers a cleaner alternative, and it contains a suite of emission lines that are sensitive to the physical conditions of the ionized gas, the hardness of the ionizing spectrum, and the chemical abundance pattern. By benchmarking these UV narrow line diagnostics against the broader galaxy population, we can test whether LRD hosts are typical, or whether the narrow lines give indication of conditions rarely seen in galaxies. For example, several LRDs have been shown to have UV nitrogen lines that point to the same nitrogen-enhanced abundances seen in a subset of early galaxies \citep{Labbe2024,Tripodi2025b,Mascia2026,Morishita2026,Papovich2026}. Whether the nitrogen lines are common in LRDs is not known, nor is it clear what mechanism produces the peculiar abundances in LRDs.


\begin{deluxetable}{cccccc}
\tablecaption{Basic properties of the four LRDs.}
\tablehead{
ID & RA (deg) & Dec (deg) & $z_{\rm spec}$ & $m_H$ & M$_{\rm UV}$
}
\startdata
SPURS-GN-2 & $189.083481$ & $+62.202583$ & $7.1863$ & $26.3$ & $-20.7$ \\
SPURS-GN-2004 & $189.032464$ & $+62.216423$ & $6.7110$ & $25.7$ & $-21.1$ \\
SPURS-GN-29 & $189.019240$ & $+62.243537$ & $7.0374$ & $26.6$ & $-20.4$ \\
CEERS-7902 & $214.983030$ & $+52.956010$ & $6.9842$ & $26.5$ & $-20.4$ \\
\enddata
\tablecomments{$H$-band magnitudes are computed from JWST/NIRCam F150W photometry.}
\label{tab:sample}
\end{deluxetable}

The rest-frame UV also tests the broad-line region (BLR) and the dense gas cocoon directly, through both emission and absorption. Existing, shallower spectra have not detected the broad permitted UV lines (C~{\small IV}, He~{\small II}, Mg~{\small II}) expected from a standard broad-line region \citep{Tang2025,Lambrides2026}, but whether this reflects absorption by dense gas or an intrinsically softer ionizing spectrum remains unclear. A dense gas envelope around the UV continuum should leave its own signature: strong damped Ly$\alpha$ absorption and a forest of metal absorption lines. The kinematics and covering fraction recovered from these lines constrain both the structure of the cocoon and the origin of the UV continuum itself.

The SPectroscopic Ultra-deep Reionization-era Survey (SPURS; GO 9214, PIs: C. Mason, D. Stark) Cycle 4 Large Program is providing the first ultra-deep, moderate-resolution grating spectra covering the rest-frame UV of LRDs. In \citet{Tang2026b}, we presented 29 hours G140M observations of two LRDs in Abell 2744, including the gravitationally lensed source Abell2744-QSO1 at $z=7.04$ (discovered by \citealt{Furtak2023}). In this paper, we extend this work to four new LRDs (Table~\ref{tab:sample}) observed with SPURS in the GOODS-North (GOODS-N; SPURS-GN-2, SPURS-GN-2004, SPURS-GN-29) and the Extended Groth Strip (EGS; CEERS-7902) fields. Using these observations, we characterize the density, temperature, and abundance pattern of the narrow-line gas; search for broad permitted UV lines (C~{\small IV}, He~{\small II}, Mg~{\small II}) that trace the broad-line region; look for signatures of massive stars in the UV continuum; and measure the interstellar absorption lines that constrain the covering fraction, kinematics, and abundance pattern of the gas surrounding the UV-emitting source.
An independent analysis of the abundances in these  LRDs has been presented in Kokorev et al. (2026b).

The organization of this paper is as follows. In Section~\ref{sec:data}, we summarize what was previously known about the four LRDs, describe the new SPURS observations, present their basic physical properties with a focus on the rest-frame optical, and characterize the NIRCam images. We then characterize the rest-frame UV emission lines of the four LRDs in Section~\ref{sec:uv_emi} and the interstellar absorption lines in Section~\ref{sec:abs}. In Section~\ref{sec:uv_diag}, we discuss the narrow UV emission line properties of LRDs in the context of broader galaxy population. We discuss the implications for LRDs with SPURS spectra in Section~\ref{sec:discussion}. Finally, we summarize our conclusions in Section~\ref{sec:summary}. Throughout the paper we adopt a $\Lambda$-dominated, flat universe with $\Omega_{\Lambda}=0.7$, $\Omega_{\rm{M}}=0.3$, and $H_0=70$~km~s$^{-1}$~Mpc$^{-1}$. All magnitudes are quoted in the AB system \citep{Oke1983} and all equivalent widths (EWs) are quoted in the rest frame.

\section{SPURS Spectroscopy of LRDs} \label{sec:data}

\begin{figure*}
\includegraphics[width=\linewidth]{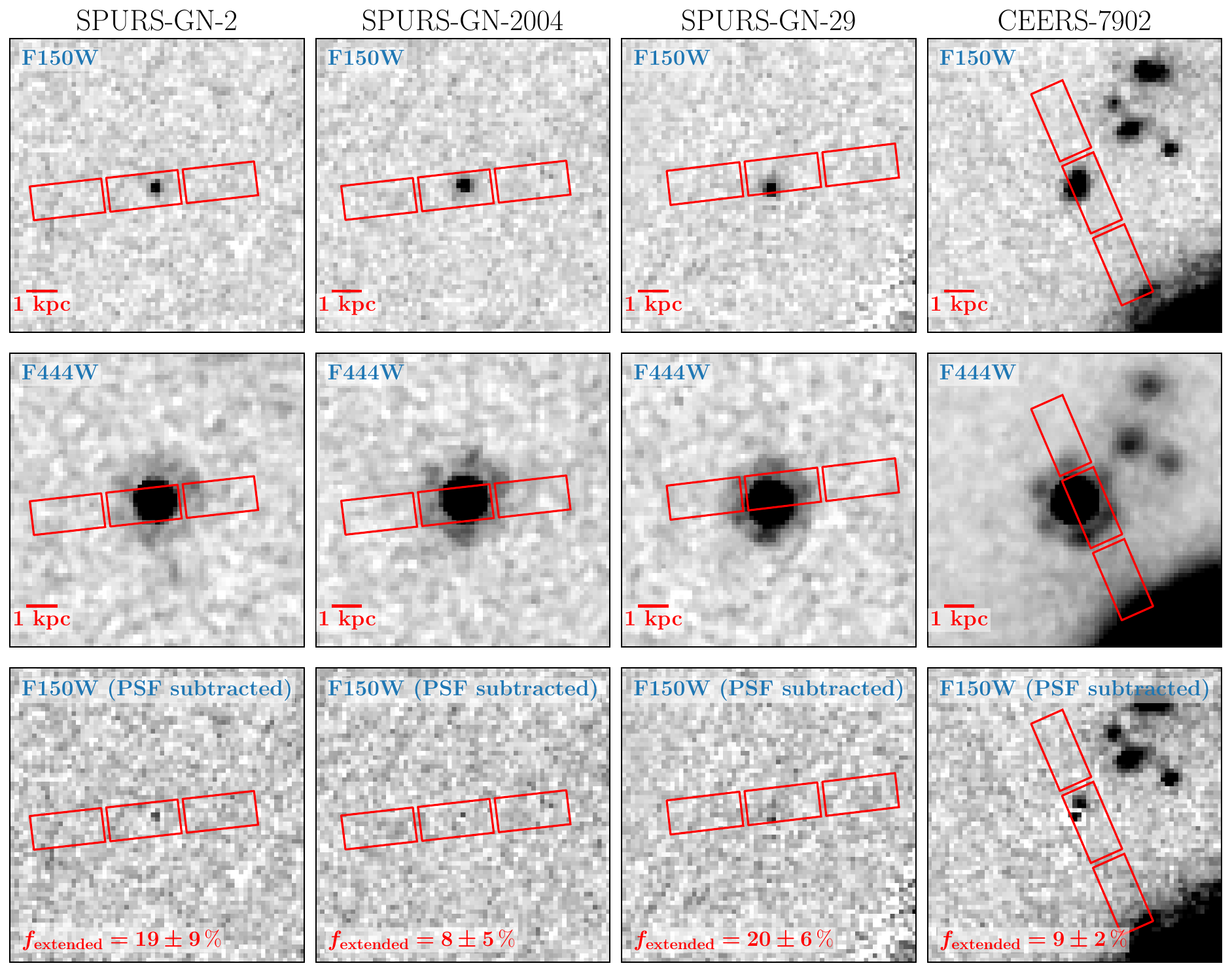}
\caption{JWST/NIRCam F150W (top), F444W (middle), and PSF-subtracted F150W images (bottom) of GN-2, GN-2004, GN-29, and CEERS-7902. Each image is presented with a size of $2''\times2''$. For the top and middle rows, we show the scale bar of $1$~kpc in the bottom left of each panel. For the bottom row, we note the light fraction from residual components after subtracting the point source in F150W. In each panel, we also overplot our NIRSpec shutters (red boxes) in SPURS observations.}
\label{fig:image}
\end{figure*}

Our sample consists of four LRDs, each selected to be bright in the rest-frame UV. 
Three of the LRDs are in GOODS-N (SPURS-GN-2, SPURS-GN-2004, SPURS-GN-29) and one is in EGS (CEERS-7902). 
We introduce the original observations of the LRDs in Section~\ref{sec:ancillary_data}, describe the SPURS observations (Figure~\ref{fig:image}) and reduction in Section~\ref{sec:spurs}, then detail the basic properties of the LRDs in Section~\ref{sec:opt} focusing primarily on the rest-frame optical spectra, and characterize their NIRCam images in Section~\ref{sec:psf}.

\subsection{Ancillary Data of LRD Sample} \label{sec:ancillary_data}

\begin{figure*}
\includegraphics[width=\linewidth]{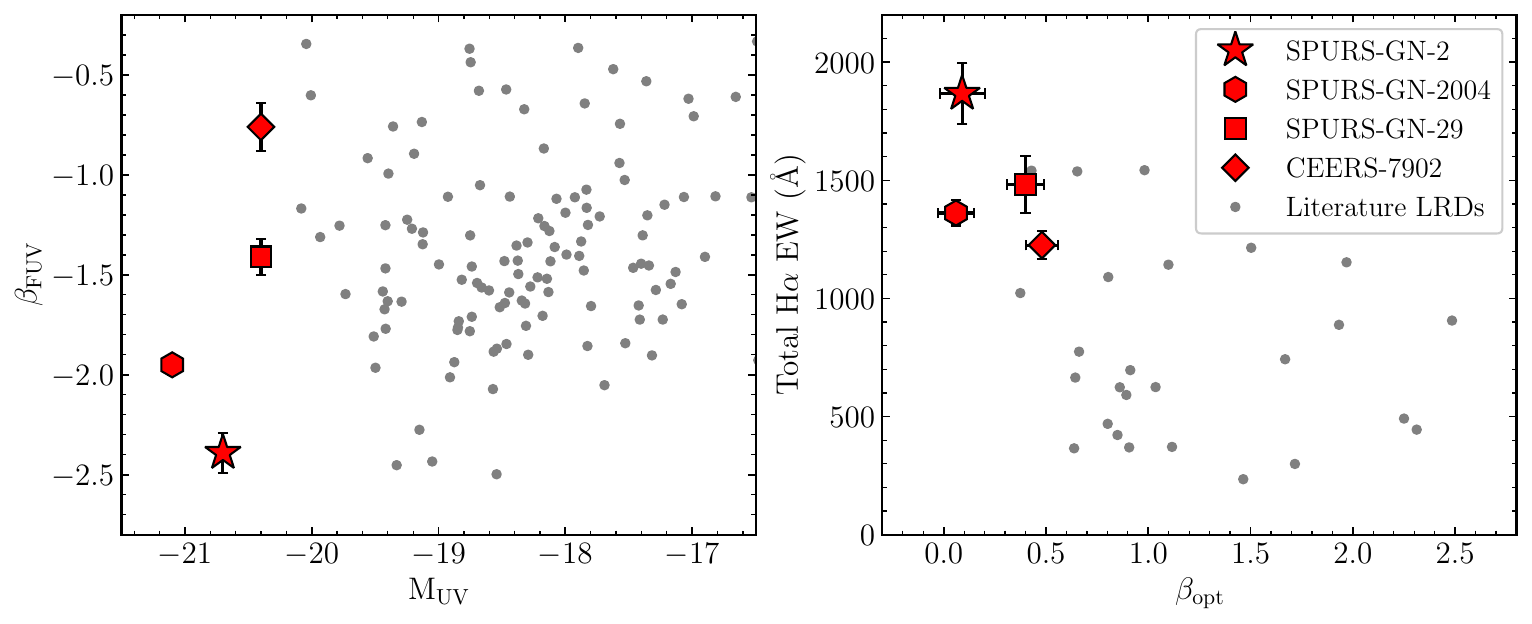}
\caption{GN-2 (red star), GN-2004 (red hexagon), GN-29 (red square), and CEERS-7902 (red diamond) on the demographics of LRDs. Left panel: absolute UV magnitude versus rest-frame FUV continuum slope. Right panel: rest-frame optical continuum slope versus total H$\alpha$ EW. In each panel, we overplot LRDs in literature \citep{deGraaff2025c,Hviding2025,Kocevski2025} as grey circles for comparison.}
\label{fig:demo}
\end{figure*}

SPURS-GN-2 (hereafter GN-2) has been imaged with JWST/NIRCam through the FRESCO \citep{Oesch2023}, CONGRESS (GO 3577, PIs: E. Egami, F. Sun), and JADES \citep{Eisenstein2026} programs. 
It is bright in the rest-frame UV continuum ($H=26.3$; Figure~\ref{fig:demo}). 
Its redshift was first identified using the NIRCam F444W grism spectrum obtained from the FRESCO survey and reported in \citet{Meyer2024}. 
The GO 4762 (PIs: S. Fujimoto, G. Brammer) program obtained a deep F410M grism spectrum (their ID 2756), which reveals broad H$\beta$ emission and narrow [O~{\small III}] emission lines \citep{Xiao2025}. 


The other two LRDs in GOODS-N were also imaged by NIRCam as part of the JADES, CONGRESS, and FRESCO surveys. 
Among the four LRDs in this paper, SPURS-GN-2004 (hereafter GN-2004) is the brightest in the rest-frame UV continuum ($H=25.7$; Figure~\ref{fig:demo}). 
The SPURS observations provide the first spectroscopic confirmation of GN-2004. The redshift of SPURS-GN-29 (hereafter GN-29) was identified using the FRESCO F444W grism spectrum and reported in \citet{Meyer2024}. 
The GO 4762 F410M grism spectrum (their ID 9094) reveals broad H$\beta$ emission and narrow [O~{\small III}] emission lines, as expected for broad-line AGN \citep{Xiao2025}. 
This LRD has the faintest rest-frame UV continuum magnitude in our sample ($H=26.6$; Figure~\ref{fig:demo}) but is still UV-bright when compared to the full population of LRDs.

CEERS-7902 was first classified as an LRD by its NIRCam photometry \citep{Labbe2023,Kocevski2025}, obtained via the CEERS Early Release Science program \citep{Finkelstein2025}.
A low-resolution prism spectrum (0.8 hrs) and a medium-resolution G395M/F290LP spectrum (0.8 hrs) were obtained from the RUBIES program \citep{deGraaff2025a}, revealing the presence of a broad H$\beta$ emission line with narrow rest-frame optical forbidden lines \citep{Wang2024,Kocevski2025}. 
The Cycle 2 GO 4287 program (PIs: C. Mason, D. Stark) observed the LRD again with NIRSpec medium resolution G395M/F290LP (1.0 hrs) and for the first time, the rest-frame UV was observed at high resolution ($R\sim2700$) with the G140H/F100LP grating (3.9 hrs).


\subsection{SPURS Observations and Reduction} \label{sec:spurs}

The SPURS NIRSpec \citep{Jakobsen2022,Boker2023} observations were conducted using the medium-resolution ($R\simeq1000$, corresponding to $\simeq300$~km~s$^{-1}$ per resolution element) grating/filter pairs G140M/F100LP, G235M/F170LP, and G395M/F290LP. 
We used the three-shutter nod pattern for dithering, which is appropriate for compact high redshift targets. 
The total on-target integration time is 29~hours, 7.9~hours, and 2.9~hours for G140M, G235M, and G395M, respectively.

We reduced the 2D NIRSpec spectra using the latest version of \texttt{msaexp}\footnote{\url{https://github.com/gbrammer/msaexp}} package \citep{Brammer2023} which is based on the standard JWST data reduction pipeline\footnote{\url{https://github.com/spacetelescope/jwst}} \citep{Bushouse2024}, following the procedures and setup described in \citet{deGraaff2025a} and \citet{Heintz2025}. 
We assumed a point source pathloss correction, motivated by the compact morphology of the four LRDs (Figure~\ref{fig:image}). 
The 1D spectra were extracted from the reduced 2D spectra using an optimal extraction \citep{Horne1986}. 
Throughout this reduction, we utilize the extended wavelength extraction available as part of \texttt{msaexp} \citep{Valentino2025} to achieve wavelength coverage beyond the nominal limits of each grating. 
This is made by picking up the full wavelength range covered by the detector, which may be contaminated by second-order spectra. 
However, we only focus on extended G395M spectra (up to $5.5\ \mu$m) to cover H$\alpha$ line at $z\simeq7$, which is not affected by higher-order contamination as contamination begins at wavelengths longer than $5.6\ \mu$m. 
Finally, we recalibrate the absolute flux of the reduced spectrum to match that of NIRCam by applying a second-order polynomial correction factor as a function of wavelength \citep{Chen2026b}. 
We derive this correction by fitting the ratio of the NIRCam photometry, where we adopt the CIRC1 photometry available from the JADES DR5 \citep[][for GN-2, GN-2004, and GN-29]{Robertson2026} and the CEERS photometric catalogs \citep[][for CEERS-7902]{Endsley2025}, to that synthesized from the coadded spectrum in broad NIRCam filters (from F115W to F444W).


\begin{figure*}
\includegraphics[width=\linewidth]{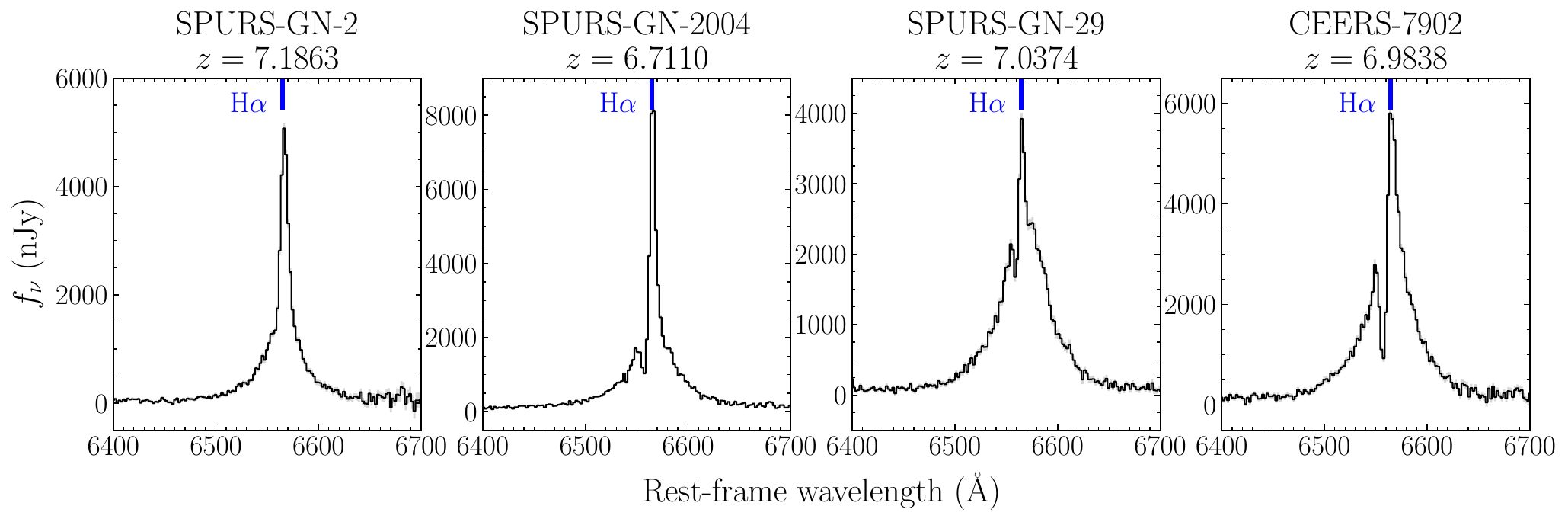}
\caption{H$\alpha$ emission lines in SPURS G395M spectra of the four LRDs. For each spectrum, the observed flux density and $1\sigma$ uncertainty are shown as black solid line and grey shaded region, respectively.}
\label{fig:ha_spec}
\end{figure*}

The G140M observations cover rest-frame wavelengths of $1220-2600$~\AA\ for GN2, $1250-3080$~\AA\ for GN-2004 (with a chip gap between $1460$ and $1600$~\AA), $1200-2110$~\AA\ for GN-29, $1210-2750$~\AA\ for CEERS-7902. 
We characterize the remainder of the near-UV ($2400-3500$~\AA) by stacking G140M and G235M.
For the G140M spectra, the median $3\sigma$ limiting flux for an unresolved emission line (in spectral direction) of a point source is $1.4\times10^{-19}$~erg~s$^{-1}$~cm$^{-2}$. 
Continua of all four LRDs are clearly detected in G140M spectra, with signal-to-noise ratio (S/N) of $5.9$ (GN-2), $10.3$ (GN-2004), $5.7$ (GN-29), and $3.5$ (CEERS-7902) per resolution element at rest-frame $1700$~\AA. 
These allow us to detect weak emission lines in the rest-frame UV with EWs of $1.3$~\AA, $0.7$~\AA, $1.2$~\AA, and $1.9$~\AA\ for GN-2, GN-2004, GN-29, and CEERS-7902, respectively. 
The median $3\sigma$ limiting flux of the G235M spectra is $1.7\times10^{-19}$~erg~s$^{-1}$~cm$^{-2}$. 
We also detect near-UV continua with S/N of $3.7$ (GN-2), $5.7$ (GN-2004), $4.3$ (GN-29), and $3.1$ (CEERS-7902) at rest-frame $2700$~\AA. 
These correspond to limiting EWs of $3.4$~\AA\ (GN-2), $2.0$~\AA\ (GN-2004), $2.9$~\AA\ (GN-29), and $4.1$~\AA\ (CEERS-7902). 
For the G395M spectra, the median $3\sigma$ limiting flux reaches $2.4\times10^{-19}$~erg~s$^{-1}$~cm$^{-2}$. 

For each source, we measure the emission lines as follows. 
For an emission line detected with S/N $>5$, we measure the line flux, centroid, width, and EW by fitting the line profile and nearby continuum with a Gaussian function. 
For lines that are close in wavelength or a line that shows multiple components, we fit the line profile with multiple Gaussians simultaneously. 
If a line is detected with lower S/N ($<5$), we compute the line fluxes by directly integrating the continuum-subtracted line profile. 
Finally, the uncertainties of line fluxes and EWs are evaluated as follows. 
We resample the flux densities of each spectrum $1000$ times by taking the observed value as the mean and the error as the standard deviation. 
Then we compute the line fluxes and EWs from the resampled spectra of each source using the same approach described above. 
We take the standard deviation of these measurements as the uncertainty. 

We derive the systemic redshifts ($z_{\rm sys}$) using strong forbidden oxygen emission lines in the rest-frame optical ([O~{\small III}]~$\lambda\lambda4959,5007$). 
For each LRD, we simultaneously fit the [O~{\small III}] doublet with Gaussian functions. 
We use the fitted line centers to compute redshifts, then average them weighted by the emission line signal-to-noise ratio to obtain the final systemic redshift. 
We obtain systemic redshifts of $z_{\rm sys}=7.1863$ for GN-2, $z_{\rm sys}=6.7110$ for GN-2004, $z_{\rm sys}=7.0374$ for GN-29, and $z_{\rm sys}=6.9838$ for CEERS-7902.

\subsection{Properties of SPURS LRDs} \label{sec:opt}

Before presenting the rest-frame UV spectra of the four LRDs, we situate the sample within the broader landscape of LRD properties established in recent compilations \citep{deGraaff2025c,Hviding2025,Kocevski2025}. 
Since the rest-frame UV spectra are presented in later sections, we focus here on the rest-frame optical spectra. 
Although several of the LRDs have prior observations from other programs (Section~\ref{sec:ancillary_data}), we adopt the SPURS grating spectra throughout for consistency with our rest-frame UV measurements.
We present the rest-frame optical spectroscopic properties of the four LRDs in Table~\ref{tab:opt_info}.
Full details of the rest-frame optical spectra are presented in Appendix~\ref{sec:full_opt}. 


\begin{deluxetable}{l|cccc}
\tablecaption{Continuum and optical spectroscopic properties of the four LRDs.}
\tablehead{
Properties & SPURS-GN-2 & SPURS-GN-2004 & SPURS-GN-29 & CEERS-7902 \\
}
\startdata
$\beta_{\rm FUV}$ & $-2.39\pm0.10$ & $-1.95\pm0.05$ & $-1.41\pm0.09$ & $-0.76\pm0.12$ \\
$\beta_{\rm NUV}$ & $-2.34\pm0.18$ & $-2.10\pm0.16$ & $-1.34\pm0.17$ & $-1.05\pm0.32$ \\
$\beta_{\rm opt}$ & $0.09\pm0.11$ & $0.06\pm0.09$ & $0.40\pm0.09$ & $0.48\pm0.08$ \\
$f_{\nu,4050}/f_{\nu,3670}$ & $1.39\pm0.39$ & $1.75\pm0.33$ & $1.90\pm0.34$ & $2.16\pm0.47$ \\
FWHM$_{\rm H\alpha,narrow}$ (km s$^{-1}$) & $353\pm9$ & $277\pm7$ & $276\pm8$ & $329\pm7$ \\
EW$_{\rm H\alpha,narrow}$ (\AA) & $504\pm18$ & $331\pm11$ & $156\pm33$ & $104\pm7$ \\
FWHM$^{\rm a}_{\rm H\alpha,broad}$ (km~s$^{-1}$) & $1395\pm101$ & $1505\pm55$ & $1639\pm90$ & $1275\pm61$ \\
 & $3523\pm228$ & $4336\pm140$ & $3664\pm154$ & $3679\pm97$ \\
EW$^{\rm b}_{\rm H\alpha,broad}$ (\AA) & $1363\pm127$ & $1120\pm52$ & $1466\pm116$ & $1266\pm58$ \\
FWHM$_{\rm H\alpha,abs}$ (km s$^{-1}$) & -- & $356\pm41$ & $258\pm85$ & $337\pm20$ \\
EW$_{\rm H\alpha,abs}$ (\AA) & -- & $-89\pm11$ & $-113\pm30$ & $-144\pm11$ \\
$\Delta v_{\rm H\alpha,abs}$ (km s$^{-1}$) & -- & $-293\pm50$ & $-140\pm48$ & $-317\pm48$ \\
$F_{\rm H\alpha,narrow}/F_{\rm H\beta,narrow}$ & $11\pm1$ & $5.9\pm0.5$ & $5.2\pm1.3$ & $7.2\pm1.0$ \\
$F_{\rm H\alpha,broad}/F_{\rm H\beta,broad}$ & $12\pm2$ & $10\pm1$ & $14\pm2$ & $11\pm1$ \\
O3 & $3.0\pm0.3$ & $7.6\pm0.5$ & $11.3\pm1.6$ & $9.0\pm1.1$ \\
O32 & $29\pm6$ & $36\pm3$ & $17\pm2$ & $16\pm3$ \\
Ne3O2 & $4.9\pm1.1$ & $2.0\pm0.2$ & $1.0\pm0.1$ & $1.7\pm0.4$ \\
\enddata
\tablecomments{H$\alpha$ properties are derived from Gaussian fits to the line profile, where the narrow H$\alpha$ emission is fitted by a Gaussian and the broad H$\alpha$ emission is described by the sum of two Gaussian functions for each LRD. More details of Gaussian fits and exponential fits are presented in Appendix~\ref{sec:balmer_fit}. \\ a: FWHMs of two broad Gaussian functions of each LRD. When adopting exponential fits, the broad FWHMs are $1180\pm95$~km~s$^{-1}$ (GN-2), $1690\pm74$~km~s$^{-1}$ (GN-2004), $1233\pm70$~km~s$^{-1}$ (GN-29), and $1485\pm59$~km~s$^{-1}$ (CEERS-7902). \\ b: EW of the sum of the two broad H$\alpha$ components of each LRD.}
\label{tab:opt_info}
\end{deluxetable}

We identify broad components of the Balmer lines in all four LRDs (Figure~\ref{fig:ha_spec}). 
We adopt two fitting techniques to recover the Balmer line profiles, one using a combination of a narrow (full width at half maximum FWHM $\lesssim500$~km~s$^{-1}$) and two broad Gaussian functions, and the other using an exponential function to capture the broad profile. 
When the Balmer line shows an absorption component, we fit the absorption line with an additional Gaussian. 
The methodology and fits are described in Appendix~\ref{sec:balmer_fit}, where we demonstrate that both methods are able to recover the Balmer profiles equally well. 
The results derived from triple Gaussian fits are presented in Table~\ref{tab:opt_info}. 
With these models, the broad component of H$\alpha$ of each LRD is described by the sum of a component with FWHM ranging from $3523$ to $4336$~km~s$^{-1}$ and an intermediate component (FWHM ranging from $1275$ to $1679$~km~s$^{-1}$).
For each LRD, the broad FWHMs of the two components span the range of those derived from single Gaussian fits to broad Balmer lines \citep[e.g.,][]{Greene2024,Matthee2024,Hviding2025,Kocevski2025}. 
When fitting the broad H$\alpha$ lines with an exponential function, we find narrower FWHMs ($1180-1690$~km~s$^{-1}$, Table~\ref{tab:ha_exp}), as is expected for electron scattering broadening scenarios \citep{Sneppen2026a}.
We also find absorption features in the H$\alpha$ lines of GN-2004, GN-29, and CEERS-7902, with no Balmer line absorption in GN-2 (Figure~\ref{fig:ha_spec}).
The H$\alpha$ absorption lines of the three LRDs are narrow (FWHM $=258-356$~km~s$^{-1}$) and blueshifted ($\Delta v=-317$ to $-140$~km~s$^{-1}$) with respect to the line center. 

We characterize the continuum of the four LRDs using our spectra, measuring the rest-frame optical and UV slopes.  
For the rest-frame optical continuum slope ($\beta_{\rm opt}$), we fit the spectra at rest-frame wavelengths of $3700-6300$~\AA\ after masking out the regions containing emission lines following \citet{deGraaff2025c,Hviding2025}. 
By definition, the four LRDs are all red in their rest-frame optical colors, but the continuum slopes span a wide range. 
GN-2 ($\beta_{\rm opt}=0.09\pm0.11$) and GN-2004 ($\beta_{\rm opt}=0.06\pm0.09$) are on the bluer side of the LRD color distribution (right panel of Figure~\ref{fig:demo}), both approaching the colors often seen in the LBD population \citep{Brazzini2026}.
GN-29 ($\beta_{\rm opt}=0.40\pm0.09$) and CEERS-7902 ($\beta_{\rm opt}=0.48\pm0.08$) are somewhat redder, yet we note LRD optical colors extend as red as $\beta_{\rm opt}=2.5$ (Figure~\ref{fig:demo}). 

We measure UV continuum slope in the far-UV (FUV) and in the near-UV (NUV), as different physical processes  may dominate in the different wavelength regimes. For example, damped Ly$\alpha$ absorption will contribute primarily to the FUV slopes, whereas the Balmer free-bound nebular continuum emission will contribute primarily to the NUV slopes. The far-UV slope is measured between $1400$~\AA\ and $2400$~\AA, and the near-UV slope is measured between $2400$~\AA\ and $3500$~\AA. 
We again mask out the regions containing emission lines or absorption lines when deriving the UV continuum slopes. 
The four LRDs also span a wide range of UV slopes. 
GN-2 ($\beta_{\rm FUV}=-2.39\pm0.10$) and GN-2004 ($\beta_{\rm FUV}=-1.95\pm0.05$) have very blue FUV slopes in the context of the full the LRD population (left panel of Figure~\ref{fig:demo}). 
GN-29 ($\beta_{\rm FUV}=-1.41\pm0.09$) and CEERS-7902 ($\beta_{\rm FUV}=-0.76\pm0.12$) have much redder  FUV continuum slopes. 
We find that the NUV continuum slope of each LRD is similar to its FUV slope. 
We derive blue NUV slopes in GN-2 ($\beta_{\rm NUV}=-2.34\pm0.18$) and GN-2004 ($\beta_{\rm NUV}=-2.10\pm0.16$), and red NUV slopes in GN-29 ($\beta_{\rm NUV}=-1.34\pm0.17$) and CEERS-7902 ($\beta_{\rm NUV}=-1.05\pm0.32$). 

We quantify the amplitude of the Balmer break as $f_{\nu,4050}/f_{\nu,3670}$, following recent literature studies \citep[e.g.,][]{deGraaff2025a}. 
GN-2 ($1.4\pm0.4$), GN-2004 ($1.8\pm0.3$), and GN-29 ($1.9\pm0.3$) show moderately strong breaks, comparable to the median for the full LRD population ($1.5$; \citealt{deGraaff2025c}). CEERS-7902 ($2.2\pm0.5$) shows the strongest break in our sample, though we note that LRD Balmer breaks can reach as large as $\simeq8$ \citep[e.g.,][]{deGraaff2025b,Naidu2026}.

We identify narrow components (FWHM $\simeq300$~km~s$^{-1}$) in the Balmer lines, and we find that the four LRDs in our sample exhibit different H$\alpha$ narrow-to-broad luminosity ratios. GN-2 and GN-2004 have comparatively strong narrow components relative to broad (broad-to-narrow ratios of $2.7$ and $3.4$), while GN-29 and CEERS-7902 are broad-dominated, with ratios of $9.4$ and $12$. If the narrow-to-broad flux ratio reflects the relative contribution of star-forming host versus AGN (or dense gas cocoon emission), GN-2 and GN-2004 should show correspondingly different UV line and continuum properties than GN-29 and CEERS-7902.

In all four LRDs, the narrow H$\alpha$/H$\beta$ flux ratios ($5.2-11$) depart significantly from expectations from case B recombination ($2.76$ for $T=2\times10^4$~K; \citealt{Osterbrock2006}). 
The narrow line spectra additionally constrain the ratio of [O~{\small III}]/[O~{\small II}] (O32) and [Ne~{\small III}]/[O~{\small II}] (Ne3O2), both of which are sensitive to the ionization parameter and electron density in the narrow line-emitting gas.
For the four LRDs, we measure elevated O32 ratios ($16-36$) and similarly large Ne3O2 ratios ($1.0-4.9$), comparable to the typical O32 ($\simeq25$) and Ne3O2 ($\simeq1$) ratios found in LRDs \citep[e.g.,][]{Nikopoulos2026}. We find that the two bluer LRDs, GN-2 and GN-2004, also have the largest O32 ($29-36$) and Ne3O2 ($2.0-4.9$) ratios in the sample, roughly a factor of $\sim2$ and up to $\sim5$ higher, respectively, than GN-29 and CEERS-7902 (O32 $=16-17$; Ne3O2 $=1.0-1.7$).

The gas-phase metallicity of the narrow line gas can be constrained from the narrow line flux ratios. The four LRDs have narrow [O~{\small III}]~$\lambda5007$/H$\beta$ (O3) flux ratios between $3.0$ and $11$, as expected for low to moderate metallicity ionized gas. For reference, these are much larger than the O3 ratio of the very low metallicity LRD Abell2744-QSO1 ($\simeq1$; \citealt{DEugenio2026,Tang2026b}) but consistent with the 
O3 values spanned by the full population of LRDs, where the $16-50-84$ percentile is $4.9-7.5-8.6$ \citep{Nikopoulos2026}.
In the SPURS rest-frame optical spectra of all the four LRDs, we detect the auroral [O~{\small III}]~$\lambda4363$ emission line (Figure~\ref{fig:opt_spec}) allowing a calculation of the electron temperature and metallicity via the direct method.
We characterize the temperatures and gas-phase oxygen abundances of the four LRDs following the methodology commonly used in the literature \citep[e.g.,][]{Isobe2025,Nikopoulos2026}. 
We use the \texttt{Python} package \texttt{PyNeb} \citep{Luridiana2015} and our methods are described in Appendix~\ref{sec:gas_method}. 
We derive high electron temperatures ($1.8-4.0\times10^4$~K) from the [O~{\small III}]$\lambda4363$/[O~{\small III}]$\lambda5007$ ratios (Table~\ref{tab:properties}). 
The direct method indicates the narrow line gas is moderately metal poor in all four LRDs ($12+\log{\rm (O/H)=7.21-8.13}$), corresponding to $Z=0.03-0.28\ Z_{\odot}$, assuming solar oxygen abundance $12+\log{\rm O/H}=8.69$ in \citealt{Asplund2021}). 
The inferred metallicities of our sample appear consistent with those derived for the narrow line gas in the majority of LRDs \citep[e.g.,][]{Isobe2025,Nikopoulos2026}.

The rest-frame optical spectrum also reveals a broad component in the [\ion{O}{3}]~$\lambda\lambda4959,5007$ profile in each of the LRDs, suggesting the presence of ionized outflow.
Following our methods in Appendix~\ref{sec:line_fit}, we fit simultaneously to the [\ion{O}{3}] doublet, considering both the single Gaussian fit and the double Gaussian fit with a narrow as well as a broad component.
We evaluate the goodness of fit using the Bayesian Information Criterion (BIC; \citealt{Schwarz1978,Liddle2007}).
We find the double Gaussian fit yields significantly better results than the single Gaussian fit in all four sources, with BIC differences of $\Delta$BIC $=$ BIC(double) $-$ BIC(single) $=-39$ to $-621$.
We present these double Gaussian fits in Figure~\ref{fig:oiii}.
We measure the largest width for the broad component in GN-2 (FWHM $=1627\pm181$\,km\,s$^{-1}$), which contributes $25\pm2\%$ of the total [\ion{O}{3}] line flux.
The other three sources have slightly smaller broad component FWHM ($1130-1285$\,km\,s$^{-1}$), which still contribute $13-21\%$ of the total line flux.


\begin{figure*}[t]
\includegraphics[width=\linewidth]{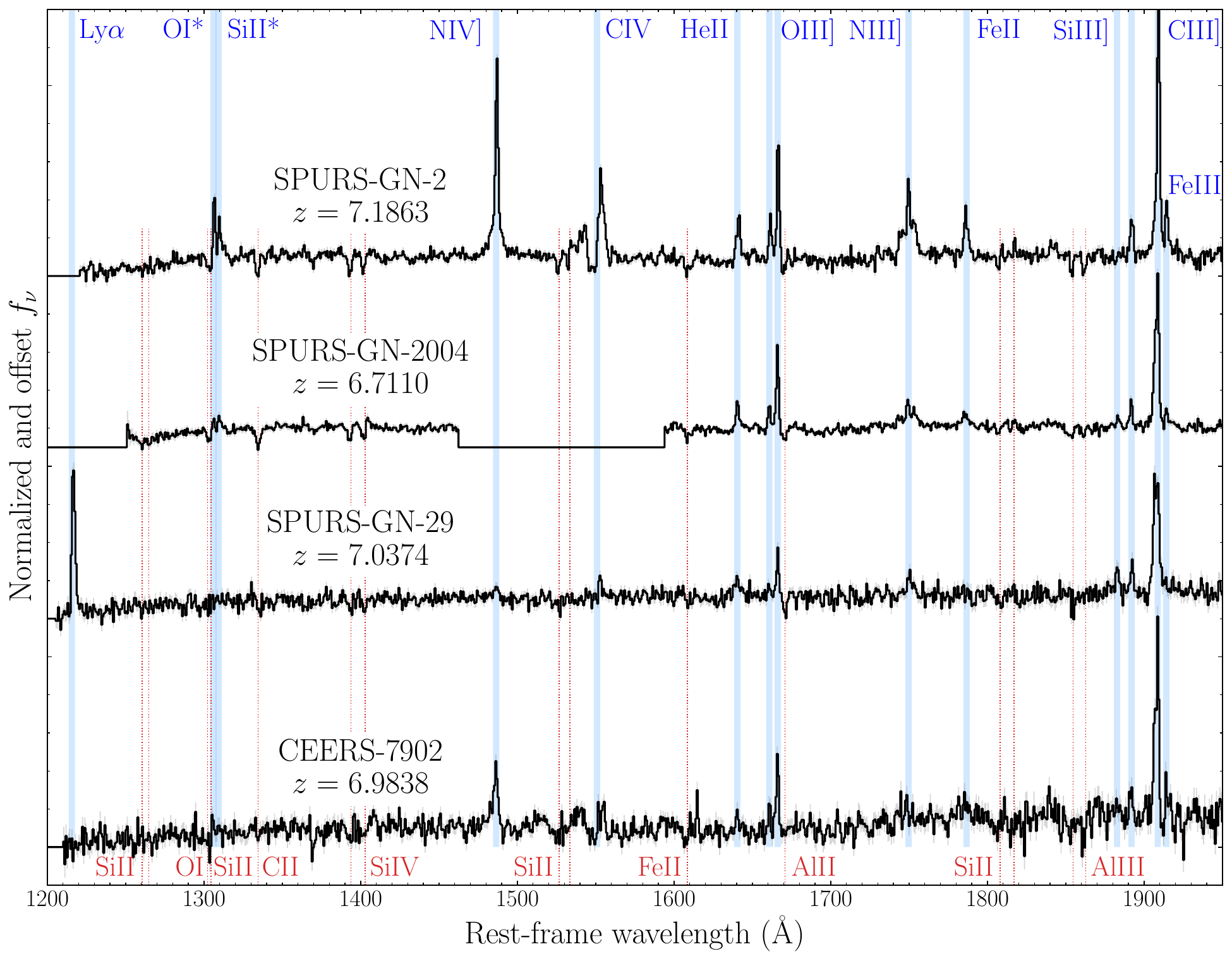}
\caption{SPURS G140M (rest-frame far-UV) spectra of GN-2, GN-2004, GN-29, and CEERS-7902. Observed fluxes are shown as black solid lines and $1\sigma$ uncertainties are shown as grey shaded regions. We highlight UV emission lines and interstellar absorption lines as blue shaded regions and red dotted lines, respectively.}
\label{fig:fuv_spec}
\end{figure*}

\subsection{NIRCam PSF Decomposition} \label{sec:psf}

While the rest-frame UV NIRCam images of all four LRDs appear to be dominated by a single compact object, we quantify whether there may be a contribution from an additional extended component that could correspond to the host galaxy. 
To do so, we use the NIRCam F150W images of GOODS-N and EGS reduced following the procedures described in \citet{Endsley2025}, together with the empirical point spread functions (PSFs) constructed from isolated stars in the same mosaics \citep{Endsley2025}. 
To decompose the point source from any possible extended component, we fit the light profile of each source with \texttt{pysersic} \citep{Pasha2023}, assuming both a point-source and a S\'ersic component whose positions, fluxes, and structural parameters are allowed to vary.
For SPURS-GN-2004, where the light profile is well fit with a single point-source component, we instead use the fit assuming a point source only.
We then subtract the fitted point-source component from the NIRCam images to obtain the PSF-subtracted images shown in the bottom panels of Figure~\ref{fig:image}. 

We find no evidence for extended components in GN-2 and GN-2004 ($<3\sigma$) and conclude the vast majority of their UV emission is unresolved. 
We detect extended components in GN-29 and CEERS-7902 and compute SNRs of 3.2 and 4.9 based on the residual flux measured with $r=0.15$\arcsec\ apertures centered at the point source positions (masking the central $3\times3$ pixels to exclude the PSF core), respectively.
However, these additional components comprise only $20\pm6\%$ (GN-29) and $9\pm2\%$ (CEERS-7902) of the total F150W flux, respectively.


\begin{deluxetable*}{l|ccc|ccc|ccc|ccc}
\setlength{\tabcolsep}{3pt}
\tablecaption{Rest-frame UV emission line fluxes ($10^{-19}$~erg~s$^{-1}$~cm$^{-2}$), EWs (\AA), and FWHMs (km~s$^{-1}$) of the four LRDs.}
\tablehead{
 & \multicolumn{3}{c|}{SPURS-GN-2} & \multicolumn{3}{c|}{SPURS-GN-2004} & \multicolumn{3}{c|}{SPURS-GN-29} & \multicolumn{3}{c}{CEERS-7902} \\
Line & Flux & EW & FWHM & Flux & EW & FWHM & Flux & EW & FWHM & Flux & EW & FWHM
}
\startdata
Ly$\alpha$ & -- & -- & -- & -- & -- & -- & $31.1\pm2.1$ & $26.4\pm1.8$ & $495\pm28$ & $<2.9$ & $<2.3$ & -- \\
N~{\scriptsize V}~$\lambda1239$ & $<1.4$ & $<2.8$ & -- & -- & -- & -- & $<2.1$ & $<1.8$ & -- & $<2.5$ & $<2.0$ & -- \\
N~{\scriptsize V}~$\lambda1243$ & $<1.5$ & $<2.8$ & -- & -- & -- & -- & $<2.0$ & $<1.7$ & -- & $<2.4$ & $<2.0$ & -- \\
O~{\scriptsize I*}~$\lambda1306$ & $8.9\pm0.5$ & $6.8\pm0.4$ & $353\pm30$ & $3.4\pm0.9$ & $1.3\pm0.4$ & $360\pm113$ & $<1.7$ & $<1.5$ & -- & $<2.2$ & $<1.8$ & -- \\
Si~{\scriptsize II*}~$\lambda1309$ & $6.7\pm0.6$ & $5.1\pm0.4$ & $416\pm78$ & $5.5\pm0.9$ & $2.0\pm0.3$ & $475\pm107$ & $<1.5$ & $<1.3$ & -- & $<2.6$ & $<2.2$ & -- \\
N~{\scriptsize IV}]$_{\rm total}$ & $33.3\pm0.8$ & $26.5\pm0.6$ & -- & -- & -- & -- & $2.1\pm0.7$ & $1.9\pm0.7$ & -- & $13.2\pm1.0$ & $11.5\pm0.9$ & -- \\
\hspace{1em}{[}N~{\scriptsize IV}]~$\lambda1483_{\rm narrow}$ & $<1.0$ & $<0.8$ & -- & -- & -- & -- & $<2.1$ & $<1.9$ & -- & $3.4\pm1.0$ & $2.9\pm0.8$ & $464\pm56$ \\
\hspace{1em}{[}N~{\scriptsize IV}]~$\lambda1483_{\rm broad}$ & $<4.6$ & $<3.7$ & -- & -- & -- & -- & -- & -- & -- & -- & -- & -- \\
\hspace{1em}N~{\scriptsize IV}]~$\lambda1486_{\rm narrow}$ & $18.2\pm0.9$ & $14.5\pm0.7$ & $357\pm13$ & -- & -- & -- & $2.1\pm1.1$ & $1.9\pm1.0$ & $524\pm228$ & $9.8\pm1.4$ & $8.6\pm1.2$ & $464\pm56$ \\
\hspace{1em}N~{\scriptsize IV}]~$\lambda1486_{\rm broad}$ & $15.1\pm2.4$ & $12.0\pm1.9$ & $1514\pm137$ & -- & -- & -- & -- & -- & -- & -- & -- & -- \\
C~{\scriptsize IV}~$\lambda1548^{\rm a}$ & $26.1\pm2.5$ & $25.6\pm2.5$ & -- & -- & -- & -- & $2.2\pm0.7$ & $2.1\pm0.6$ & $318\pm70$ & $7.9\pm3.8$ & $8.0\pm3.9$ & -- \\
He~{\scriptsize II}~$\lambda1640$ & $4.5\pm1.2$ & $5.6\pm1.5$ & $468\pm91$ & $6.2\pm1.5$ & $3.5\pm0.8$ & $424\pm67$ & $4.1\pm1.2$ & $4.1\pm1.2$ & $929\pm211$ & $1.6\pm0.9$ & $2.0\pm1.2$ & $291\pm127$ \\
O~{\scriptsize I}]~$\lambda1641$ & $1.8\pm0.7$ & $2.2\pm0.9$ & $340\pm16$ & $<3.0$ & $<1.6$ & -- & $<1.6$ & $<1.7$ & -- & $<2.4$ & $<2.9$ & -- \\
O~{\scriptsize III}]~$\lambda1661$ & $4.2\pm0.5$ & $5.3\pm0.6$ & $340\pm16$ & $5.2\pm0.7$ & $3.0\pm0.4$ & $334\pm17$ & $1.1\pm0.4$ & $1.1\pm0.4$ & $319\pm39$ & $2.1\pm0.6$ & $2.7\pm0.7$ & $286\pm33$ \\
O~{\scriptsize III}]~$\lambda1666$ & $11.3\pm0.7$ & $14.6\pm0.9$ & $340\pm16$ & $15.6\pm1.1$ & $9.0\pm0.7$ & $334\pm17$ & $3.9\pm0.6$ & $4.0\pm0.7$ & $319\pm39$ & $5.9\pm0.9$ & $7.4\pm1.2$ & $286\pm33$ \\
N~{\scriptsize III}]$_{\rm total}$ & $18.7\pm0.7$ & $21.8\pm0.8$ & -- & $15.2\pm0.9$ & $7.9\pm0.5$ & -- & $4.5\pm0.8$ & $5.0\pm0.8$ & -- & $3.9\pm1.0$ & $4.1\pm1.0$ & -- \\
\hspace{1em}N~{\scriptsize III}]~$\lambda1746-1754^{\rm b}$ & $5.4\pm0.2$ & $6.3\pm0.2$ & -- & $6.9\pm0.4$ & $3.6\pm0.2$ & -- & $4.5\pm0.8$ & $5.0\pm0.8$ & -- & $3.9\pm1.0$ & $4.1\pm1.0$ & -- \\
\hspace{1em}N~{\scriptsize III}]$_{\rm broad}$ & $13.3\pm3.0$ & $15.5\pm3.5$ & $1797\pm219$ & $8.3\pm4.0$ & $4.3\pm2.1$ & $2271\pm649$ & -- & -- & -- & -- & -- & -- \\
Fe~{\scriptsize II}~$\lambda1786$ & $5.4\pm0.5$ & $6.0\pm0.5$ & $381\pm60$ & $5.6\pm0.8$ & $3.1\pm0.4$ & $789\pm154$ & $<1.6$ & $<1.8$ & -- & $5.4\pm0.9$ & $5.5\pm1.0$ & $2189\pm621$ \\
Si~{\scriptsize III}]~$\lambda1883$ & $<0.9$ & $<1.2$ & -- & $1.8\pm0.6$ & $1.1\pm0.3$ & $302\pm47$ & $2.6\pm0.6$ & $3.1\pm0.7$ & $396\pm59$ & $<2.7$ & $<2.6$ & -- \\
Si~{\scriptsize III}]~$\lambda1892$ & $3.4\pm0.7$ & $4.7\pm1.0$ & $373\pm57$ & $4.1\pm0.9$ & $2.6\pm0.6$ & $302\pm47$ & $2.8\pm0.6$ & $3.3\pm0.7$ & $396\pm59$ & $2.5\pm1.2$ & $2.5\pm1.2$ & $410\pm140$ \\
C~{\scriptsize III}]$_{\rm total}$ & $29.5\pm0.7$ & $42.5\pm1.0$ & -- & $34.3\pm0.8$ & $22.5\pm0.5$ & -- & $17.7\pm0.7$ & $21.5\pm0.9$ & -- & $17.8\pm0.7$ & $17.8\pm0.7$ & -- \\
\hspace{1em}{[}C~{\scriptsize III}]~$\lambda1907_{\rm narrow}$ & $<1.4$ & $<2.0$ & -- & $11.6\pm0.8$ & $7.6\pm0.5$ & $312\pm12$ & $8.5\pm1.1$ & $10.3\pm1.3$ & $394\pm39$ & $6.5\pm0.7$ & $6.5\pm0.7$ & $243\pm13$ \\
\hspace{1em}{[}C~{\scriptsize III}]~$\lambda1907_{\rm broad}$ & $<2.7$ & $<3.9$ & -- & -- & -- & -- & -- & -- & -- & -- & -- & -- \\
\hspace{1em}C~{\scriptsize III}]~$\lambda1909_{\rm narrow}$ & $17.8\pm0.4$ & $25.7\pm0.6$ & $305\pm10$ & $22.7\pm1.2$ & $14.9\pm0.8$ & $312\pm12$ & $9.2\pm1.1$ & $11.2\pm1.3$ & $394\pm39$ & $11.3\pm0.9$ & $11.3\pm0.9$ & $243\pm13$ \\
\hspace{1em}C~{\scriptsize III}]~$\lambda1909_{\rm broad}$ & $9.8\pm2.6$ & $14.1\pm3.7$ & $2271\pm371$ & -- & -- & -- & -- & -- & -- & -- & -- & -- \\
{[}Ne~{\scriptsize IV}]~$\lambda2422$ & $<1.2$ & $<1.9$ & -- & $<1.5$ & $<1.4$ & -- & $<1.6$ & $<2.5$ & -- & $<2.5$ & $<3.7$ & -- \\
{[}Ne~{\scriptsize IV}]~$\lambda2424$ & $<1.2$ & $<1.8$ & -- & $<1.6$ & $<1.5$ & -- & $<1.4$ & $<2.3$ & -- & $<2.3$ & $<3.4$ & -- \\
Mg~{\scriptsize II}~$\lambda2796$ & $2.1\pm0.4$ & $4.3\pm0.9$ & $262\pm65$ & $<1.6$ & $<1.6$ & -- & $<1.6$ & $<2.9$ & -- & $<2.3$ & $<3.2$ & -- \\
Mg~{\scriptsize II}~$\lambda2803$ & $<1.3$ & $<2.7$ & -- & $<1.6$ & $<1.6$ & -- & $<1.7$ & $<3.0$ & -- & $<2.7$ & $<3.7$ & -- \\
{[}Ne~{\scriptsize V}]~$\lambda3427$ & $<1.2$ & $<4.6$ & -- & $<1.4$ & $<2.5$ & -- & $<1.6$ & $<4.2$ & -- & $<2.8$ & $<5.8$ & -- \\
\enddata
\tablecomments{For non-detections, $3\sigma$ upper limits of line fluxes and EWs are provided. \\ a: Integrated line flux and EW of C~{\scriptsize IV} doublet. \\ b: Integrated line flux and EW of narrow components of N~{\scriptsize III}] quintuplet.}
\label{tab:uv_lines}
\end{deluxetable*}

To constrain the size of the unresolved component itself, we repeat the \textsc{pysersic} decomposition with the point source replaced by a PSF-convolved exponential profile ($n=1$ S\'ersic), while retaining the second S\'ersic component, if present, so that the extended light (e.g.\ the neighboring component of CEERS-7902) does not bias the size of the compact component.
The recovered effective radii of the compact component are $<80$~pc (0.5~pixel) in all four sources, which is below the nominal PSF F150W FWHM of $\simeq1.6$~pixel listed in JDox, as expected for PSF-convolved fits at high S/N.
We therefore conclude that the rest-frame UV emission in the NIRCam images, and thus in our observed spectra, primarily arises from the central point source ($r_e \lesssim 80$~pc) in these LRDs.

\section{Rest-frame UV Emission Lines} \label{sec:uv_emi}

The SPURS observations provide robust constraints on faint rest-frame UV emission lines across all four LRDs. In this section, we describe the UV emission line properties and their implications for the physical conditions of the LRDs. Figure~\ref{fig:fuv_spec} shows the far-UV spectra of the sample, and we show the UV emission line measurements in Table~\ref{tab:uv_lines}. We will present more detailed views of individual features as they arise in the discussion below.


\begin{figure*}
\includegraphics[width=\linewidth]{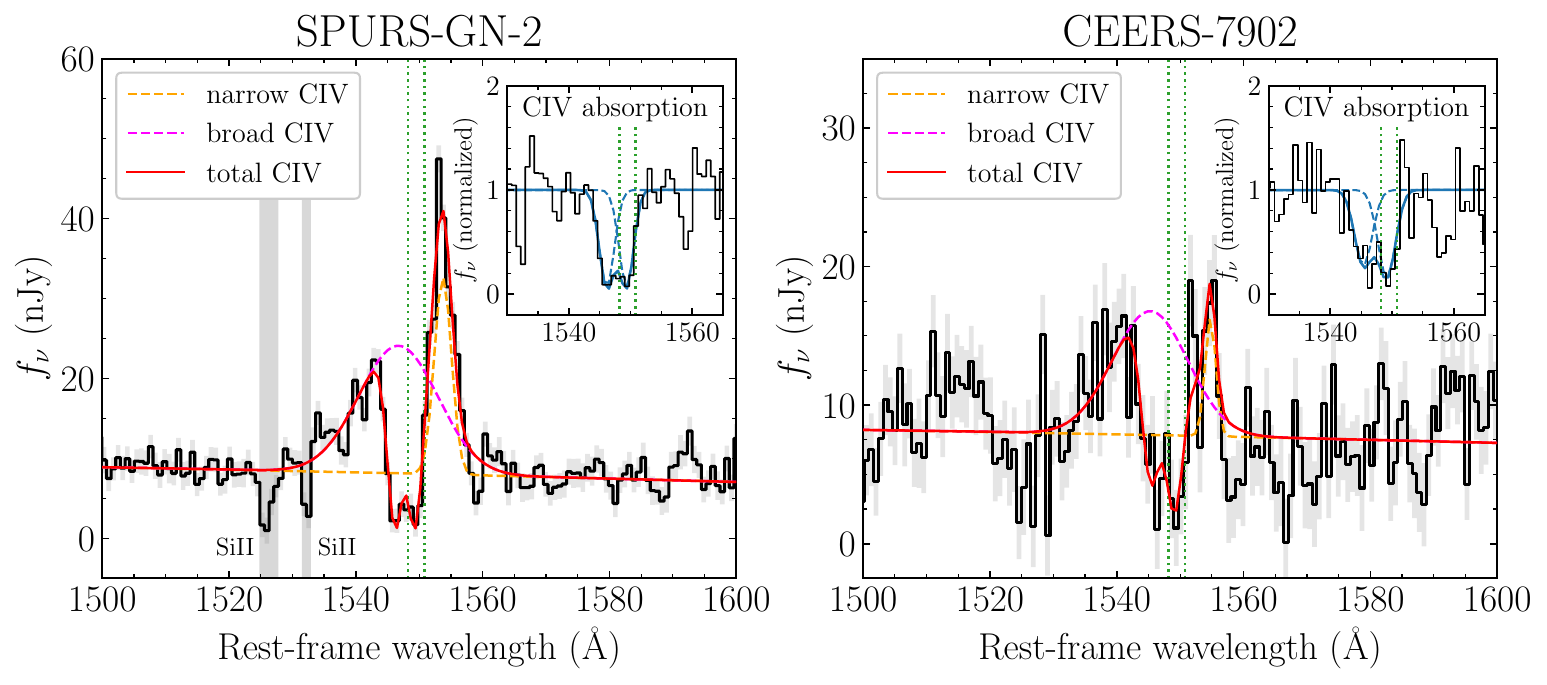}
\caption{Fits to the C~{\scriptsize IV} profiles of GN-2 (left) and CEERS-7902 (right). Each C~{\scriptsize IV} profile can be fitted by a combination of four Gaussians: a broad emission (magenta dashed line), a narrow emission (orange dashed line), and two absorption components. Each absorption component is shown as the blue dashed line in the upper right corner of each panel, with the sum of the two shown as a blue solid line. The total C~{\scriptsize IV} model profile of each object is shown as a red solid line. Green dotted lines show the line centers of C~{\scriptsize IV} doublet. For GN-2, we mask out the regions containing Si~{\scriptsize II} absorption lines.}
\label{fig:civ}
\end{figure*}


\begin{deluxetable*}{l|cccc|cccc}
\tablecaption{C~{\scriptsize IV} profile fitting results of GN-2 and CEERS-7902. For each object, we list the line flux ($10^{-19}$~erg~s$^{-1}$~cm$^{-2}$), EW (\AA), FWHM (km~s$^{-1}$), and velocity offset (km~s$^{-1}$) of each component.}
\tablehead{
 & \multicolumn{4}{c|}{SPURS-GN-2} & \multicolumn{4}{c}{CEERS-7902} \\
Line & Flux & EW & FWHM & $\Delta v$ & Flux & EW & FWHM & $\Delta v$
}
\startdata
C~{\scriptsize IV}$^{\rm a}_{\rm narrow}$ & $10.4\pm1.2$ & $10.5\pm1.2$ & $540\pm57$ & $+568\pm71$ & $2.3\pm1.4$ & $2.4\pm1.5$ & $389\pm189$ & $+417\pm73$ \\
C~{\scriptsize IV}$^{\rm b}_{\rm broad}$ & $31.3\pm2.5$ & $30.7\pm2.5$ & $2830\pm191$ & $-293\pm71$ & $16.7\pm3.8$ & $17.0\pm3.9$ & $2695\pm507$ & $-571\pm73$ \\
C~{\scriptsize IV}~$\lambda1548_{\rm abs}$ & $-8.0\pm1.4$ & $-7.9\pm1.3$ & $513\pm64$ & $-353\pm39$ & $-4.7\pm1.5$ & $-4.8\pm1.5$ & $570\pm114$ & $-497\pm46$ \\
C~{\scriptsize IV}~$\lambda1550_{\rm abs}$ & $-7.6\pm1.3$ & $-7.4\pm1.3$ & $513\pm64$ & $-353\pm39$ & $-6.3\pm1.5$ & $-6.4\pm1.5$ & $570\pm114$ & $-497\pm46$ \\
\enddata
\tablecomments{a: The velocity offset of the narrow emission peak is calculated relative to the line center of C~{\scriptsize IV}~$\lambda1550$. \\ b: The velocity offset of the broad emission peak is calculated relative to the line center of C~{\scriptsize IV}~$\lambda1548$.}
\label{tab:civ}
\end{deluxetable*}

\subsection{C~{\small IV} Emission and Absorption} \label{sec:c4}

The C~{\small IV}~$\lambda\lambda1548,1550$ resonant profile presents a mixture of emission and absorption requiring moderate resolution spectroscopy to disentangle. 
Three of the four SPURS LRDs (GN-2, GN-29, CEERS-7902) have C~{\small IV} spectral coverage. 
We present the C~{\small IV} profiles of GN-2 and CEERS-7902 in Figure~\ref{fig:civ} and discuss each of the three LRDs with C~{\small IV} observations below.

GN-2 provides our most detailed view of the C~{\small IV}~$\lambda\lambda1548,1550$ profile. 
The most prominent feature is a narrow emission line (FWHM $=540\pm57$~km~s$^{-1}$) centered at a velocity of $+568\pm71$~km~s$^{-1}$ relative to the C~{\small IV}~$\lambda1550$ centroid. 
Such redshifted emission is common in  C~{\small IV} profiles owing to resonant scattering through an outflowing medium \citep[e.g.,][]{Berg2019,Senchyna2022}. 
There is also an absorption trough spanning $1400$~km~s$^{-1}$, between rest-frame wavelengths of $\sim1544$~\AA\ and $1550$~\AA. 
The absorption lines are blueshifted, with a velocity offset of $-353\pm39$~km~s$^{-1}$ with respect to the line center of each component of the doublet. 
This is consistent with there being significant resonant scattering of C~{\small IV} photons from an outflowing highly ionized medium. 
At shorter wavelengths ($1529-1543$~\AA), we see a blue wing of emission, corresponding to a broad component with a FWHM of $2830\pm191$~km~s$^{-1}$.
Overall, the C~{\small IV} profile in GN-2 is well fit by a combination of Gaussian functions reproducing the absorption profile and the two emission profiles (left panel of Figure~\ref{fig:civ}). 
The best-fit parameters are presented in Table~\ref{tab:civ}. The fit suggests that the broad component is slightly blueshifted ($-293\pm71$~km~s$^{-1}$) with respect to line center of C~{\small IV}~$\lambda1548$.

\begin{figure}
\includegraphics[width=\linewidth]{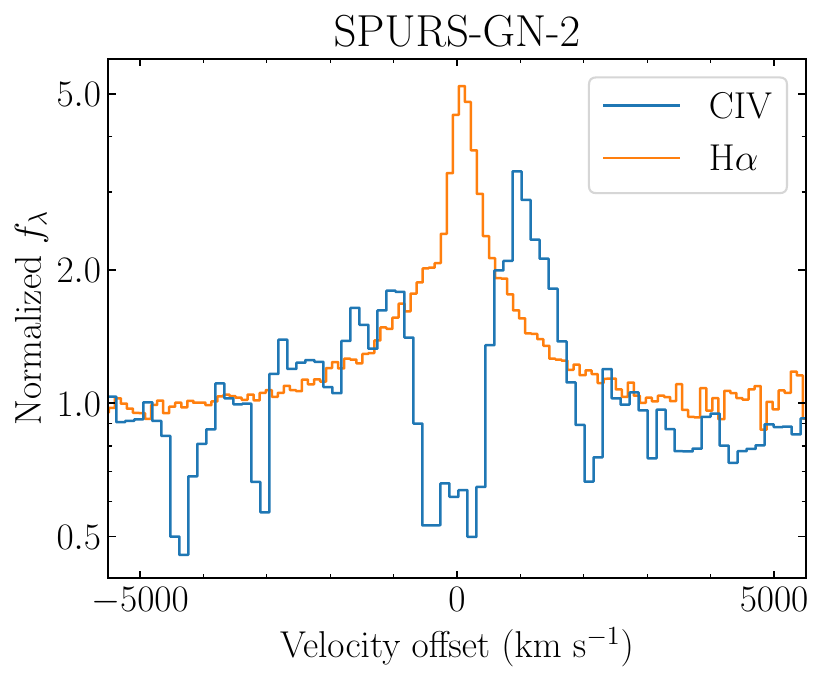}
\caption{Comparison between broad C~{\scriptsize IV} (blue) and broad H$\alpha$ (orange) emission line profiles of GN-2. Flux densities are normalized to the peak flux of each broad component.}
\label{fig:civ_ha}
\end{figure}

We compare the broad C~{\small IV} profile of GN-2 to the broad Balmer emission line profile. The broad C~{\small IV} line width appears qualitatively similar to that of H$\alpha$ (Figure~\ref{fig:civ_ha}). The FWHM of broad C~{\small IV} ($2830$~km~s$^{-1}$) is comparable to broad H$\alpha$, which is fit as the combination of two broad Gaussians with FWHM of $1395$ and $3523$~km~s$^{-1}$ (Section~\ref{sec:opt}).
Using the fit described above, we find that the total broad C~{\small IV}/H$\beta$ flux ratio is $0.71\pm0.09$. 
However, the C~{\small IV} absorption trough removes a significant component of the broad C~{\small IV} line. 
As a result, the observed broad C~{\small IV}/H$\beta$ flux ratio is $0.36\pm0.07$. 
Both derived flux ratios are considerably lower than those of typical BL AGNs ($\sim3$; e.g., \citealt{Francis1991,Brotherton2001,VandenBerk2001}). 
So while we find evidence of a broad C~{\small IV} profile in GN-2, it appears to be significantly attenuated relative to the general population of BL AGNs.

The C~{\small IV} profile of CEERS-7902 (right panel of Figure~\ref{fig:civ}) resembles that of GN-2. It has a broad blue emission wing, an absorption trough, and narrow redshifted emission. We fit this profile with the same model used for the C~{\small IV} line of GN-2. The broad emission component has a FWHM of $2695\pm507$~km~s$^{-1}$ and is centered $571\pm73$~km~s$^{-1}$ blueward of C~{\small IV}~$\lambda1548$. The narrow emission component is redshifted by $+417\pm73$~km~s$^{-1}$ relative to C~{\small IV}~$\lambda1550$. Both absorption components are blueshifted by $-497\pm46$~km~s$^{-1}$ relative to the line center of their respective doublet component. 
The observed broad C~{\small IV}/H$\beta$ flux ratio of CEERS-7902 is $0.04\pm0.02$, a factor of $9$ smaller than in GN-2. Excluding the absorption component increases the broad C~{\small IV} flux, raising the ratio to $0.13\pm0.03$. Even in this case, the ratio is considerably lower than typical for BL AGNs.

The C~{\small IV} emission is weakest in GN-29 (EW $=2.1\pm0.6$~\AA). 
We detect a narrow (FWHM $=318\pm70$~km~s$^{-1}$) emission line close to the red component of the doublet, with a peak flux redshifted by $+396\pm72$~km~s$^{-1}$ relative to the line center. 
We do not detect any emission line close to the blue component, placing a $3\sigma$ limiting C~{\small IV}~$\lambda1548$/C~{\small IV}~$\lambda1550$ flux ratio of $<0.9$. 
This is below the theoretical value ($2$; e.g., \citealt{Flower1979}), as expected from resonant scattering through outflowing gas.
We do not detect a broad C~{\small IV} component in GN-29. If the observed broad-to-narrow C~{\small IV} flux ratio matches that of GN-2, the expected broad-line flux is $2\times10^{-20}$~erg~s$^{-1}$~cm$^{-2}$ with absorption. Assuming broad C~{\small IV} shares the FWHM of the broad Balmer lines, our $3\sigma$ limiting flux is $3.6\times10^{-19}$~erg~s$^{-1}$~cm$^{-2}$. This expected flux falls below our detection limit, so we cannot rule out broad C~{\small IV} in GN-29 if it has a broad-to-narrow flux ratio similar to GN-2.

In summary, our results suggest that broad C~{\small IV} profiles are present in a subset of LRDs, with line widths comparable to the broad Balmer lines. We detect broad C~{\small IV} most clearly in GN-2, the optically bluest LRD in our sample. The redder LRD CEERS-7902 also shows broad C~{\small IV}, but with a considerably smaller broad C~{\small IV}/H$\beta$ flux ratio. The similarity between the C~{\small IV} and H$\alpha$ broad-line widths hints at a common broadening origin for the two lines. If so, this suggests that broad C~{\small IV} is at least partially transmitted in a subset of LRDs, perhaps indicating non-uniform covering of dense neutral gas. While our sample is small, we note that we find the largest broad C~{\small IV}/H$\beta$ ratio in GN-2, the LRD with the bluest rest-frame optical colors.


\begin{figure*}
\centering
\includegraphics[width=0.95\linewidth]{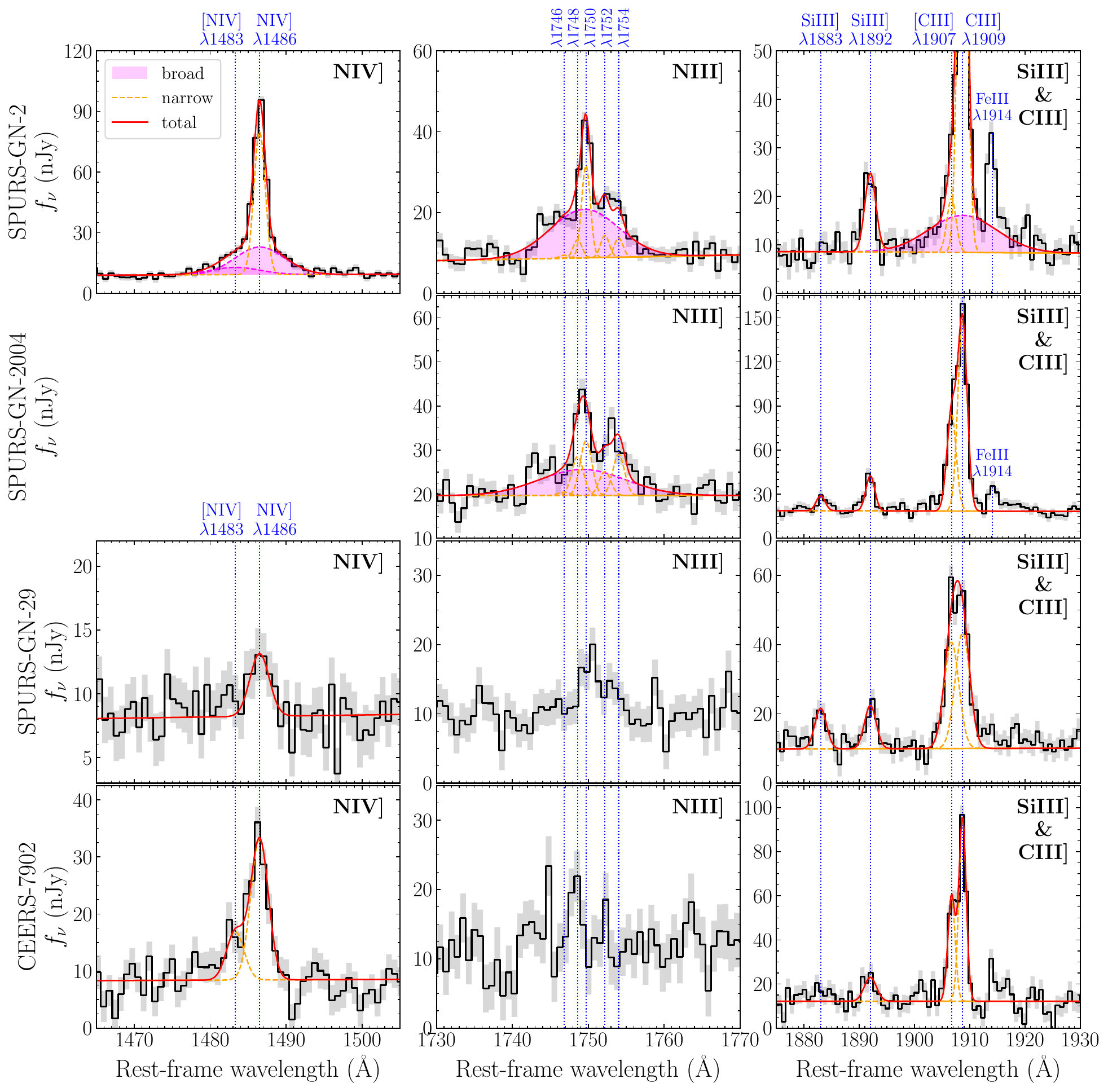}
\caption{Density-sensitive UV emission lines (N~{\scriptsize IV}], N~{\scriptsize III}], Si~{\scriptsize III}], C~{\scriptsize III}]) of the four LRDs. We overplot the best-fit line profile to each spectrum: the narrow (FWHM $<500$~km~s$^{-1}$) components are shown as orange dashed lines, the broad components are shown as magenta shaded regions, and the sum of the best-fit profile is shown as a red solid line. Because the N~{\scriptsize III}] lines of GN-29 and CEERS-7902 have relatively low S/N ($4-6$), we do not resolve the individual components of their N~{\scriptsize III}] quintuplets.}
\label{fig:fuv_lines}
\end{figure*}

\subsection{C~{\small III}], Si~{\small III}], O~{\small III}] Emission} \label{sec:c3si3o3}

The rest-frame UV hosts a suite of emission lines sensitive to the physical conditions of the ionized gas. The SPURS observations cover the [C~{\small III}], C~{\small III}]~$\lambda\lambda1907,1909$ (hereafter C~{\small III}]) and Si~{\small III}]~$\lambda\lambda1883,1892$ (hereafter Si~{\small III}]) doublets, both of which are sensitive to electron density. The spectra also probe the auroral O~{\small III}]~$\lambda\lambda1661,1666$ emission lines, which can be combined with [O~{\small III}]~$\lambda5007$ to constrain the electron temperature.    

We first describe the density-sensitive lines, focusing on GN-2, the optically bluest LRD in our sample. We showed in Section~\ref{sec:opt} that this LRD has the largest narrow-to-broad ratio for the Balmer lines, and in Section~\ref{sec:c4} that it has the most prominent broad C~{\small IV}. We detect strong narrow emission from C~{\small III}], centered on the red ($\lambda1909$) component of the doublet. Despite the very strong C~{\small III}]~$\lambda1909$, we detect no [C~{\small III}]~$\lambda1907$ emission. Since the critical density of [C~{\small III}]~$\lambda1907$ ($7\times10^4$~cm$^{-3}$) is lower than that of C~{\small III}]~$\lambda1909$ ($1\times10^9$~cm$^{-3}$), this suggests the narrow-line gas in GN-2 may reach extremely high densities. The Si~{\small III}] doublet suggests a similar picture: we detect the red component ($\lambda1892$), with no emission seen in the blue $\lambda1883$ component. The critical density of Si~{\small III}]~$\lambda1883$ ($4\times10^4$~cm$^{-3}$) is lower than that of Si~{\small III}]~$\lambda1892$ ($3\times10^{10}$~cm$^{-3}$), so this is again consistent with large electron densities. We will quantify the implied densities below. Finally, we note an additional emission line at rest-frame $1914$~\AA, consistent with the expected position of Fe~{\small III}~$\lambda1914$.

To characterize the doublet line ratios, we model the C~{\small III}] and Si~{\small III}] line profiles in GN-2 by fitting multiple Gaussians, with the centroids fixed to the rest-frame wavelengths of each doublet. We first attempt to fit C~{\small III}] with two narrow lines. We find that there is excess emission on the blue and red side of the doublet which we attribute to a broad C~{\small III}] component (Figure~\ref{fig:fuv_lines}). To capture the broad C~{\small III}] emission, we re-fit the line profile with two narrow Gaussians and two broad Gaussians, each with centroids fixed to the rest-frame wavelengths of the doublet components. We mask the Fe~{\small III}~$\lambda1914$ line during the fit. We use the Bayesian Information Criterion to quantify whether the addition of the broad components is preferred, as is common in the literature \citep[e.g.,][]{DEugenio2025,Chen2026b,Matthee2026}.  
Details of the fits are presented in Appendix~\ref{sec:line_fit}. We do find the model with narrow and broad components is strongly preferred. 
This model well reproduces the observed C~{\small III}] profile (Figure~\ref{fig:fuv_lines}). 
The fit to the narrow C~{\small III}]~$\lambda1909$ implies a large rest-frame equivalent width (EW $=25.7\pm0.6$~\AA), and the non-detection of [C~{\small III}]~$\lambda1907$ suggests a large C~{\small III}]/[C~{\small III}] flux ratio ($>13$ at $3\sigma$). The broad component is centered on the C~{\small III}]~$\lambda1909$ component with a FWHM of $2271\pm371$~km~s$^{-1}$. The total EW of the broad component is $14\pm4$~\AA, comprising $35\%$ of the total C~{\small III}] flux. 
We do not find the broad [C~{\small III}]~$\lambda1907$ component, placing a $3\sigma$ limiting broad C~{\small III}]/[C~{\small III}] flux ratio of $>3.6$. We additionally fit the Si~{\small III}] doublet with two Gaussians. In this case, we do not have the S/N to detect broad components. Our flux measurements 
indicate a large Si~{\small III}]~$\lambda1892$/$\lambda1883$ ratio ($>3.7$ at $3\sigma$).

\begin{figure*}
\includegraphics[width=\linewidth]{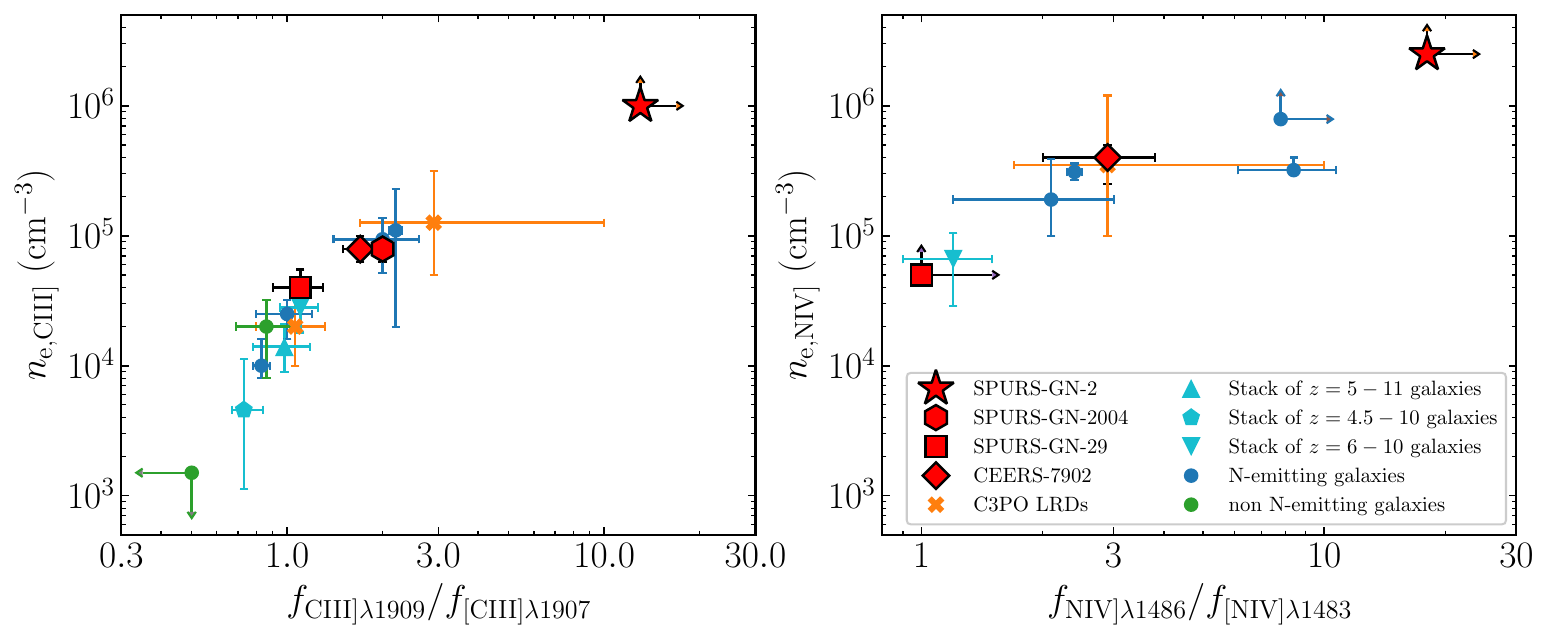}
\caption{Electron densities ($n_{\rm e}$) of four LRDs (GN-2: red star; GN-2004: red hexagon; GN-29: red square; CEERS-7902: red diamond) inferred from C~{\scriptsize III}]~$\lambda1909$/[C~{\scriptsize III}]~$\lambda1907$ (left) and N~{\scriptsize IV}]~$\lambda1486$/[N~{\scriptsize IV}]~$\lambda1483$ (right) flux ratios. For comparison, we overplot C3PO LRDs \citep[][orange cross symbols]{Papovich2026}, average densities inferred from composite spectra of galaxies ($z=5-11$, \citealt{Topping2025a}, cyan triangle; $z=4.5-10$, \citealt{Umeda2026}, cyan pentagon; $z=6-10$, \citealt{SinghRai2026}, cyan upside down triangle), individual N~{\scriptsize IV}] or N~{\scriptsize III}] emitters (blue circles; \citealt{Topping2024,Topping2025a,Chen2026b}) and non nitrogen emitters (green circles; \citealt{Topping2025a}).}
\label{fig:density}
\end{figure*}

We compute electron densities of GN-2 from the doublet ratios using \texttt{PyNeb}. Details of the density calculations are provided in Appendix~\ref{sec:gas_method}. Reproducing the C~{\small III}]/[C~{\small III}] flux ratios requires very high electron densities in GN-2, suggesting a lower bound of $>1.0\times10^6$~cm$^{-3}$ at $3\sigma$. We find that the Si~{\small III}] ratio is also consistent with very large densities ($>5.0\times10^5$~cm$^{-3}$). The C~{\small III}] density of GN-2 is much higher than that probed by C~{\small III}] in other star-forming galaxies (left panel of Figure~\ref{fig:density}; e.g., \citealt{Topping2025b,SinghRai2026,Umeda2026}). One possibility is that we are seeing something akin to Godzilla in the Sunburst Arc, a lensed, compact source where C~{\small III}] and Si~{\small III}] similarly imply densities of $\gtrsim10^6$~cm$^{-3}$ \citep{Vanzella2020,Pascale2024,Choe2025}, comparable to the dense Weigelt blobs surrounding the luminous blue variable $\eta$ Carinae ($\sim 10^7-10^8$~cm$^{-3}$; \citealt{Davidson1995}). Another possibility is that we are seeing narrow emission from gas clouds with densities approaching those commonly found in the broad-line region ($\gtrsim10^9$~cm$^{-3}$).


\begin{deluxetable*}{lcccc}
\tablecaption{Ionized gas properties inferred from narrow emission lines of the four LRDs.}
\tablehead{
Property & SPURS-GN-2 & SPURS-GN-2004 & SPURS-GN-29 & CEERS-7902
}
\startdata
$n_{\rm e}$(N~{\scriptsize IV}])/cm$^{-3}$ & $>2.5\times10^6$ & -- & $>2.5\times10^4$ & $4.0^{+1.0}_{-1.5}\times10^5$ \\
$n_{\rm e}$(C~{\scriptsize III}])/cm$^{-3}$ & $>1.0\times10^6$ & $7.9^{+2.1}_{-1.6}\times10^4$ & $4.0^{+1.5}_{-1.0}\times10^4$ & $7.9^{+2.1}_{-1.6}\times10^4$ \\
$n_{\rm e}$(Si~{\scriptsize III}])/cm$^{-3}$ & $>5.0\times10^5$ & $7.9^{+7.9}_{-2.9}\times10^4$ & $2.0^{+2.0}_{-1.2}\times10^4$ & $>2.5\times10^4$ \\
$T_{\rm e}$(O$^{2+}$)/K & $3.5^{+0.1}_{-0.2}\times10^4$ & $2.0^{+0.2}_{-0.1}\times10^4$ & $1.8^{+0.2}_{-0.1}\times10^4$ & $4.0^{+1.1}_{-0.9}\times10^4$ \\
$12+\log{\rm (O/H)}$ & $7.21^{+0.04}_{-0.04}$ & $7.52^{+0.12}_{-0.10}$ & $8.13^{+0.10}_{-0.10}$ & $7.41^{+0.10}_{-0.05}$ \\
$Z/Z_{\odot}$ & $0.03^{+0.01}_{-0.01}$ & $0.07^{+0.02}_{-0.01}$ & $0.28^{+0.08}_{-0.06}$ & $0.05^{+0.02}_{-0.01}$ \\
N/O & $0.35^{+0.03}_{-0.02}$ & $0.14^{+0.01}_{-0.02}$ & $0.25^{+0.06}_{-0.06}$ & $0.49^{+0.11}_{-0.09}$ \\
(N/O)/(N/O)$_{\odot}$ & $2.5^{+0.3}_{-0.1}$ & $1.0^{+0.1}_{-0.1}$ & $1.8^{+0.4}_{-0.4}$ & $3.5^{+0.8}_{-0.6}$ \\
N/C & $2.3^{+0.1}_{-0.1}$ & $0.9^{+0.1}_{-0.1}$ & $0.7^{+0.2}_{-0.1}$ & $1.7^{+0.3}_{-0.3}$ \\
(N/C)/(N/C)$_{\odot}$ & $9.9^{+0.5}_{-0.6}$ & $4.0^{+0.4}_{-0.3}$ & $3.2^{+0.6}_{-0.6}$ & $7.1^{+1.3}_{-1.1}$ \\
\enddata
\tablecomments{The O$^{2+}$ temperatures provided are derived using the less dust-sensitive [O~{\scriptsize III}]~$\lambda4363$/[O~{\scriptsize III}]~$\lambda5007$ flux ratios. Using the O~{\scriptsize III}]~$\lambda1666$/[O~{\scriptsize III}]~$\lambda5007$ flux ratios (uncorrected for dust), we derive O$^{2+}$ temperatures of $2.9^{+0.1}_{-0.1}\times10^4$~K (GN-2), $1.8^{+0.1}_{-0.1}\times10^4$~K (GN-2004), $1.2^{+0.1}_{-0.1}\times10^4$~K (GN-29), and $1.6^{+0.1}_{-0.1}\times10^4$~K (CEERS-7902).}
\label{tab:properties}
\end{deluxetable*}

We also detect the C~{\small III}] doublets in the other three LRDs (Figure~\ref{fig:fuv_lines}), each of which reveals large integrated EWs ($18-23$~\AA). For reference, the median C~{\small III}] EW in composite spectra of star forming galaxies at $z\simeq7$ is $\simeq8$~\AA\ \citep[e.g.,][]{Roberts-Borsani2024,SinghRai2026,Tang2026a}, well below what is found in the LRDs in our sample. We fit the C~{\small III}] line profiles using the same approach as we applied to GN-2. For these three LRDs, we do not find evidence for a broad C~{\small III}] component. Our fits constrain the C~{\small III}]/[C~{\small III}] flux ratios: $2.0\pm0.2$ in GN-2004, $1.1\pm0.2$ in GN-29, $1.7\pm0.2$ in CEERS-7902, none of which reach the extremely large ratio seen in GN-2 ($>13$). 
The three LRDs also reveal detection of the Si~{\small III}] doublet. 
We measure Si~{\small III}]~$\lambda1892$/Si~{\small III}]~$\lambda1883$ flux ratios of $2.3\pm0.9$ in GN-2004, $1.1\pm0.3$ in GN-29, and $>0.9$ ($3\sigma$) in CEERS-7902. 
Following the same techniques we used for GN-2, we derive electron densities from the doublet ratios. The C~{\small III}]-based densities are $7.9^{+2.1}_{-1.6}\times10^4$~cm$^{-3}$ (GN-2004), $4.0^{+1.5}_{-1.0}\times10^4$~cm$^{-3}$ (GN-29), and $7.9^{+2.1}_{-1.6}\times10^4$~cm$^{-3}$ (CEERS-7902).
The Si~{\small III}] flux ratios reveal a consistent picture, with derived electron densities of $7.9^{+7.9}_{-2.9}\times10^4$~cm$^{-3}$ (GN-2004), $2.0^{+2.0}_{-1.2}\times10^4$~cm$^{-3}$ (GN-29), and $>2.5\times10^4$~cm$^{-3}$ (CEERS-7902). 
The electron densities of these LRDs are not as large as GN-2, but they are still elevated with respect to the typical C~{\small III}] density of $z\sim7$ galaxies ($\simeq1\times10^4$~cm$^{-3}$, left panel of Figure~\ref{fig:density}; e.g., \citealt{Topping2025b,SinghRai2026,Umeda2026}). These results suggest that the physical conditions of the narrow line gas in LRDs are distinct from star forming galaxies at similar redshifts. 

The O~{\small III}]~$\lambda\lambda1661,1666$ doublet allows us to calculate the electron temperatures. 
We recover the doublet in all four LRDs and measure $\lambda1661:\lambda1666$ ratios of $0.3-0.4$, consistent with expectations from theoretical transition probabilities ($0.4$; \citealt{FroeseFischer1985}). We measure electron temperatures from the ratio of the O~{\small III}]~$\lambda1666$ and [O~{\small III}]~$\lambda5007$ fluxes, assuming the densities derived from C~{\small III}]/[C~{\small III}] flux ratios. The temperatures are derived using \texttt{PyNeb} (see the Appendix~\ref{sec:gas_method} for more details). In our initial measurements, we do not apply dust correction for the O~{\small III}]~$\lambda1666$/[O~{\small III}]~$\lambda5007$ flux ratio. For GN-2 and GN-2004, the temperatures derived from O~{\small III}] ($1.8-2.9\times10^4$~K) are consistent with that inferred from [O~{\small III}]~$\lambda4363$ ($2.0-3.5\times10^4$~K). For GN-29 and CEERS-7902, the temperatures derived based on O~{\small III}] ($1.2-1.6\times10^4$~K) are lower than those inferred from [O~{\small III}]~$\lambda4363$ ($1.8-4.0\times10^4$~K). Since these are the two reddest LRDs in our sample, we suggest that one possible explanation is dust.  
We can estimate the reddening that would be required to bring the O~{\small III}]-based temperature in line with the [O~{\small III}]~$\lambda4363$ temperature.
To do so, we calculate the O~{\small III}]~$\lambda1666$ flux that we would have expected using the [O~{\small III}]~$\lambda4363$-based temperature. The comparison of this expected value with that which we observe allows us to estimate the dust attenuation required to make the two measurements consistent. 
Assuming the Small Magellanic Cloud (SMC) extinction law of \citet{Gordon2003}, we find that only modest dust attenuation is required for GN-29 ($A_V=0.46$) and CEERS-7902 ($A_V=0.75$). 
Adopting the flatter extinction law of \citet{Cardelli1989}, however, yields substantially larger attenuation values of $A_V=0.99$ and $1.62$ for GN-29 and CEERS-7902, respectively.


\subsection{N~{\small IV}] and N~{\small III}] Emission} \label{sec:n4n3}

The [N~{\small IV}], N~{\small IV}]~$\lambda\lambda1483,1486$ doublet (hereafter N~{\small IV}]) has been detected in a subset of the galaxies and AGN observed by JWST. Three of the LRDs in our sample (GN-2, GN-29, CEERS-7902) have spectra that cover N~{\small IV}]. All four LRD spectra cover the lower-ionization N~{\small III}] quintuplet, comprising five emission lines in the range $1746-1754$~\AA.

We first consider the N~{\small IV}] complex in GN-2, given its distinct spectral properties within our sample (see Section~\ref{sec:opt}, \ref{sec:c4}, and \ref{sec:c3si3o3}). N~{\small IV}] emission is well detected in its spectrum  (Figure~\ref{fig:fuv_spec}). The narrow emission is centered on N~{\small IV}]~$\lambda1486$, with no corresponding feature at [N~{\small IV}]~$\lambda1483$. A broad N~{\small IV}] component is also present. We fit the N~{\small IV}] profile of GN-2 with two narrow and two broad Gaussians (see Appendix~\ref{sec:line_fit} for details). The resulting model reproduces the profile well (Figure~\ref{fig:fuv_lines}). The narrow N~{\small IV}]~$\lambda1486$ emission is very strong (EW $=14.5\pm0.7$~\AA), and the non-detection of the blue component indicates a large N~{\small IV}]/[N~{\small IV}] flux ratio ($>18$ at $3\sigma$). 
The broad N~{\small IV}] component in GN-2 has FWHM $=1514\pm137$~km~s$^{-1}$ comprising $45\%$ of the total N~{\small IV}] flux and is centered on N~{\small IV}]~$\lambda1486$. 

We note that similar broad N~{\small IV}] components were recently discovered in several nitrogen emitters that are not LRDs, including the luminous galaxy GN-z11 \citep{Chen2026b}. The origin of this broad component has been debated, with dense Wolf-Rayet (WR)-like winds from very massive stars and AGN-driven outflows both proposed as possible explanations. Our detection of a comparable broad N~{\small IV}] component in GN-2 extends this phenomenon to the LRD population, possibly lending support to an AGN-driven origin for the broad N~{\small IV}] emission, at least in this source.

The N~{\small IV}]/[N~{\small IV}] doublet ratio provides an additional constraint on the electron density in GN-2. We compute the density using \texttt{PyNeb}, following the methods described in Appendix~\ref{sec:gas_method}. The flux ratio of the narrow lines ($>18$ at $3\sigma$) points to an extremely large density ($>2.5\times10^6$~cm$^{-3}$) for the nitrogen-emitting gas. While UV nitrogen lines commonly trace large densities \citep{Topping2024,Topping2025a,Chen2026b,SinghRai2026}, this measurement places GN-2 as the highest density yet seen in a narrow N~{\small IV}] emitter (right panel of Figure~\ref{fig:density}). Along with the C~{\small III}] and Si~{\small III}] constraints in Section~\ref{sec:c3si3o3}, this result builds on the picture that the narrow rest-frame UV lines in GN-2 emerge from extremely dense gas that has almost never before been seen in typical star forming galaxies. 

We also detect N~{\small IV}] in CEERS-7902 and GN-29. The line properties are measured in the same manner as for GN-2, but in these cases we do not recover broad profiles at the current S/N. 
CEERS-7902 presents N~{\small IV}] with an integrated EW of $11.5\pm0.9$~\AA\ and a large N~{\small IV}]/[N~{\small IV}] ratio ($2.9\pm0.9$). The N~{\small IV}] in GN-29 is weaker and has a lower S/N. We measure an integrated EW $=1.9\pm0.7$~\AA. 
The emission is centered on the red component, but with our current sensitivity limits, we can only place a rough lower bound on the N~{\small IV}]/[N~{\small IV}] ratio ($>1.0$ at $3\sigma$). 
We compute N~{\small IV}]-based electron densities from the doublet ratios. The results reveal a large density in $4.0^{+1.0}_{-1.5}\times10^5$~cm$^{-3}$ in CEERS-7902. In GN-29, the limit does not strongly constrain the density  ($>2.5\times10^4$~cm$^{-3}$). While the N~{\small IV}] density in CEERS-7902 is not as large as that in GN-2, it is larger than the N~{\small IV}]-based density derived from the composite spectrum of $z=6-10$ galaxies ($7\times10^4$~cm$^{-3}$, \citealt{SinghRai2026}; right panel of Figure~\ref{fig:density}).

We detect N~{\small III}] emission in all four LRDs (Figure~\ref{fig:fuv_spec}). 
GN-2 and GN-2004 have high S/N ($27$ for GN-2, $17$ for GN-2004) allowing us to model individual lines within the N~{\small III}]~$\lambda1746-1754$ quintuplet. 
As discussed in Appendix~\ref{sec:line_fit}, we also find evidence for a broad N~{\small III}] component in these two objects. We model the observed profiles with five narrow Gaussians and one broad Gaussian. 
Following \citet{Chen2026b}, the five narrow components are centered at the rest-frame wavelengths of the quintuplet lines, share a common line width, and have fixed flux ratios of $f_{\rm NIII]\lambda1746}/f_{\rm NIII]\lambda1752}=0.14$ and $f_{\rm NIII]\lambda1748}/f_{\rm NIII]\lambda1754}=0.95$, corresponding to their theoretical values. The centroid and width of the broad component are left free. 

We show the resulting model of the N~{\small III}] quintuplet in Figure~\ref{fig:fuv_lines} and line measurements in Table~\ref{tab:uv_lines}. As described in Appendix~\ref{sec:line_fit}, the fits prefer a broad N~{\small III}] component in GN-2 (FWHM $=1797\pm219$~km~s$^{-1}$) and GN-2004 (FWHM $=2271\pm649$~km~s$^{-1}$). This suggests a fast-moving component of nitrogen that includes both N~{\small IV}] and N~{\small III}]. One complication in interpreting the broad N~{\small III}] profile is the potential presence of Fe~{\small II}~$\lambda1743$ emission. To test the potential impact of Fe~{\small II} on our fits, we simultaneously fit the N~{\small III}] profile with five narrow Gaussians and an additional Gaussian for the Fe~{\small II}~$\lambda1743$ component. We are able to recover the N~{\small III}] line profile, provided the Fe~{\small II}~$\lambda1743$ component is very strong, with a flux of $3.9\pm0.9\times10^{-19}$ (GN-2) and $2.6\pm0.9\times10^{-19}$~erg~s$^{-1}$~cm$^{-2}$ (GN-2004). In this scenario, the N~{\small III}] broad profile would be weaker. However, we consider this unlikely: the implied Fe~{\small II}~$\lambda1743$ emission would be well above the values expected given the measured flux of Fe~{\small II}~$\lambda1786$ and the typical $\lambda1743/\lambda1786$ flux ratios ($\sim0.01$ based on the ratio between Einstein A coefficients). We thus conclude the broad N~{\small III}] component is the most likely explanation for the observed profile, but we caution that there remains some uncertainty in its interpretation. Using the narrow line decompositions of the quintuplet, we can constrain the electron density of the N~{\small III}]-emitting gas. However, in practice we find that the line ratios do not unambiguously constrain the density, and as a result we focus our analysis on densities derived from the doublet ratios discussed above.

Finally, we characterize the N~{\small III}] emission in the other two LRDs (GN-29, CEERS-7902). These sources, the two redder LRDs in our sample, show weak N~{\small III}] emission with S/N $=4-6$. The sensitivity of these spectra is not sufficient to meaningfully characterize the presence of broad components. 
The integrated N~{\small III}] EWs are $5.0\pm0.8$~\AA\ (GN-29) and $4.1\pm1.0$~\AA\ (CEERS-7902), comparable to those derived for the narrow lines in GN-2 ($6.3\pm0.2$~\AA) and GN-2004 ($3.6\pm0.2$~\AA). 

The detections of N~{\small IV}] and N~{\small III}] together with the other rest-frame UV emission lines allow us to infer the relative nitrogen abundances in the four LRDs. 
We derive N/O and N/C ratios following the procedures described in \citet{Martinez2025}. Here we summarize the results, but our detailed methodology is described in Appendix~\ref{sec:gas_method}.
We find that the four LRDs have nitrogen-enhanced abundance patterns, with N/O $=0.35^{+0.03}_{-0.02}$ (GN-2), $0.14^{+0.01}_{-0.02}$ (GN-2004), $0.25^{+0.06}_{-0.06}$ (GN-29), and $0.49^{+0.11}_{-0.09}$ (CEERS-7902). 
These correspond to super-solar N/O ratios ($1.0-3.5$~N/O$_{\odot}$). We also derive large N/C ratios, $2.3^{+0.1}_{-0.1}$ (GN-2), $0.9^{+0.1}_{-0.1}$ (GN-2004), $0.7^{+0.2}_{-0.1}$ (GN-29), and $1.7^{+0.3}_{-0.3}$ (CEERS-7902), corresponding to $3.2-9.9$~N/C$_{\odot}$. 
The relative abundances are comparable to the N~{\small IV}] emitters in the galaxy population at high redshift \citep[e.g.,][]{Castellano2024,Schaerer2024,Senchyna2024,Topping2024,Topping2025a,Navarro-Carrera2025}. This suggests some similarity between the LRDs and the broader class of nitrogen emitters. We will come back to generalize this result in Section~\ref{sec:N_line}.

\subsection{Mg~{\small II} Emission} \label{sec:mg2}

Mg~{\small II}~$\lambda\lambda2796,2803$ is a resonant doublet which often appears broad in the spectra of AGN. 
In AGN, it arises in the low-ionization broad-line region alongside the broad Balmer series emission lines \citep[e.g.,][]{Netzer1980,VandenBerk2001}. 
Additionally, we may expect to detect narrow Mg~{\small II} from narrow line regions in the LRDs. As Mg~{\small II} can be resonantly scattered by dense cool $T\sim10^4$~K gas, the emergent profile and doublet ratio of narrow lines are sensitive to gas kinematics, dust attenuation and column density \citep[e.g.,][]{Erb2012,Henry2018,Chisholm2020}.

In GN-2, we detect a narrow Mg~{\small II}~$\lambda2796$ emission line along with an absorption feature. 
The peak flux of Mg~{\small II}~$\lambda2796$ is redshifted by $+300\pm66$~km~s$^{-1}$, while the centroid of Mg~{\small II} absorption is blueshifted by $-264\pm74$~km~s$^{-1}$ with respect to the line center of Mg~{\small II}~$\lambda2796$, consistent with a P-Cygni profile. 
We measure a Mg~{\small II}~$\lambda2796$ EW of $4.3\pm0.9$~\AA. 
We do not detect the Mg~{\small II}~$\lambda2803$ component, placing a $3\sigma$ lower limit of $>1.5$ to the Mg~{\small II}~$\lambda2796$/Mg~{\small II}~$\lambda2803$ flux ratio. 
A P-Cygni profile indicates continuum radiation is scattered by outflowing, optically thick, Mg~{\small II} gas \citep{Prochaska2011,Xu2023a,Chang2024}. 

We do not detect broad Mg~{\small II} in any of the four LRDs. Assuming the broad Mg~{\small II} FWHM is the same as the broad H$\beta$ and integrating over $2\times$ FWHM$_{\rm broad}$, we place $3\sigma$ limits on the broad Mg~{\small II}/H$\beta$ flux ratio for the four LRDs ($<0.06-0.10$). These are far below those of typical BL AGNs ($\simeq1.5$; e.g., \citealt{Francis1991,Brotherton2001,VandenBerk2001}). The absence of broad Mg~{\small II} may reflect a combination of factors, reviewed in \citet{Tang2026b}. In particular, owing to the frequency dependence of the photoionization cross-section, Mg~{\small II} photons are more likely to be absorbed than C~{\small IV} photons as they pass through a dense layer of $n=2$ hydrogen. This wavelength-dependent opacity offers a natural explanation for why broad C~{\small IV} can be partially transmitted through a sightline in which broad Mg~{\small II} is fully suppressed.

\begin{figure}
\includegraphics[width=\columnwidth]{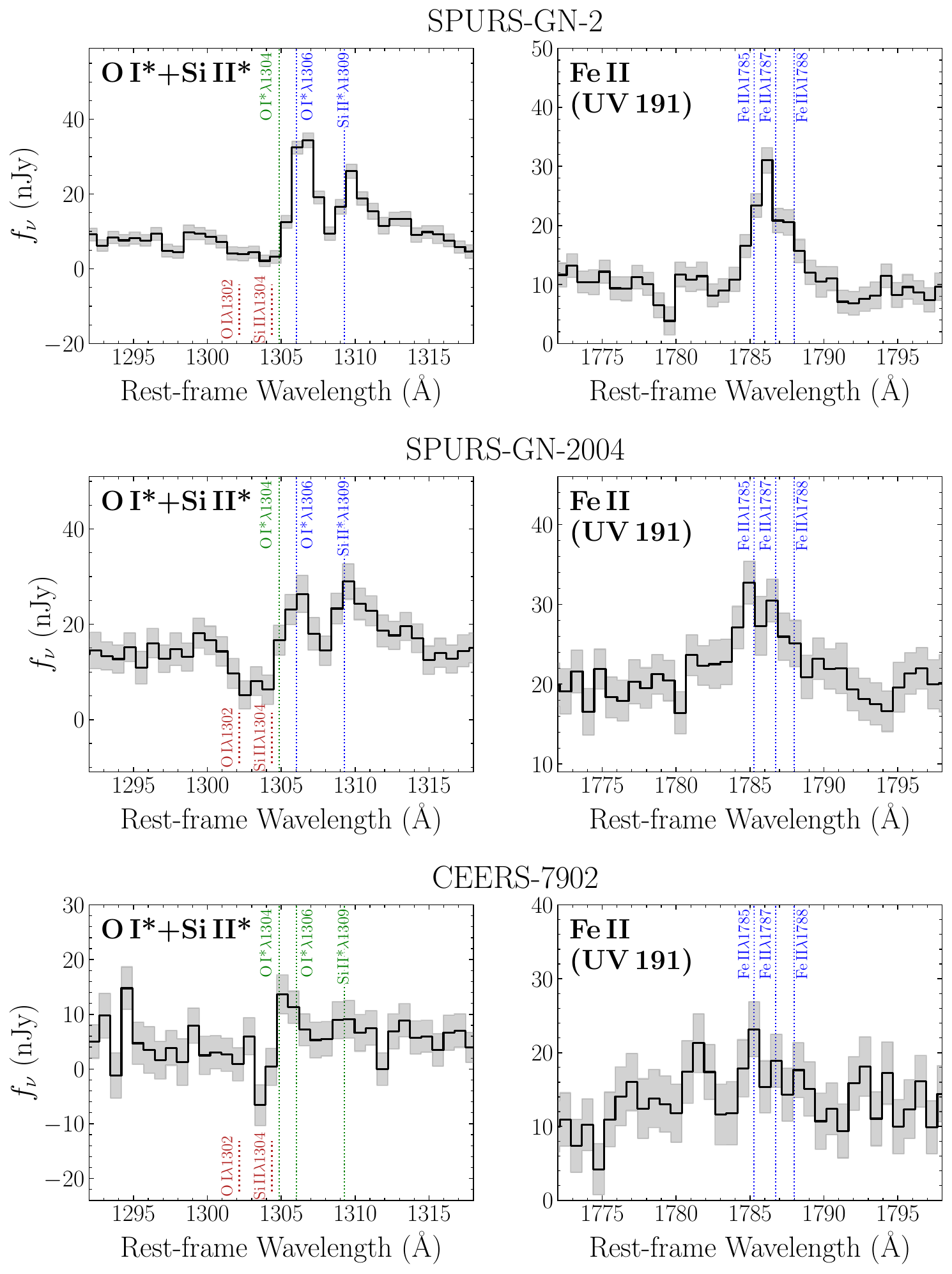}
\caption{O~{\scriptsize I}*~$\lambda1306$+Si~{\scriptsize II}*~$\lambda1309$ (left panels), and Fe~{\scriptsize II} UV 191 (right panels) fluorescent emission lines of GN-2, GN-2004, and CEERS-7902. We mark each detected emission line in blue, undetected emission lines in green, and interstellar absorption in red.}
\label{fig:fluorescent_lines}
\end{figure}

\begin{figure*}
\includegraphics[width=\linewidth]{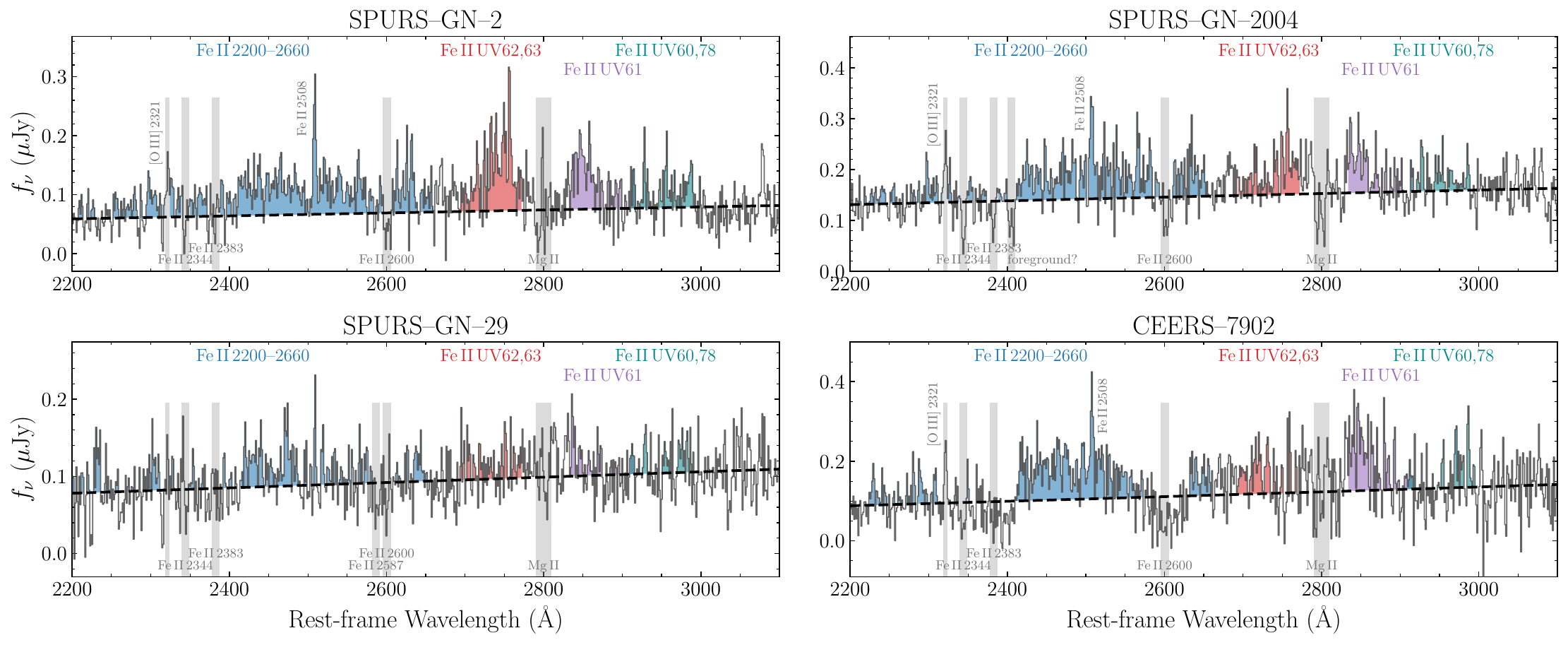}
\caption{The NUV \ion{Fe}{2} complexes of the four LRDs. The colored shaded regions show the emission components associated with the unresolved \ion{Fe}{2}~$\lambda2200-2600$ blend, and the \ion{Fe}{2}~UV62,63, \ion{Fe}{2}~UV61, and \ion{Fe}{2}~UV60,78 complexes. The black dashed line shows the continuum model from our \texttt{fantasy} fit.
The gray shaded bands mark individual emission lines and absorption features seen in our spectra.}
\label{fig:nuv_feii}
\end{figure*}

\subsection{O~{\small I}, Fe~{\small II}, and Si~{\small II} Emission} \label{sec:low_ion}

The LRDs also show a suite of low-ionization UV emission lines that provide evidence for dense neutral gas absorbing UV photons. 
We detect both \ion{O}{1} and \ion{Fe}{2} emission lines, that are increasingly detected in LRDs \citep[e.g.,][]{Labbe2024,DEugenio2025,Tripodi2025a,Tang2026b,Torralba2026b}, and report the first detections of \ion{Si}{2} fine-structure emission in LRDs.
We show the detected lines in Figure~\ref{fig:fluorescent_lines} and list their derived flux, EW, and FWHM in Table~\ref{tab:uv_lines}.

We detect strong \ion{O}{1} emission in two of the four LRDs.
GN-2 and GN-2004 both show a \ion{O}{1}*~$\lambda1306$ fine-structure emission line (EW~$=6.8\pm0.4$~\AA\ and EW~$=1.3\pm0.4$~\AA\ respectively).
In GN-2 we also clearly detect the rare semi-forbidden line \ion{O}{1}]~$\lambda1641$ (EW~$=2.2\pm0.9$~\AA), which we resolve from \ion{He}{2}~$\lambda1640$ emission.
\ion{O}{1}]~$\lambda1641$ emission has not, to our knowledge, been previously reported in star-forming galaxies, but is seen in symbiotic star outflows \citep{Shore2010} and has been detected in the Godzilla cluster in the Sunburst Arc \citep{Pascale2024,Choe2025}, and implies extremely high \ion{O}{1} column densities.
Both \ion{O}{1}*~$\lambda1306$ and \ion{O}{1}]~$\lambda1641$ arise from decays from the $3s\,^3S^0$ state, which is populated by continuum absorption at 1302~\AA\ by oxygen atoms in the ground state, and boosted by Ly$\beta$ fluorescence of the $3d\,^3D^0$ excited state -- requiring high \ion{H}{1} column densities to trap Ly$\beta$ \citep[e.g.,][]{Kwan1981,Matsuoka2007,Shore2010}.
To estimate the implied column density, we consider the conditions required to detect strong \ion{O}{1}]~$\lambda1641$ emission.
The relative rates of spontaneous emission in the two channels differ by $A_{1306}/A_{1641}\sim10^5$, 
thus our detection of $\lambda1641$ emission at comparable strength to $\lambda1306$ emission implies a very high optical depth in the ground state to suppress \ion{O}{1}~$\lambda1302,1304,1306$ emission: $\tau \gtrsim 10^5$, consistent with the strong absorption we detect at 1302~\AA\ (see Section~\ref{sec:abs}).
Conservatively assuming a Doppler parameter $b=10$~km~s$^{-1}$ this implies a ground state OI column density $N_{\rm OI}\gtrsim3\times10^{18}\,{\rm cm}^{-2}$.
Assuming solar abundance ratios and $0.03-0.28\,Z_\odot$ metallicity (see Section~\ref{sec:opt}) suggests an extremely high \ion{H}{1} column density around the Ly$\beta$-emitting region: $N_{\rm HI}\gtrsim 10^{22.5-23.0}\,{\rm cm}^{-2}$.

In GN-2 and GN-2004 we also detect strong \ion{Si}{2}*~$\lambda1309$ fine-structure emission (EW~$=5.1\pm0.4$~\AA\ and $2.0\pm0.3$~\AA, respectively). 
This line typically arises from pumping by the UV continuum: photons absorbed by the \ion{Si}{2}~$\lambda1304$ resonance transition preferentially decay to the fine-structure level rather than back to the ground state, producing $\lambda1309$ emission.
In the optically thick limit, we expect the emission EW to be equal to the absorption EW, however in star-forming galaxies the emission EW is typically weaker, likely due to scattering out of slit spectrographs into an extended halo and absorption by dust \citep[e.g.,][]{Shapley2003,Jones2012}.
Strikingly, in GN2, the \ion{Si}{2}*~$\lambda1309$ emission EW is stronger than the resonant line absorption (see Section~\ref{sec:abs}): suggesting low dust opacity and an additional excitation source.
One possibility is that photons emitted as \ion{O}{1}*~$\lambda1304$ are absorbed by the ground state \ion{Si}{2}~$\lambda1304$.
As the lines are separated by only 100~km~s$^{-1}$, moderately broad \ion{O}{1} emission could thus boost the excitation of \ion{Si}{2}.

We detect permitted \ion{Fe}{2} emission in both the FUV and NUV of the LRD spectra.
In the UV, GN-2, GN-2004 and CEERS-7902 all show strong emission in the \ion{Fe}{2} UV 191 multiplet at $1785,1787,1788$~\AA.
The total multiplet flux \ion{Fe}{2}~$\lambda1786$/broad H$\beta$ ratio is $0.11\pm0.02$, $0.08\pm0.01$, and $0.07\pm0.01$ for GN-2, GN-2004, and CEERS-7902, respectively. These ratios are $2-4\times$ larger than the typical type~I AGN value of $0.03$ \citep{VandenBerk2001}. 
\ion{Fe}{2}~$\lambda1786$ emission can be enhanced by Ly$\alpha$ fluorescence \citep{Johansson1984}, which may explain the high \ion{Fe}{2}~$\lambda1786$/broad H$\beta$ ratio.
In GN-2 and GN-2004, we also detect \ion{Fe}{3}~$\lambda1914$ emission, which can also be pumped by Ly$\alpha$ \citep{Johansson2000}. This line has previously been reported in $\eta$ Carinae \citep{Johansson2000} and the Sunburst arc \citep{Vanzella2020}.
Similarly to the fluorescence excitation channel for \ion{O}{1} emission described above, pumping these transitions requires a high \ion{H}{1} column density to trap Ly$\alpha$ photons \citep[e.g.,][]{Sigut2003,Sigut2004}.

We also detect a forest of strong permitted NUV \ion{Fe}{2} emission in all four LRDs, which we plot in Figure~\ref{fig:nuv_feii}.
These lines have previously been reported in prism spectra of LRDs \citep{DEugenio2025,Ando2026,Perez-Gonzalez2026,Torralba2026b}, but are resolved here with grating spectroscopy for the first time.
Strong NUV \ion{Fe}{2} is commonly seen in AGN, where it is linked to high accretion rates and is often accompanied by signatures of outflows in broad C~{\small IV} \citep[][]{Vestergaard2001,Leighly2004}, and has also been detected in $\eta$ Car \citep[e.g.,][]{Viotti1989}.
In both contexts, the emission is thought to be powered by a combination of continuum and Ly$\alpha$ fluorescence, as well as collisional excitation \citep[e.g.,][]{Johansson2001,Sigut2003}.

In AGN, permitted \ion{Fe}{2} is typically broad, and thought to arise in the BLR  \citep[e.g.][]{Wills1985,Vestergaard2001,Barth2013}.
However, the emission we detect is significantly narrower than the broad Balmer lines (FWHM$\approx188-464$~km~s$^{-1}$ for isolated emission lines in GN-2, GN-2004 and CEERS-7902).
To characterize the strength of the \ion{Fe}{2} emission we measure their EWs by directly integrating their continuum-normalized profiles. 
We follow \citet{Ando2026} and jointly fit the NUV continuum and emission components. 
Specifically, we fit a power-law continuum, $f_\nu = A\,(\lambda / 2500\,{\rm \AA})^{\alpha}$, together with the \citet{Vestergaard2001} UV \ion{Fe}{2} template, and the \ion{Fe}{3}\,UV47 template over $2000-3500$\,\AA\ using \texttt{fantasy} \citep{Ilic2023}.
We normalize the spectrum by the fitted power-law continuum, and integrate the continuum-normalized spectrum over the windows $2200-2660$~\AA, $2692.8-2772.7$~\AA, $2833.4-2917.5$~\AA, and $2907.9-3002.6$~\AA\ (with \ion{Mg}{2} masked over a window of 10~\AA) to measure the equivalent widths of the four NUV \ion{Fe}{2} complexes: the unresolved \ion{Fe}{2}~$\lambda2200-2600$ blend, and the \ion{Fe}{2}~UV62,63, \ion{Fe}{2}~UV61, and \ion{Fe}{2}~UV60,78 multiplet groups, where we adopt the extents of each multiplet in the line list of \citet{Ilic2023}.
We report the measured EW in Table~\ref{tab:nuv_feii} in Appendix~\ref{sec:nuv_feii}.

\begin{figure*}
\includegraphics[width=\linewidth]{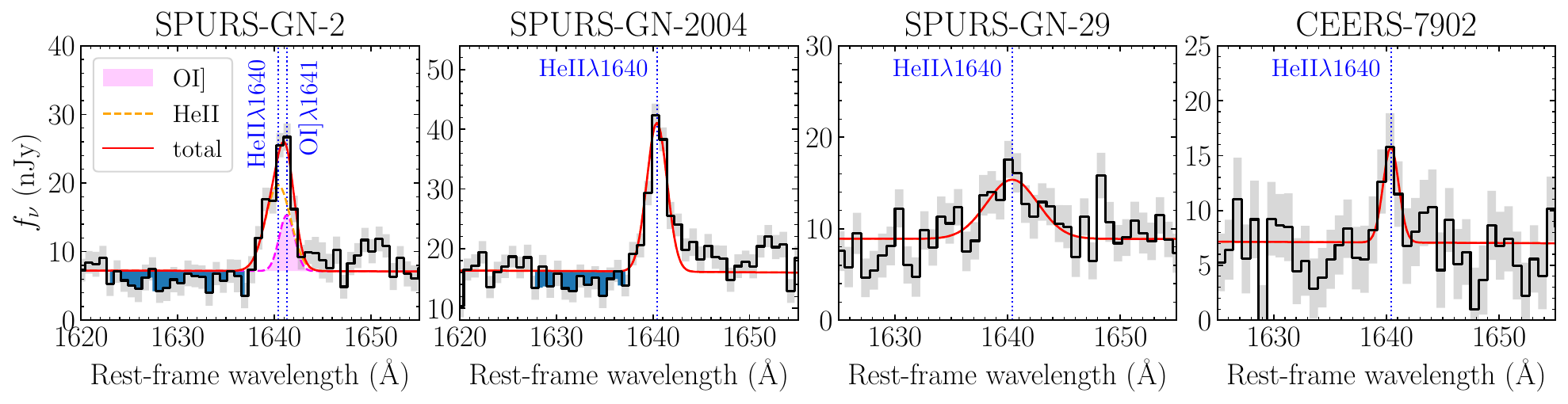}
\caption{\ion{He}{2} emission line profiles of the four LRDs. In GN-2, we also detect the \ion{O}{1}]~$\lambda1641$ emission line, which is shown as the magenta shaded region. The \ion{He}{2} emission can be described by a single narrow Gaussian except for GN-29, where we measure a significantly broader line width (FWHM $=929\pm211$~km~s$^{-1}$). We find tentative blueshifted \ion{He}{2} absorption features in GN-2 and GN-2004, which are highlighted by blue shaded regions.}
\label{fig:heii}
\end{figure*}

GN-2 and CEERS-7902 show exceptionally strong NUV \ion{Fe}{2} emission, with total equivalent widths of $356\pm24$~\AA\ and $194\pm26$~\AA, respectively, placing them among the strongest NUV \ion{Fe}{2} emission reported in LRDs to date \citep[e.g.,][]{Labbe2024,Ando2026,Torralba2026b}.
The remaining two LRDs show weaker NUV \ion{Fe}{2} emission, with total equivalent widths of $141\pm12$~\AA\ for GN-2004 and $72\pm16$~\AA\ for GN-29.
Three of the LRDs (GN-2, GN-2004, and CEERS-7902) also show extremely strong narrow \ion{Fe}{2}~$\lambda2508$ emission (EW $=6-10$~\AA), a feature which has also been detected in $\eta$ Car \citep{Davidson1995}. 
This line can also be pumped by Ly$\alpha$ emission, however, the relevant \ion{Fe}{2} transition is offset from the Ly$\alpha$ resonance by $+630$\,km~s$^{-1}$ \citep{Johansson1984,Johansson1993,Hamann2012}: the detection of this line implies broad Ly$\alpha$ emission reaches a dense region containing \ion{Fe}{2} in these three LRDs.

\subsection{He~{\small II} Emission} \label{sec:he2}

The He~{\small II}~$\lambda1640$ line provides a sensitive measure of the ionizing spectrum associated with LRDs. Above we have shown that broad C~{\small IV} is present in several LRDs, suggesting that if the ionizing spectrum is hard enough, we might also see hints of comparably broad He~{\small II} emission. Additionally, we may expect strong emission from the ionizing sources in the narrow line regions of LRDs. While previous work has constrained the strength of He~{\small II}~$\lambda4686$ \citep{DEugenio2025,Brazzini2026,Ji2026a,Wang2026}, the UV line is $\simeq8\times$ stronger, making it easier to constrain in individual sources. We detect He~{\small II}~$\lambda1640$ in all four LRDs and describe the range of line profiles below. 

GN-2 displays a complex asymmetric He~{\small II} profile (Figure~\ref{fig:heii}) due to the blending with the O~{\small I}]~$\lambda1641$ emission line presented in Section~\ref{sec:low_ion}.
To decompose the He~{\small II} and O~{\small I}] emission, we fit the entire emission line profile with two Gaussians.
We fix the centroids to the rest-frame wavelengths of He~{\small II} and O~{\small I}], and we fix the line width of O~{\small I}] equal to that of O~{\small III}]~$\lambda1666$.
We find the model reproduces the emission well and is significantly preferred over a single Gaussian including only He~{\small II} (Appendix~\ref{sec:line_fit}).
We derive a He~{\small II}~$\lambda1640$ EW of $5.6\pm1.5$~\AA\ from the deblended fit. 
We find the He~{\small II} is broader (FWHM $=468\pm91$~km~s$^{-1}$) than (semi-)forbidden UV emission lines (FWHM $\simeq340$~km~s$^{-1}$), but much narrower than the broad C~{\small IV} or broad Balmer emission lines. 
We also note tentative evidence for blueshifted \ion{He}{2}~$\lambda1640$ P-Cygni absorption (EW $=-3.3\pm0.8$~\AA) extending to $\approx1624$~\AA\ (Figure~\ref{fig:heii}), corresponding to a velocity of $\approx-2900$~km~s$^{-1}$, which suggests the presence of a high-velocity wind.
While \ion{Fe}{2}* transitions falling in this window may contribute weakly, these discrete, unresolved features can reproduce neither the full velocity extent nor the smooth, continuous shape of the blueshifted \ion{He}{2} absorption we find.

In the other three LRDs, we do not find evidence for a significant contribution from O~{\small I}]~$\lambda1641$, so we characterize their lines with single Gaussian fits.
We find none of the LRDs have He~{\small II} emission with line widths approaching the broad components of the Balmer emission lines or C~{\small IV}.
GN-29 (EW $=4.1\pm1.2$~\AA) presents the broadest He~{\small II} emission line in the sample, with a FWHM of $929\pm211$~km~s$^{-1}$, though its He~{\small II} emission is still significantly narrower than its broad Balmer emission lines. 
Such a broad He~{\small II} profile has also been seen in several $z\sim6-11$ galaxies \citep[e.g.,][]{Chen2026b,Yang2026} and can be explained via a significant contribution from strong stellar winds in WR or very massive star (VMS) populations \citep[e.g.,][]{Schaerer1996,Chandar2004,Brinchmann2008,Nanayakkara2019,Senchyna2021}, which we discuss further in Section~\ref{sec:vms}.
GN-2004 (EW $=3.5\pm0.8$~\AA) has a He~{\small II} FWHM of $424\pm67$~km~s$^{-1}$, which is slightly broader than its forbidden emission lines (FWHM $\simeq330$~km~s$^{-1}$). 
As in GN-2, it shows a tentative blueshifted P-Cygni absorption (rest-frame EW $=-1.6\pm0.8$~\AA), though extending to a lower maximum velocity of $\approx2200$~km~s$^{-1}$ ($\approx1628$~\AA).
We additionally see a broad redshifted excess over the single-Gaussian model (up to 1648~\AA; Figure~\ref{fig:heii}), suggestive of P-Cygni emission, although the BIC still favors the single Gaussian over the narrow$+$broad two-component fit.
In CEERS-7902, He~{\small II} is marginally detected (EW $=2.0\pm1.2$~\AA), but it also has a narrow profile with FWHM ($=291\pm127$~km~s$^{-1}$) similar to that of forbidden lines (FWHM $\simeq290$~km~s$^{-1}$). The range of narrow He~{\small II}~$\lambda1640$ EWs in the four LRDs ($2.0-5.6$~\AA) is larger than the average He~{\small II} EW measured in the composite spectrum of $6-10$ galaxies ($1.7$~\AA; \citealt{SinghRai2026}). 
Such strong He~{\small II} has been identified in a few individual galaxies \citep[e.g.,][]{Bunker2023,Topping2024,Topping2025a,Chen2026b}.

While we do not detect broad He~{\small II} with similar FWHMs to the broad Balmer lines in the four LRDs, the deep SPURS spectra allow us to place upper limits on broad He~{\small II} emission.
To derive upper limits, we assume the broad He~{\small II} emission FWHM is the same as broad H$\beta$ (FWHM$_{\rm broad}=3000-4000$~km~s$^{-1}$) and integrate the error spectrum in quadrature over a spectral window spanning $2\times$ FWHM$_{\rm broad}$.
We place $3\sigma$ limiting fluxes of $3.1-5.1\times10^{-19}$~erg~s$^{-1}$~cm$^{-2}$ for broad He~{\small II} emission for the four LRDs. 
These correspond to broad He~{\small II}$\lambda1640$/H$\beta$ flux ratios of $<0.036-0.070$, significantly below the typical value seen in type I AGNs ($\simeq0.5$; e.g., \citealt{Francis1991,Brotherton2001,VandenBerk2001}). Given the detection of broad C~{\small IV}, the absence of broad He~{\small II} may suggest a soft ionizing spectrum.

\subsection{Ly$\alpha$ Emission} \label{sec:lya}

Due to its high cross-section for resonant scattering, Ly$\alpha$ emission is a particularly sensitive tracer of the column density and kinematics of the neutral hydrogen through which it scatters, as well as of dust attenuation \citep[e.g.,][]{Neufeld1990,Verhamme2006}. Broad Ly$\alpha$ emission is commonly seen in AGN, arising in the same region as broad Balmer lines. Broad Ly$\alpha$ has been reported in two LRDs with deep grating spectroscopy \citep{Morishita2026,Tang2026b}, but most LRDs observed to date show narrow Ly$\alpha$ emission, more consistent with formation outside of a BLR \citep{Torralba2026a,Kageura2026,Geris2026}. Our observations do not cover Ly$\alpha$ in two of the LRDs (GN-2, GN-2004). We do not detect Ly$\alpha$ in CEERS-7902, which also shows among the strongest Ly$\alpha$-pumped \ion{Fe}{2} emission features in the sample, however, we do recover it in GN-29.

The rest-frame UV spectrum of GN-29 reveals a strong Ly$\alpha$ emission line (S/N $=15$, EW $=26\pm2$~\AA). 
The Ly$\alpha$ peak flux is redshifted by $+284\pm92$~km~s$^{-1}$ with respect to the line center, as commonly seen in Ly$\alpha$-emitting galaxies at $z\sim7$ \citep[e.g.,][]{Tang2024,Jones2025}. 
The FWHM of Ly$\alpha$ ($495\pm28$~km~s$^{-1}$) is wider than that of narrow (semi-)forbidden emission lines (FWHM $\simeq320$~km~s$^{-1}$), but significantly narrower than the broad Balmer lines.
We estimate the Ly$\alpha$ escape fraction ($f_{\rm esc,Ly\alpha}$) using the dust-corrected narrow H$\alpha$ line to predict the intrinsic Ly$\alpha$ luminosity ($L^{\rm int}_{\rm Ly\alpha}=8.7\times L_{\rm H\alpha}$; e.g., \citealt{Hayes2015,Henry2015}). 
The narrow H$\alpha$/H$\beta$ flux ratio ($5.2\pm1.3$) is larger than the intrinsic value assuming case B recombination ($2.76$ for $T=2\times10^4$~K; \citealt{Osterbrock2006}). 
This suggests substantial dust attenuation to the narrow-line gas assuming the SMC extinction law ($A_V=1.7$). 
We then derive a Ly$\alpha$ escape fraction of $0.010\pm0.002$. 
If we assume the modest dust attenuation derived in Section~\ref{sec:c3si3o3} ($A_V=0.46$), we will obtain a Ly$\alpha$ escape fraction of $0.024\pm0.006$. 

\subsection{Very High Ionization Line Emission} \label{sec:high_ion}

Very high ionization emission lines in the rest-frame UV (N~{\small V}, [Ne~{\small IV}], [Ne~{\small V}]) are often seen in the spectra of AGN, probing hard radiation fields with ionizing photon energies $>64-97$~eV. These lines have been identified in a handful of LRDs \citep{Labbe2024,Tang2025,Tang2026b,Treiber2025}, but remain undetected in the majority of the LRD population. The SPURS dataset provides our most stringent constraints yet on very high ionization lines in LRDs. We search for N~{\small V}~$\lambda\lambda1239,1243$, [Ne~{\small IV}]~$\lambda\lambda2422,2424$, and [Ne~{\small V}]~$\lambda3427$ emission in the deep G140M and G235M spectra of the four LRDs.

We do not detect any of the [Ne~{\small IV}] or [Ne~{\small V}] emission lines in the SPURS spectra of the four LRDs. 
With the G140M spectra, we place $3\sigma$ limiting EWs of $<1.4-3.7$~\AA\ for each individual component of [Ne~{\small IV}]. 
[Ne~{\small V}] is shifted to G235M spectra. 
We place $3\sigma$ limiting EWs of $<2.5-5.8$~\AA\ for the [Ne~{\small V}]~$\lambda3427$ emission lines.
These EW limits are similar to those measured from a handful of LRDs with the deepest rest-frame UV NIRSpec grating coverage \citep{Tang2025}.

The N~{\small V} doublet falls within the G140M spectra of all four LRDs. 
For GN-2 and GN-29, we do not detect N~{\small V}, placing $3\sigma$ limiting EWs of $<1.7-2.8$~\AA\ on each individual component of the doublet. 
However, GN-2 shows a tentative (S/N $=2.1$) emission line peaked at rest-frame $1248$~\AA, offset by $+1320$~km~s$^{-1}$ from \ion{N}{5}~$\lambda1243$, with a broad FWHM of $1258\pm530$~km~s$^{-1}$ and an EW of $1.5\pm0.7$~\AA\ (adopting the continuum from the power-law fit of the FUV continuum).
As we will discuss below (see Section~\ref{sec:abs_dens}), this wavelength region is strongly impacted by damped Ly$\alpha$ absorption from high column density \ion{H}{1} gas, requiring this feature to have a very large intrinsic EW ($\approx8$~\AA) and a very broad profile extending to $\approx2000$~km~s$^{-1}$ from the line center if it corresponds to \ion{N}{5}.
The N~{\small V} lines of GN-2004 fall at the short-wavelength end of the G140M spectrum, where the sensitivity is very low (Figure~\ref{fig:dla}), so we are unable to place any constraint on them. Previously, an emission feature near the position of N~{\small V}~$\lambda1239$ was reported in CEERS-7902 from the Cycle 2 GO 4287 program (EW $=27.9\pm3.4$~\AA; \citealt{Tang2025}). No such emission line is found in the SPURS G140M spectrum; the non-detection yields a $3\sigma$ limiting EW ($<2.0$~\AA) well below the previously reported value. It is possible that the slightly varied spatial positions covered between the Cycle 2 and Cycle 4 observations account for the difference. Alternatively, it is conceivable that the emission feature in the earlier program is spurious.
By checking the individual exposures of the GO-4287 data, reprocessed following the procedures described in Section~\ref{sec:spurs}, we find that the emission feature previously seen is present only in the last of the three exposures, while the same spatial position is saturated in the second exposure. This suggests that it may be possible that the emission feature was due to persistence from a saturated cosmic-ray hit.

\section{Interstellar Absorption Lines} \label{sec:abs}

The SPURS spectra provide the first constraints on UV interstellar absorption lines and resolved Ly$\alpha$ damping wings in LRDs.
These absorption features trace the interstellar medium (ISM) and the circumgalactic medium (CGM) gas surrounding the UV emitting regions in the LRDs.
In Section~\ref{sec:abs_meas}, we describe our measurements of the UV absorption lines. 
We describe the Ly$\alpha$ damping wings and constraints on the absorbing gas column density in Section~\ref{sec:abs_dens}.
We discuss the implications of fine-structure absorption line detections for the size of the absorbing region in Section~\ref{sec:abs_fs}, and explore constraints on aluminum abundances in Section~\ref{sec:abs_Al}.


\begin{figure*}
\includegraphics[width=\linewidth]{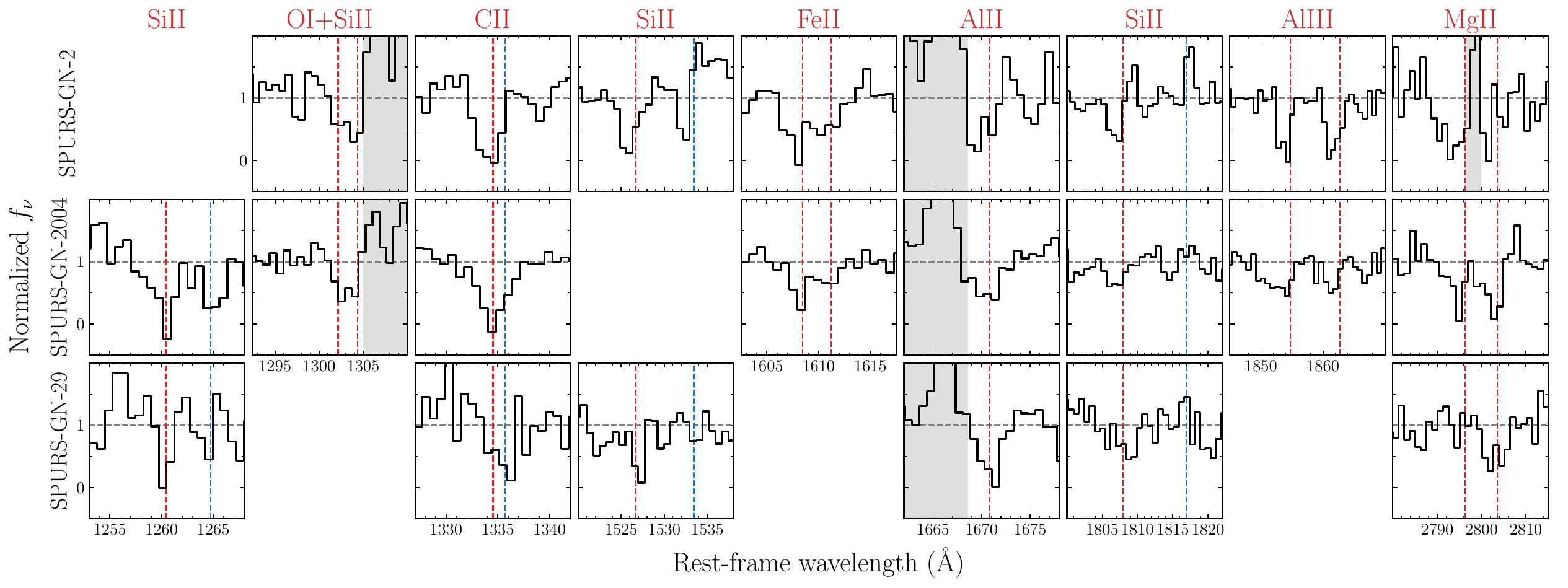}
\caption{The strongest UV and NUV LIS absorption lines in the three LRDs with the highest continuum SNR. Red dashed lines show the line centers expected from systemic redshifts of detected LIS absorption lines.
Blue vertical lines also show the wavelengths for the fine-structure lines.}
\label{fig:abs}
\end{figure*}

\subsection{Absorption Line Measurements} \label{sec:abs_meas}

The rest-frame UV continua of all the four LRDs are clearly detected in the SPURS spectra (median S/N $=4-6$ per resolution element for GN-2, GN-2004, GN-29 and S/N $=2$ for CEERS-7902), enabling us to measure multiple absorption lines.
We detect a suite of both low- (LIS; Si~{\small II}~$\lambda1260$, O~{\small I}~$\lambda1302$+Si~{\small II}~$\lambda1304$, C~{\small II}~$\lambda1334$, Si~{\small II}~$\lambda1526$, Fe~{\small II}~$\lambda\lambda1608,1611$, Al~{\small II}~$\lambda1670$, Si~{\small II}~$\lambda1808$, Fe~{\small II}~$\lambda2344$, Fe~{\small II}~$\lambda2383$, Fe~{\small II}~$\lambda2600$, Mg~{\small II}~$\lambda\lambda2796,2803$), intermediate- (Al~{\small III}~$\lambda\lambda1854,1862$), and high-ionization (HIS; Si~{\small IV}~$\lambda\lambda1393,1402$, C~{\small IV}~$\lambda\lambda1548,1550$) interstellar absorption lines across the four LRDs, several of which are clearly seen in Figure~\ref{fig:fuv_spec} and Figure~\ref{fig:nuv_feii}.

In Figure~\ref{fig:abs} we show zoom-ins of the strongest individual low and intermediate ionization absorption lines \ion{O}{1}, \ion{Si}{2}, \ion{Fe}{2}, \ion{Al}{2}, \ion{Al}{3} and \ion{Mg}{2} in each LRD, which we will discuss in more detail below.
We also detect low-ionization fine-structure absorption at $>3\sigma$ in two of our LRDs (Si~{\small II}*~$\lambda1533$ in GN-2, and C~{\small II}*~$\lambda1335$ in GN-2004), as well as several $2-3\sigma$ detections (Si~{\small II}*~$\lambda1265$, Si~{\small II}*~$\lambda1817$ in GN-2). 
In star-forming galaxies, fine-structure absorption lines are usually much weaker than resonant absorption lines \citep[e.g.,][]{James2014,Xu2023b}, and at high redshift are most commonly seen in emission \citep[e.g.,][]{Shapley2003,Jones2012,Kornei2013}. 
However, in the LRDs we find the fine-structure absorption lines are comparable in EW to the resonant lines, indicating a high column density in the excited states, which we discuss in detail in Section~\ref{sec:abs_fs}.

To quantify the strength and kinematics of the absorption lines, we fit each transition with a Gaussian profile on the continuum-normalized spectrum.
For absorption lines below rest-frame 2000~\AA, we use a linear continuum fit over the line-free pixels in a window (typically $150-200$~\AA\ in the rest-frame) around each transition, masking nearby emission lines during the fit.
For absorption lines in the NUV, we adopt the power-law continuum obtained in Section~\ref{sec:low_ion}.
To improve the recovery of the velocity profile given the low SNR of some of the absorption profiles, we fit transitions in similar ionization states simultaneously.
Thus, in the FUV, we fit each of the LIS (\ion{O}{1}, \ion{Si}{2}, \ion{C}{2}, \ion{Al}{2}, \ion{Fe}{2}), intermediate-ionization (\ion{Al}{3}), and HIS (\ion{Si}{4}, \ion{C}{4}) lines in the FUV with a shared peak velocity offset from systemic and intrinsic line width (which we convolve with the NIRSpec LSF, assuming $R=1300$, though we note our conclusions are unchanged if we assume the wavelength-dependent resolution), while the depth of each line is fitted independently. 
In the NUV, covered by G235M, we jointly fit Fe~{\small II} and Mg~{\small II} lines in the same way.
We derive uncertainties by perturbing the spectrum by the error array and repeating the continuum and line fits 100 times.
From these fits we constrain the absorption EW, peak velocity offset, and resolution-deconvolved peak covering fraction (discussed below), which we report in Table~\ref{tab:abs_lines}.
For closely separated lines, we report a total EW.
For non-detections, we report their $3\sigma$ EW upper limits obtained from the joint fit.

We first consider the strength of the absorption lines, which provide insights into the covering fraction and column density of gas around the UV emitting regions. 
In star-forming galaxies at $z\sim7$, the highest oscillator strength LIS absorption lines measured in stacks, and individual spectra from SPURS, are typically weak \citep[median EW $\approx -1$~\AA; e.g.,][]{Glazer2025,KeerthiVasan2026}.
By contrast, we find the LIS absorption lines in the LRDs all have large EWs, with medians for each LRD ranging from $-2.2$~\AA\ to $-1.2$~\AA\ (computed based on measurements for single FUV lines at $<2000$~\AA\ only).
These values are more typical of, or stronger than, those seen in lower redshift galaxies with weak Ly$\alpha$ emission, where strong LIS absorption lines are thought to indicate a high covering fraction of neutral gas ($z\simeq1-3$; e.g., \citealt{Shapley2003,Steidel2010,Jones2012,Saldana-Lopez2023}).
In contrast to other galaxies in the SPURS sample and at lower redshift, we also detect strong ($<-1$~\AA) absorption in \ion{Si}{2}~$\lambda1808$ and \ion{Fe}{2}~$\lambda1611$ in three of the LRDs.
These transitions have optical depths two orders of magnitude below most other \ion{Si}{2} and \ion{Fe}{2} transitions detected in our spectra, and are normally extremely weak in galaxies \citep[e.g.,][]{Jones2018}. 
The large EW implies very high column densities, which we will discuss below. Moreover it indicates that there is widespread saturation in all of the well-detected LIS absorption lines. 
We also detect strong Al~{\small III}~$\lambda\lambda1854,1862$ absorption in the LRDs, most clearly seen in GN-2 and GN-2004 (\ion{Al}{3} EW $=-2.2\pm0.3$, $-2.1\pm0.5$~\AA, respectively).
These values exceed typical Al~{\small III} strengths in star-forming galaxies \citep[e.g.,][]{Jones2018}, implying a high column density of Al in moderately ionized \ion{H}{2} gas which we discuss more in Section~\ref{sec:abs_Al}.
Finally, the HIS absorption lines have comparable EW to the LIS absorption lines (median EW $=-2.0$~\AA), and are again at the high end of the range reported in star-forming galaxies at $z\sim3-5$ \citep{Shapley2003,Pahl2020}.

The velocity peak of the lines provides insights into the bulk kinematics of the absorbing gas.
In star-forming galaxies, both LIS and HIS absorption lines typically show comparable blueshifted velocities of $\sim-100$ to $-200$~km~s$^{-1}$, indicating multi-phase outflows \citep[e.g.,][]{Shapley2003,Steidel2010,Chisholm2015,Xu2022,KeerthiVasan2023}.
In the LRDs, we detect blueshifted absorption in the HIS lines (\ion{Si}{4}) of GN-2, GN-2004, and CEERS-7902, peaking at $\Delta v=-231\pm32$, $-204\pm37$, and $-242\pm92$~km~s$^{-1}$, respectively.
Notably, the HIS absorption lines in GN-2 and GN-2004 show significantly higher peak velocities than the LIS absorption lines ($-153\pm23$~km~s$^{-1}$ and $-58\pm30$~km~s$^{-1}$).
We detect outflows in the LIS lines in GN-2 and CEERS-7902 ($-153\pm23$ and $-298\pm74$\,km~s$^{-1}$, respectively), while strikingly, both GN-2004 and GN-29 show significant LIS absorption close to systemic velocities, with $\Delta v\approx-53\pm30$~km~s$^{-1}$ and $+39\pm21$~km~s$^{-1}$ respectively, lower than typically seen in star-forming galaxies at lower redshifts \citep[though three galaxies with similarly low LIS velocities have been reported in a sample from SPURS at $z\sim5-9$;][]{KeerthiVasan2026}.
Interestingly, the HIS absorption lines in GN-29 are also consistent with peaking close to systemic velocities ($\Delta v\approx-52\pm54$~km~s$^{-1}$), which is not typical for star-forming galaxies at high redshifts.
Higher resolution spectroscopy will be required to more precisely determine the velocity profile of the absorbing gas, but these results indicate a diversity of outflow velocities in LRDs, which are not typical for star-forming galaxies.

The physical interpretation of the depth of the absorption lines depends on their optical depth: in optically thick gas this is a sensitive probe of the gas covering fraction, while it traces column density in optically thin gas \citep[e.g.,][]{Shapley2003,Jones2013}. 
We investigate whether the rest-frame UV absorption lines are saturated by comparing the relative EWs of transitions in the same ions (\ion{Si}{2}, \ion{Si}{4}, \ion{C}{4}): in optically thin gas, these EWs should scale as $Nf\lambda^2$, where $N$ is the ion column density, $f$ is the oscillator strength, and $\lambda$ is the rest-frame wavelength.
For all four LRDs we find the EW ratios deviate strongly from the optically thin predictions in both the LIS and HIS lines, indicating saturation.
Thus we interpret the residual intensity of the absorption lines as primarily tracing the covering fraction rather than column density of the absorbing gas, with the absorption depth providing a lower limit on the column density.
Therefore, we also report the peak covering fraction ($C_f\approx1-I(v=v_{\rm peak})/I_0$, where $I$ and $I_0$ are observed intensity and continuum intensity, respectively), corrected for instrumental resolution, in Table~\ref{tab:abs_lines}.
In star-forming galaxies, typical LIS absorption line strengths decrease towards higher redshifts \citep{Pahl2020,Glazer2025,KeerthiVasan2026}, which has been interpreted as an evolution towards lower neutral gas covering fraction based on data with moderate spectral resolution \citep[$R\sim5000$; e.g.,][]{Jones2013,Leethochawalit2016}.
To compare the covering fractions in the LRDs, we consider the lines with highest optical depth, which we expect to be strongly saturated, in regions free from emission lines (\ion{Si}{2}~$\lambda\lambda1260,1526$, \ion{Si}{4}~$\lambda1393$, \ion{Al}{2}~$\lambda1670$, \ion{Al}{3}~$\lambda1854$).
We find very high covering fractions in both the LIS and HIS lines (median $C_f=0.7-1.0$ for each LRD), implying the vast majority of the UV-emitting regions in these LRDs are surrounded by both low and high ionization gas.
These covering fractions are at the high end of the $z\sim5-9$ galaxy sample from SPURS (spanning $C_f\approx0.2-0.9$; \citealt{KeerthiVasan2026}).
In particular, GN-2 and GN-29 show covering fractions of the UV continuum consistent with unity.

In summary, the interstellar absorption lines in the LRDs are not typical of star-forming galaxies at similar or lower redshifts. 
The lines reveal a diversity of kinematics, with two LRDs showing the bulk of low ionization gas at velocities close to systemic. They suggest a high column density and high covering fraction of dense neutral (\ion{H}{1}) gas around the UV emitting region, which we quantify more below.


\begin{deluxetable*}{l|ccc|ccc|ccc|ccc}
\tablecaption{Rest-frame UV interstellar absorption line EWs (\AA), velocity offsets (km~s$^{-1}$) and intrinsic covering fraction ($C_f$) of the four LRDs.}
\tablehead{
 & \multicolumn{3}{c|}{SPURS-GN-2} & \multicolumn{3}{c|}{SPURS-GN-2004} & \multicolumn{3}{c|}{SPURS-GN-29} & \multicolumn{3}{c}{CEERS-7902} \\
Line                                                             & EW                  & $\Delta v$   & $C_f$\tablenotemark{a} & EW                            & $\Delta v$  & $C_f$\tablenotemark{a} & EW                       & $\Delta v$  & $C_f$\tablenotemark{a} & EW                  & $\Delta v$  & $C_f$\tablenotemark{a}
}
\startdata
Si~{\scriptsize II}~$\lambda1260$                                & $-1.1\pm0.5$        & $-153\pm23$  & $0.7\pm0.3$            & $-1.7\pm0.4$                  & $-58\pm30$  & $1.0\pm0.1$            & $-1.2\pm0.3$             & $+39\pm21$  & $1.0\pm0.2$            & $>-1.6$             & --          & -- \\
Si~{\scriptsize II}*~$\lambda1265$                               & $>-1.5$             & --           & --                     & $-1.6\pm0.6$                  & $-58\pm30$  & $0.9\pm0.2$            & $>-1.1$                  & --          & --                     & $>-3.3$             & --          & -- \\
O~{\scriptsize I}+Si~{\scriptsize II}~$\lambda\lambda1302,1304$  & $-1.8\pm0.4$        & $-153\pm23$  & $0.7\pm0.1$            & $-1.7\pm0.5$                  & $-58\pm30$  & $0.5\pm0.1$            & $-1.2\pm0.5$             & $+39\pm21$  & $0.7\pm0.2$            & $-3.5\pm1.4$        & $-298\pm74$ & $0.8\pm0.1$ \\
C~{\scriptsize II}+C~{\scriptsize II}*~$\lambda\lambda1334,1335$ & $-2.9\pm0.6$        & $-153\pm23$  & $0.9\pm0.1$            & $-3.2\pm0.5$                  & $-58\pm30$  & $0.9\pm0.1$            & $-1.9\pm0.7$             & $+39\pm21$  & $0.8\pm0.2$            & $>-2.2$             & --          & -- \\
Si~{\scriptsize IV}~$\lambda1393$                                & $-2.3\pm0.3$        & $-231\pm32$  & $1.0\pm0.0$            & $-2.0\pm0.2$                  & $-204\pm37$ & $0.8\pm0.1$            & $-1.4\pm0.4$             & $-52\pm54$  & $0.9\pm0.2$            & $-2.2\pm0.6$        & $-242\pm92$ & $0.8\pm0.2$ \\
Si~{\scriptsize IV}~$\lambda1402$                                & $-2.0\pm0.2$        & $-231\pm32$  & $0.9\pm0.1$            & $-1.5\pm0.3$                  & $-204\pm37$ & $0.6\pm0.1$            & $-1.5\pm0.4$             & $-52\pm54$  & $0.9\pm0.2$            & $>-2.5$             & --          & -- \\
Si~{\scriptsize II}~$\lambda1526$                                & $-1.9\pm0.2$        & $-153\pm23$  & $1.0\pm0.1$            & --\tablenotemark{d}           & --          & --                     & $-1.4\pm0.2$             & $+39\pm21$  & $1.0\pm0.1$            & $-1.9\pm0.7$        & $-298\pm74$ & $0.7\pm0.3$ \\
Si~{\scriptsize II}*~$\lambda1533$                               & $-1.5\pm0.3$        & $-153\pm23$  & $0.8\pm0.2$            & --\tablenotemark{d}           & --          & --                     & $>-1.0$                  & --          & --                     & $>-1.0$             & --          & -- \\
C~{\scriptsize IV}~$\lambda1548$                                 & --\tablenotemark{b} & --           & --                     & --\tablenotemark{d}           & --          & --                     & $>-1.2$                  & --          & --                     & --\tablenotemark{b} & --          & -- \\
C~{\scriptsize IV}~$\lambda1550$                                 & --\tablenotemark{b} & --           & --                     & --\tablenotemark{d}           & --          & --                     & $>-0.3$                  & --          & --                     & --\tablenotemark{b} & --          & -- \\
Fe~{\scriptsize II}~$\lambda1608$                                & $-2.0\pm0.3$        & $-153\pm23$  & $1.0\pm0.1$            & $-1.5\pm0.3$                  & $-58\pm30$  & $0.7\pm0.1$            & $-0.8\pm0.4$             & $+39\pm21$  & $0.6\pm0.2$            & $-2.4\pm0.7$        & $-298\pm74$ & $0.8\pm0.3$ \\
Fe~{\scriptsize II}~$\lambda1611$                                & $-1.5\pm0.3$        & $-153\pm23$  & $0.7\pm0.2$            & $-0.9\pm0.3$                  & $-58\pm30$  & $0.4\pm0.1$            & $>-1.1$                  & --          & --                     & $>-0.7$             & --          & -- \\
Al~{\scriptsize II}~$\lambda1670$                                & $-2.1\pm0.3$        & $-153\pm23$  & $1.0\pm0.1$            & $-1.7\pm0.3$                  & $-58\pm30$  & $0.7\pm0.1$            & $-1.5\pm0.3$             & $+39\pm21$  & $1.0\pm0.0$            & $>-3.2$             & --          & -- \\
Si~{\scriptsize II}~$\lambda1808$                                & $-1.9\pm0.4$        & $-153\pm23$  & $0.8\pm0.2$            & $-1.0\pm0.3$                  & $-58\pm30$  & $0.4\pm0.1$            & $-1.2\pm0.3$             & $+39\pm21$  & $0.7\pm0.2$            & $>-2.5$             & --          & -- \\
Si~{\scriptsize II}*~$\lambda1817$                               & $>-0.9$             & --           & --                     & $-0.5\pm0.2$\tablenotemark{c} & $-385\pm41$ & $1.0\pm0.0$            & $>-1.0$                  & --          & --                     & $-1.7\pm0.7$        & $-298\pm74$ & $0.5\pm0.2$ \\
Al~{\scriptsize III}~$\lambda1854$                               & $-2.2\pm0.3$        & $-167\pm22$  & $1.0\pm0.1$            & $-2.1\pm0.5$                  & $-275\pm57$ & $0.6\pm0.1$            & $-1.8\pm0.7$             & $+62\pm62$  & $0.8\pm0.2$            & $-2.1\pm0.9$        & $-303\pm59$ & $0.8\pm0.2$ \\
Al~{\scriptsize III}~$\lambda1862$                               & $-2.2\pm0.4$        & $-167\pm22$  & $1.0\pm0.0$            & $-1.4\pm0.4$                  & $-275\pm57$ & $0.4\pm0.1$            & $>-0.7$                  & --          & --                     & $-2.6\pm0.8$        & $-303\pm59$ & $1.0\pm0.2$ \\
Fe~{\scriptsize II}~$\lambda2344$                                & $-2.1\pm0.4$        & $-147\pm30$  & $1.0\pm0.0$            & $-2.9\pm0.5$                  & $+0\pm40$   & $1.0\pm0.1$            & $-1.6\pm0.6$             & $+53\pm64$  & $1.0\pm0.3$            & $-5.3\pm1.3$        & $-111\pm98$ & $0.7\pm0.2$ \\
Fe~{\scriptsize II}~$\lambda2383$                                & $>-2.5$             & --           & --                     & $-2.1\pm0.5$                  & $+0\pm40$   & $0.7\pm0.2$            & $>-1.8$                  & --          & --                     & $-4.4\pm1.6$        & $-111\pm98$ & $0.6\pm0.2$ \\
Fe~{\scriptsize II}~$\lambda2600$                                & $-2.3\pm0.5$        & $-147\pm30$  & $1.0\pm0.1$            & $-2.1\pm0.8$                  & $+0\pm40$   & $0.6\pm0.2$            & $-1.7\pm0.7$             & $+53\pm64$  & $1.0\pm0.2$            & $-8.0\pm2.3$        & $-111\pm98$ & $0.9\pm0.1$ \\
Mg~{\scriptsize II}~$\lambda2796$                                & $-5.3\pm1.5$        & $-316\pm118$ & $1.0\pm0.1$            & $-2.8\pm0.6$                  & $-138\pm38$ & $0.8\pm0.2$            & $>-0.9$                  & --          & --                     & $-2.9\pm0.8$        & $-434\pm55$ & $1.0\pm0.1$ \\
Mg~{\scriptsize II}~$\lambda2803$                                & $-4.2\pm1.4$        & $-316\pm118$ & $0.8\pm0.3$            & $-3.4\pm0.6$                  & $-138\pm38$ & $1.0\pm0.1$            & $-3.3\pm1.0$             & $-57\pm124$ & $0.7\pm0.2$            & $-2.8\pm0.8$        & $-434\pm55$ & $1.0\pm0.2$ \\
\enddata
\tablecomments{EW, $\Delta v$ and $C_f$ are reported for lines detected at $>2\sigma$; otherwise the $3\sigma$ upper limit on the EW is given (negative EW denotes absorption). \\ a: The covering fractions reported are resolution-deconvolved peak covering fractions. \\ b: C~{\scriptsize IV} absorption in SPURS-GN-2 and CEERS-7902 is measured separately given its complicated profile. \\ c: Si~{\scriptsize II}*~$\lambda1817$ in SPURS-GN-2004 is fitted separately from the other low-ionization lines because an apparent P-Cygni emission component shifts its velocity centre. \\ d: These lines are not covered, falling in the detector gap.}
\label{tab:abs_lines}
\end{deluxetable*}

\subsection{Ly$\alpha$ Damping Wings and Column Density} \label{sec:abs_dens}

All four of the LRDs show significant downturns in their UV continuum around the wavelength of Ly$\alpha$, including absorption troughs and suppression extending to $\gtrsim1300$~\AA\ in the rest frame.
Such turnovers are expected due to Ly$\alpha$ damping wing absorption if neutral hydrogen with very large column densities covers the majority of the UV emitting regions.
We note two-photon nebular continuum emission also produces downturns in the UV continuum around 1300~\AA\ \citep[][]{Schaerer2002,Raiter2010,Cameron2024}. However, comparing the spectra to nebular continuum models we find two-photon continuum produces a more gradual turnover and no absorption trough, inconsistent with the observed spectra. Thus in the following we consider only the impact of the Ly$\alpha$ damping wing.

To estimate the \ion{H}{1} column density, we fit the spectral region around Ly$\alpha$ in each source following the approach of \citet{Mason2026} \citep[see also][]{Chen2026a,KeerthiVasan2026}.
We first model the intrinsic continuum as a power law, fit to the observed spectrum over rest-frame $1350-1800$~\AA, masking both IS absorption and nebular emission lines.
We model Ly$\alpha$ attenuation local to the galaxy using a Voigt absorption profile with column density ($18<\log (N_{\rm HI}/{\rm cm}^{-2})<24$) centered at the systemic redshift, assuming a covering fraction of unity, motivated by the large covering fractions inferred from the metal absorption lines, noting that our conclusions are unchanged if we also fit for the \ion{H}{1} covering fraction.
In GN-29, where we detect Ly$\alpha$, we additionally model the emergent Ly$\alpha$ emission as a Gaussian, leaving its centroid, width, and EW as free parameters. As the Ly$\alpha$ emission is narrow, it must propagate without significant resonant scattering in the very high column density gas \citep[implying either a patchy/clumpy absorbing medium or that the line is produced in an outer region, e.g.,][]{Torralba2026a,Tang2026b}; thus we add the emission component on top of the damped Ly$\alpha$ (DLA) and continuum.
We then apply the IGM damping wing attenuation \citep[following][]{Miralda-Escude1998,Barkana2002,Mesinger2008}, fitting for the distance from the source to the nearest neutral IGM region and assuming the IGM is fully neutral beyond this. 
We fit the spectra from the beginning of the spectral coverage for each LRD ($\approx1210-1260$~\AA) up to 1450~\AA\ in the rest frame, deriving constraints and uncertainties on the \ion{H}{1} column density using the Markov chain Monte Carlo ensemble sampler \texttt{emcee} \citep{Foreman-Mackey2013}.

\begin{figure}
\includegraphics[width=\columnwidth]{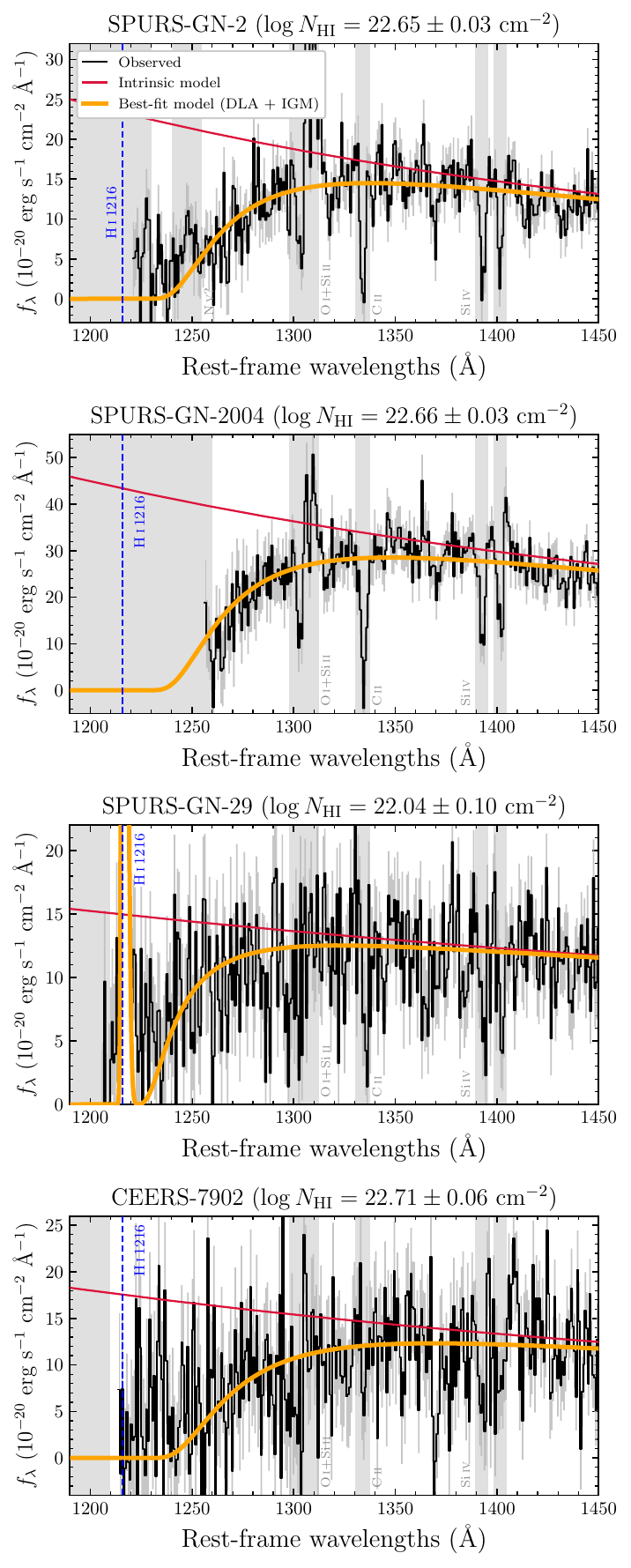}
\caption{Fit to the Ly$\alpha$ profile. The black line shows the observed spectrum with the error spectrum shown in gray. We show the best-fit absorption line model in orange and the intrinsic power-law continuum model in red. The low-ionization absorption lines are masked in gray and excluded from the fit.}
\label{fig:dla}
\end{figure}

Figure~\ref{fig:dla} shows the best-fit model to the Ly$\alpha$ absorption profile, as well as the intrinsic continuum model, for each LRD.
In all cases, we find very large column densities ($N_{\rm HI} \approx10^{22.0-22.7}$~cm$^{-2}$) are required to reproduce the observed profiles, dominating over the IGM damping wing.
The high column densities in our LRD sample are significantly larger (by $>10\times$) than what is typically inferred in $z\gtrsim5$ galaxies from both low-resolution NIRSpec prism spectra and medium-resolution grating spectra \citep{Umeda2026,Heintz2025,Mason2026,Chen2026a,KeerthiVasan2026}.

These high \ion{H}{1} column densities are also consistent with the detections of absorption in low oscillator strength transitions: \ion{Si}{2}~$\lambda1808$ ($f=0.0022$) in GN-2, GN-2004 and GN-29 and \ion{Fe}{2}~$\lambda1611$ ($f=0.0014$) in GN-2 and GN-2004.
A precise estimate of the column density is challenging given that the \ion{Si}{2} and \ion{Fe}{2} lines are likely all saturated, and there is uncertainty due to the intrinsic linewidth and dust depletion.
However, we can estimate a conservative lower limit on the \ion{Si}{2} and \ion{Fe}{2} column densities assuming the 1808~\AA\ and 1611~\AA\ transitions are optically thin and no dust depletion of either element.
In this case, we estimate $N > 1.13\times10^{20}\, W_\lambda/f\lambda^2$\,cm$^{-2}$, resulting in $N_{\rm SiII} > 10^{16}$\,cm$^{-2}$ and $N_{\rm FeII} > 10^{16}$\,cm$^{-2}$ for all LRDs where these lines are detected.
Assuming solar abundance ratios and the gas-phase metallicity implied by the emission lines ($Z\approx0.03-0.28 Z_\odot$), this also implies very high HI column densities $N_{\rm HI} > 10^{21.3-22.5}$\,cm$^{-2}$.

\subsection{Fine-structure absorption} \label{sec:abs_fs}

In addition to the ground-state interstellar absorption lines, in GN-2 and GN-2004 we detect absorption lines from several fine-structure states, with comparable strength to the resonant absorption lines. 
Fine-structure absorption requires a population in an excited level of the ion's ground term, which can only compete against rapid spontaneous decay if the gas is collisionally excited or radiatively pumped by an intense local UV field \citep{Bahcall1968,Prochaska2006,James2014}. 
In star-forming galaxies these transitions are typically weak or seen only in emission, and are thought to trace diffuse ($n_e\sim10$~cm$^{-3}$) gas over kiloparsec scales \citep{Jones2012,Xu2023b}.
Strong fine-structure absorption is however often seen in gamma ray burst (GRB) afterglows, where the levels are thought to be excited by the strong radiation field of the GRB within $\lesssim100$~pc \citep{Prochaska2006}. 
While some of these lines (Figure~\ref{fig:abs}) also show likely P-Cygni emission components, the detection of strong fine-structure absorption in these LRDs suggests densities and/or radiation fields that are not typical of low redshift star-forming galaxies.

To investigate the conditions that produce these lines, we first estimate the excited-state population fraction implied by these detections. 
We focus on \ion{Si}{2} and \ion{Fe}{2} transitions, for which we detect multiple absorption lines, and are sensitive tracers of gas conditions \citep{Prochaska2006}.
As the detected lines in both ground-state and excited state transitions are saturated in our spectra, we consider an order of magnitude approach, leaving more detailed analysis to future work. 
From the fine-structure EWs, we estimate conservative lower limits on the excited state column densities, assuming the lines are optically thin, using the lowest oscillator strength lines which are robustly detected and free from blending with ground-state lines.
The best line for this purpose is \ion{Si}{2}~$\lambda1817$ ($f=0.0016$) in GN-2004 (EW $=-0.5\pm0.2$~\AA), from which we estimate 
$N_{\rm SiII^*}\gtrsim1.6\times10^{16}$~cm$^{-2}$.
As the ground-state transitions in \ion{Si}{2} all appear saturated we estimate the expected ground-state column density assuming the damped Ly$\alpha$ and metal resonant and fine-structure absorption all arise in the same gas.
The \ion{H}{1} column density from the damping wing fit implies ground-state columns 
$N_{\rm SiII}\approx3\times10^{17}$~cm$^{-2}$, assuming $Z=0.2\,Z_\odot$ and solar abundance ratios.
This implies $n_{\rm excited}/n_{\rm ground}\gtrsim0.04$ for \ion{Si}{2}, which is four times higher than the mean \ion{Si}{2}* level population inferred in the CLASSY sample at $z\sim0$ \citep{Xu2023a}.
We note this is likely to be underestimated given saturation, suggesting potentially significant excitation of the fine-structure states relative to what is seen at low redshifts.

We consider excitation by both radiative pumping and collisional excitation, though a combination of both effects may contribute. 
To assess radiative pumping, we compute the strength of the local UV radiation field at a distance $r$ from the UV-emitting source: $G = \lambda L_\lambda/(4\pi r^2)$, expressed relative to the mean Galactic interstellar field, $G_0$ \citep{Habing1968}, following \citet{Prochaska2006} and \citet{Xu2023b}. 
Based on the excitation calculations of \citet{Prochaska2006}, producing $n_{\rm excited}/n_{\rm ground}>0.04$ for \ion{Si}{2} requires $G_0>10^{4}$.
Using the rest-frame UV luminosities of GN-2004, we find this is achieved provided the absorbing gas is within just $<300$~pc of the UV-emitting source.
Reproducing the same excited fraction through collisional excitation requires $n_e>30$~cm$^{-3}$.
Assuming $N_{\rm HI}\approx10^{22.6}$~cm$^{-2}$, this also implies the gas must be compact ($\lesssim400$~pc).
Thus, both excitation channels imply the absorbing gas must be compact, confined to $<400$~pc from the UV-emitting source, to produce the strong fine-structure absorption.
Such a compact region is comparable to the unresolved UV continuum ($\lesssim 80$~pc).

\subsection{Aluminum Abundances in LRDs} \label{sec:abs_Al}

Given the nitrogen enhancement in all four LRDs reported above (Section~\ref{sec:n4n3}), we might expect other anomalous abundance signatures associated with enrichment in dense nuclear clusters.
In particular, \citet{Marques-Chaves2024} suggested aluminum enhancement is a promising additional test (see also \citealt{Chisholm2026}).
Aluminum enhancements are expected from the MgAl chain in very hot hydrogen burning, and large aluminum abundance ratios can be a signature of nucleosynthesis under conditions likely only achieved in the most massive stars (VMS or SMS) and in massive AGBs undergoing hot bottom burning \citep[][]{Prantzos2007,bastianMultipleStellarPopulations2018,Higgins2023,Higgins2025}.
Indeed, \citet{Topping2024} recently reported a tentative Al~{\small III}~$\lambda1854$ emission feature in a N~{\small IV}]-emitting galaxy at $z=7.04$, implying $>30\times$ solar Al-enhancement if confirmed. 
While we do not detect aluminum emission features in our LRD sample, we do detect markedly strong Al~{\small II}~$\lambda1670$ and Al~{\small III}~$\lambda\lambda1854,1862$ absorption (Section~\ref{sec:abs_meas}), motivating an exploration of potential aluminum enhancement from the absorption lines.

\begin{figure}
\includegraphics[width=\columnwidth]{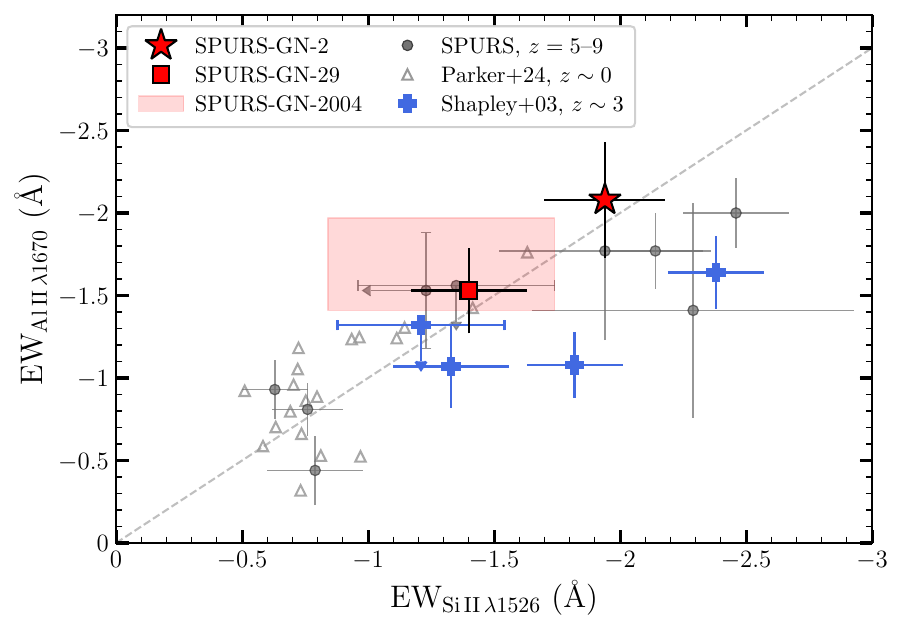}
\caption{Al~{\scriptsize II}~$\lambda1670$ absorption line EWs vs. Si~{\scriptsize II}~$\lambda1526$ absorption line EWs of the three LRDs with absorption line detections (GN-2, GN-2004, GN-29). For GN-2004, which lacks Si~{\scriptsize II}~$\lambda1526$ coverage, we plot the expected range spanning from the Si~{\scriptsize II}~$\lambda1304$ EW to the Si~{\scriptsize II}~$\lambda1260$ EW as a red shaded region. For comparison, we overplot EWs of Lyman break galaxies at $z\sim3$ (blue crosses; \citealt{Shapley2003}), nearby star-forming galaxies from CLASSY (open triangles; \citealt{Parker2024}), and SPURS galaxies at $z\simeq5-9$ (grey circles; \citealt{Chen2026a,KeerthiVasan2026}).}
\label{fig:Al}
\end{figure}

As discussed in Section~\ref{sec:abs_meas}, we find all the well-detected absorption lines are affected by saturation, making robust estimates of column densities and abundance patterns very uncertain. We caution that apparent optical depth methods provide lower limits on column density in such cases, and the correction factors can be large \citep[e.g.,][]{Savage1991,Pettini2002,Quider2009}. 
To assess a potential aluminum enhancement in the LRDs, we consider the relative strength of Al~{\small II}~$\lambda1670$ to Si~{\small II}~$\lambda1526$ absorption EW in the LRDs and the rest of the SPURS sample. These specific transitions are chosen to mitigate uncertainties arising from multiple sources. 
We assess Al/Si rather than Al/Mg because Al~{\small II} and Si~{\small II} are both accessible in the high SNR G140M spectra for a large fraction of the SPURS sample, and trace the same odd-even nucleosynthetic signature.
The ratio of Al~{\small II}~$\lambda1670$ to Si~{\small II}~$\lambda1526$ is chosen because, for a solar gas-phase abundance pattern, the line strengths are expected to be similar and with low variation. Specifically, EW(Al~{\small II}~$\lambda1670$)/EW(Si~{\small II}~$\lambda1526$) $\approx 1.47$ in the optically thin limit, decreasing toward 1.09 as the lines saturate \citep[based on oscillator strengths from][]{Morton1991}. 
Dust depletion, which preferentially removes aluminum from the gas phase relative to silicon \citep[e.g.,][]{Jenkins2009,Konstantopoulou2022}, can weaken Al~{\small II} relative to Si~{\small II}. 
On the other hand, an increased ratio of EW(Al~{\small II}~$\lambda1670$)/EW(Si~{\small II}~$\lambda1526$) may suggest super-solar Al/Si ratio. 
We also caution that ionization effects can lead to increased EW(Al~{\small II}~$\lambda1670$)/EW(Si~{\small II}~$\lambda1526$), as the abundance ratio of Al$^+$/Si$^+$ can increase by potentially $\gtrsim 0.5$ dex \citep{Howk1999,Vladilo2001} in cases where there is significant contribution from ionized gas (i.e., a phase bearing H$^+$, Al$^+$, and Si$^+$). 
Given the high EW \ion{Al}{3} absorption we detect in the SPURS LRDs, the ionization effects may be significant.
Thus it is important to consider line saturation, dust depletion, and ionization effects alongside the abundance patterns. 

In Figure~\ref{fig:Al} we plot the Al~{\small II}~$\lambda1670$ and Si~{\small II}~$\lambda1526$ EW for the LRDs in this paper, along with the rest of the SPURS sample (which will be presented fully in upcoming work).
As we lack coverage of Si~{\small II}~$\lambda1526$ in GN-2004, we instead use the EW of Si~{\small II}~$\lambda1260,\lambda1304$ as upper and lower limits on the EW respectively (as $\tau_{1302} < \tau_{1526} \ll \tau_{1260}$).
We also show measurements from lower redshifts for comparison, from \citet{Shapley2003} ($z\sim3$) and CLASSY \citep[$z\sim0$;][]{Parker2024}.
The SPURS sample generally lies along the one-to-one line, with the LRDs lying towards the upper end of both Al~{\small II}~$\lambda1670$ and Si~{\small II}~$\lambda1526$ EW, consistent with the high covering fraction and column densities discussed in Section~\ref{sec:abs}.
Notably the Al~{\small II}/Si~{\small II} EW ratio in the SPURS sample is higher than in the \citet{Shapley2003} stacks, indicating physical differences in the absorbing gas. The lower Al~{\small II}/Si~{\small II} EW ratio $<1$ in $z\sim3$ stacks implies that at least some of the absorption is unsaturated {\it and} has a sub-solar Al$^+$/Si$^+$ abundance in the gas phase. We reiterate that a sub-solar gas-phase Al/Si is consistent with dust depletion effects (or, of course, intrinsically sub-solar Al/Si ratio). 

A natural explanation for the higher Al~{\small II}/Si~{\small II} EW ratios in the SPURS sample compared to $z\sim3$ stacks is that lines are fully saturated due to high low-ion column densities (as described in Section~\ref{sec:abs_meas}). 
In general the line ratios suggest that \ion{Si}{2}~$\lambda1526$ is essentially fully saturated, and consequently the ratios shown in Figure~\ref{fig:Al} imply that \ion{Al}{2}~$\lambda1670$ is likewise saturated. 
In this case, we cannot robustly recover the Al/Si abundance from this line ratio. 
We find GN-2004 may be consistent with an offset toward higher Al~{\small II}~$\lambda1670$ than the rest of the sample, suggestive of an enhanced Al/Si abundance in this source.
However, without a direct measurement of the Si~{\small II}~$\lambda1526$ EW, this interpretation remains tentative: a spectrum with coverage of this line will be required to make a more definitive statement.

We now consider the challenges and future prospects for more reliable abundance patterns of Al and other elements based on interstellar absorption in LRDs. 
First, it is essential to measure unsaturated absorption lines, and to account for hidden saturation effects. 
Next, dust depletion effects can be assessed by comparing elements of common nucleosynthetic origin with different depletion rates. Alternatively the effects can be mitigated by comparing elements with very low (such as O and S) or similar depletions (such as Si and Mg; e.g., \citealt{Konstantopoulou2022}). 
Finally, ionization effects can be either modeled based on measurements from multiple ionization states, mitigated by comparing ions with similar ionization corrections, or perhaps most simply by analyzing systems where weak intermediate-ion absorption such as \ion{Al}{3} indicates that corrections to the low ion abundances are minimal \citep[e.g.,][]{Howk1999}. 
Our SPURS sample analysis suggests that precise Al abundances may be difficult to constrain in LRDs due to the large optical depth of Al~{\small II}~$\lambda1670$ (and lack of suitable weaker transitions), and potentially significant ionization corrections indicated by the strong \ion{Al}{3} absorption.
However, lower limits of Al relative to other elements can be obtained if suitably weak lines can be identified. 
Relatively weak lines may also provide interesting constraints on other chemical abundance patterns once saturation effects are accounted for \citep[e.g.,][]{Quider2009}. 
We observe potentially not fully saturated absorption from such weak lines as \ion{Si}{2}~$\lambda1808$ and \ion{Fe}{2}~$\lambda1611$, which are sensitive to $\alpha$/Fe ratio for example. 
Additional low-oscillator strength transitions from species such as \ion{S}{2}, \ion{Ni}{2}, and \ion{Zn}{2} may prove interesting for further analysis.

\section{Rest-frame UV Line Diagnostics in the Broader LRD Population} \label{sec:uv_diag}

We have investigated the UV spectroscopic properties of a sample of four LRDs. In this section, we place these results in context using spectra from the public archive. In Section~\ref{sec:N_line}, we quantify how common N~{\small IV}] and N~{\small III}] emission is in LRDs, and in Section~\ref{sec:ciii_ew} we characterize the range of C~{\small III}] EWs in LRDs. In both subsections, we compare these measurements to those of star-forming galaxies at the same redshifts, with the goal of establishing whether the narrow-line properties of LRDs are typical of the broader galaxy population.

\subsection{Nitrogen Enhanced Gas in LRDs} \label{sec:N_line}


\begin{deluxetable*}{ccccccccc}
\tablecaption{LRDs at $z>4$ with new rest-frame UV spectroscopic constraints obtained from SPURS and DIVER programs}
\tablehead{
Program & ID & RA (deg) & Dec (deg) & $z_{\rm spec}$ & M$_{\rm UV}$ & N~{\scriptsize IV}] EW (\AA) & N~{\scriptsize III}] (\AA) & C~{\scriptsize III}] EW (\AA)
}
\startdata
SPURS & A2744-52 & $3.602778$ & $-30.419355$ & $4.725$ & $-18.1$ & -- & $<3.7$ & $16\pm1$ \\
SPURS & A2744-1069 & $3.583813$ & $-30.374517$ & $5.145$ & $-19.4$ & -- & $2.4\pm0.6$ & $11\pm1$ \\
SPURS & EGS-39 & $214.985672$ & $+52.956233$ & $5.200$ & $-19.4$ & -- & $<3.7$ & $18\pm2$ \\
SPURS & EGS-43 & $215.066442$ & $+52.941884$ & $6.271$ & $-19.6$ & $<7.9$ & $10\pm2$ & $<10$ \\
SPURS & EGS-45 & $214.990972$ & $+52.916527$ & $5.682$ & $-18.7$ & $6.0\pm1.0$ & $<4.6$ & $24\pm3$ \\
SPURS & EGS-187 & $215.020802$ & $+52.917789$ & $6.380$ & $-18.7$ & $<4.3$ & $<6.7$ & $<5.2$ \\
SPURS & GN-500 & $189.094351$ & $+62.198975$ & $4.880$ & $-18.2$ & -- & $<4.5$ & $19\pm3$ \\
DIVER & 1008671 & $189.159025$ & $+62.260221$ & $4.416$ & $-18.6$ & $<4.2$ & $<4.4$ & $<3.0$ \\
DIVER & 1020621 & $189.125779$ & $+62.287404$ & $4.681$ & $-18.7$ & $<23$ & $<21$ & $<16$ \\
DIVER & 1033320 & $189.159764$ & $+62.295924$ & $4.486$ & $-18.6$ & $<7.4$ & $<8.6$ & $<5.0$ \\
DIVER & 1034620 & $189.179302$ & $+62.292533$ & $5.191$ & $-20.2$ & $<7.6$ & $<15$ & $<15$ \\
\enddata
\tablecomments{N~{\scriptsize IV}], N~{\scriptsize III}], and C~{\scriptsize III}] EWs are the integrated EWs of N~{\scriptsize IV}] doublet, N~{\scriptsize III}] quintuplet, and C~{\scriptsize III}] doublet, respectively. For non-detections, $5\sigma$ upper limits of EWs are provided. For the 4 LRDs in DIVER observations (S. Cai et al. in prep), H$\alpha$ emission lines have been characterized with grism spectra \citep{Matthee2024,Zhang2026}, and here we list the rest-frame UV emission lines measured from the new G140M spectra.}
\label{tab:new_lrd}
\end{deluxetable*}

Each of the four LRDs considered in this paper has N~{\small IV}] or N~{\small III}] detections that suggest nitrogen-enriched (high N/O or N/C) abundance patterns (Section~\ref{sec:n4n3}). The apparent ubiquity of the UV nitrogen emission in the LRDs is striking compared to the rarity of such detections in the broader galaxy population \citep[e.g.,][]{Topping2025a}. If nitrogen-enhanced abundance patterns are more common in LRDs, it would not only hold clues as to the physics of the narrow-line gas in LRDs, but it would also add to our understanding of the physics that drives the nitrogen enhancements. Of course, with a sample of just four LRDs, definitive conclusions are not possible. In this section, we consider the archive to characterize how commonly N~{\small IV}] or N~{\small III}]-emitting LRDs appear in the full LRD population at $z>4$. We will then compare against similar literature on the incidence of nitrogen emission in the full galaxy population at $z>4$.

To achieve this goal, we construct a large sample of LRDs at $z>4$ with publicly available NIRSpec spectra and constrain the strengths of N~{\small IV}] and N~{\small III}] emission lines. 
To maximize the sample size, we take advantage of LRDs with either grating spectra or prism spectra covering the rest-frame UV wavelengths. 
For LRDs with NIRSpec grating spectra, we identify 20 objects which have been presented in the literature \citep{Harikane2023,Maiolino2024b,Kocevski2025,Tang2025,Juodzbalis2026b,Papovich2026}. 
We also utilize the DIVER (GO 8018, PI: X. Lin) dataset whose G140M spectra are publicly available. 
The DIVER survey adds 5 LRDs to our database (which will be presented by the DIVER team in S. Cai et al. in prep). 
H$\alpha$ emission lines of these 5 LRDs have been characterized with NIRCam grism spectra \citep{Matthee2024,Zhang2026}, and we measure rest-frame UV emission lines using the new G140M spectra.
The DIVER G140M spectrum in 1 of 5 LRDs has been reported in \citet{Mascia2026}. 
We then add 12 LRDs with SPURS grating spectra, including Abell2744-QSO1 \citep{Tang2026b}. 
We report 11 LRDs with new rest-frame UV spectroscopic constraints identified from SPURS and DIVER programs in Table~\ref{tab:new_lrd}. 
For LRDs with prism spectra, we utilize the sample described in \citet{deGraaff2025c}. 
This adds 85 more LRDs at $z>4$, after cross-matching LRDs with grating observations and removing repeat sources. 
Overall these yield a sample of 122 LRDs at $z>4$ with rest-frame UV spectroscopic coverage, including 37 with grating observations and 85 with only prism observations. 


\begin{figure*}
\includegraphics[width=\linewidth]{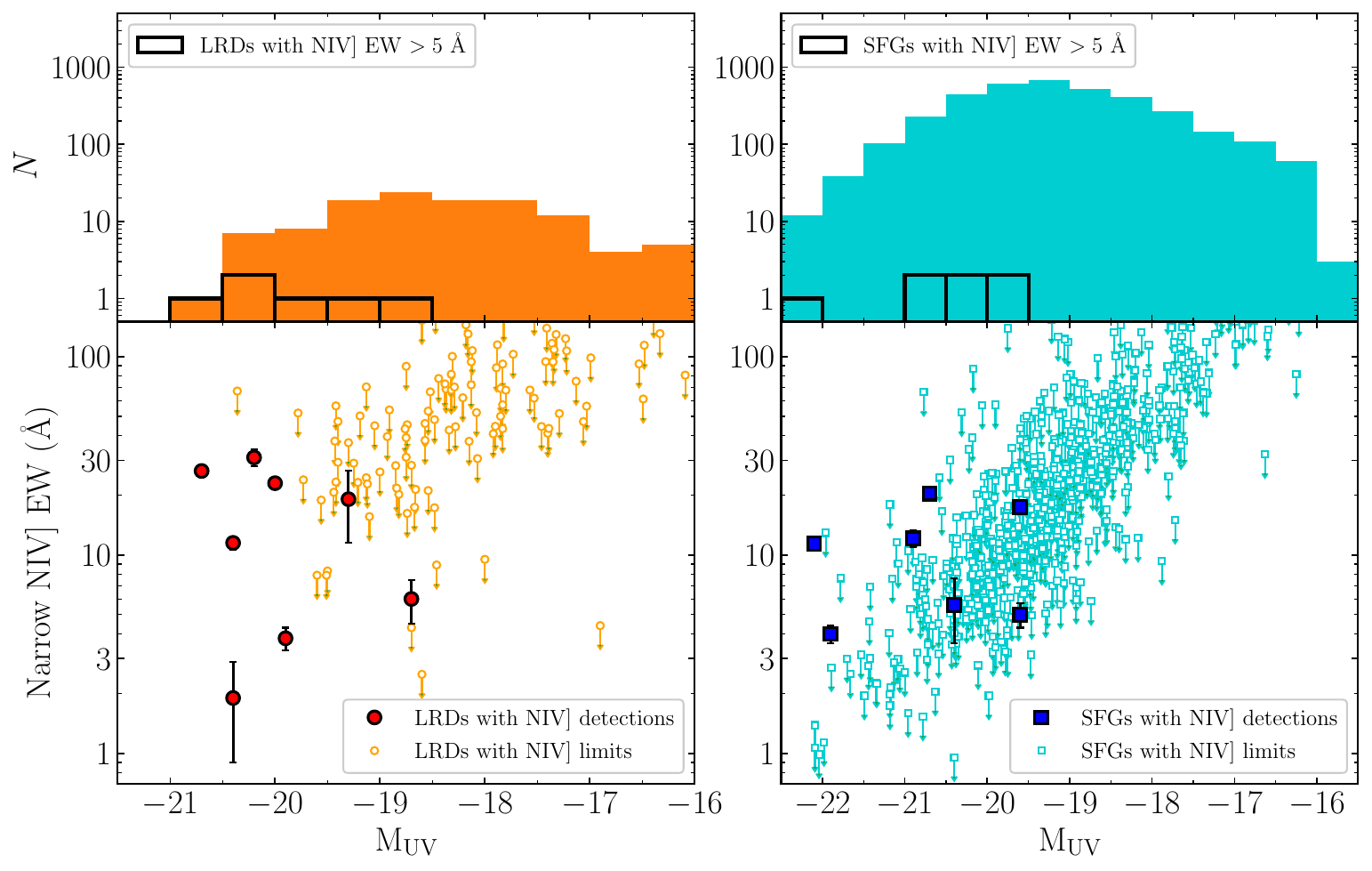}
\caption{N~{\scriptsize IV}] EW versus M$_{\rm UV}$ of LRDs (left panel) and full galaxy population (right panel) at $z>4$. LRDs with N~{\scriptsize IV}] detections are shown as red circles, and star-forming galaxies (SFGs) are shown as blue squares \citep{Castellano2024,Maiolino2024a,Schaerer2024,Topping2024,Topping2025a,Chen2026b}. The $5\sigma$ upper limits of non-detections are shown as orange open circles for LRDs (cyan open squares for galaxies). On the top of each panel, we provide the M$_{\rm UV}$ distributions, highlighting those with N~{\scriptsize IV}] EW $>5$~\AA\ with black colors.}
\label{fig:N_emitter}
\end{figure*}

The 122 LRDs span a redshift range from $4.05-9.29$. 
We use the NIRSpec spectra of these 122 LRDs reduced in a self-consistent way. 
For the 110 LRDs that are not in the SPURS sample, we utilize their spectra reduced by the DAWN JWST Archive (DJA; \citealt{Heintz2024,deGraaff2025a}) with \texttt{msaexp}. 
We reduce the SPURS spectra of the remaining 12 LRDs with \texttt{msaexp} following the same procedures as DJA (see Section~\ref{sec:spurs}). 
To derive the absolute UV magnitudes, we also collect the NIRCam photometry of these 122 LRDs from DJA \citep{Valentino2025}. 
The M$_{\rm UV}$ of these 122 LRDs are shown in Figure~\ref{fig:N_emitter}, with a median value of $-18.4$ and the $16-84$th percentiles of $-19.4$ to $-17.3$. 

We characterize the N~{\small IV}] and N~{\small III}] emission lines in these 122 LRDs by visually inspecting their 2D spectra. Details of the nitrogen emitters are listed in Table~\ref{tab:new_lrd}.  
We identify 7 LRDs showing N~{\small IV}] emission with doublet S/N $>5$, and an additional LRD with N~{\small IV}] with S/N $=3$ (GN-29, see Section~\ref{sec:n4n3}). 
We detect N~{\small III}]~$\lambda1746-1754$ emission with integrated S/N $>5$ in 9 LRDs. 
There are 5 LRDs with spectra that detect both N~{\small IV}] and N~{\small III}]. 
These form a sample of 12 LRDs with detections of either N~{\small IV}] or N~{\small III}]. 
Four of these twelve LRDs are presented in this paper, and the UV nitrogen line detections of another five LRDs have been reported previously in literature (UNCOVER-45924, \citealt{Labbe2024}; CANUCS-LRD-z8.6, \citealt{Tripodi2025b,Morishita2026}; C3PO-45290, C3PO-46403, \citealt{Papovich2026}; GN-16813, \citealt{Mascia2026}). 
The N~{\small IV}] or N~{\small III}] emission lines of the remaining three LRDs are newly-identified from SPURS program (SPURS-A2744-1069, SPURS-EGS-43, SPURS-EGS-45; Table~\ref{tab:new_lrd}), and we show their N~{\small IV}] or N~{\small III}] spectra in Figure~\ref{fig:new_lrd_uv} in appendix. 
We will present the new N~{\small IV}] and N~{\small III}] detections in more detail in a future paper. 
The N~{\small IV}] or N~{\small III}] EWs of these 12 LRDs are measured following the same methodology described in Section~\ref{sec:spurs}. 

The vast majority (110) of the LRDs do not show N~{\small IV}] and N~{\small III}] detections in their NIRSpec spectra. 
However, given the widely varying continuum strengths and sensitivity limits among different programs, the EW limits of non-detections span a wide range. 
For LRDs with grating spectra, we estimate the flux limits of each component of N~{\small IV}] doublet and N~{\small III}] quintuplet by integrating the error spectrum over two resolution elements ($600$~km~s$^{-1}$ for $R\simeq1000$; $222$~km~s$^{-1}$ for $R\simeq2700$) in quadrature. 
For LRDs with low-resolution ($R\sim100$) prism spectrum only, we estimate the total flux limit of N~{\small IV}] and N~{\small III}] by integrating the error spectrum over two resolution elements ($6000$~km~s$^{-1}$) in quadrature as it is impossible to resolve the individual components. 
We measure the underlying continuum from the spectrum and compute the EW limit. 
For the  N~{\small IV}] doublets of these 110 LRDs, we derive $16-50-84$th percentiles of $5\sigma$ limiting EWs of $9$~\AA, $48$~\AA, and $115$~\AA, respectively. 
The $16-50-84$th percentiles of $5\sigma$ limiting N~{\small III}] EWs are $13$, $71$, and $173$~\AA. 
This demonstrates that many of the LRDs do not have deep enough spectra to place a useful limit on the N~{\small IV}] or N~{\small III}] EW. 

To put robust constraints on the fraction of LRDs presenting UV nitrogen emission lines, we focus on the subset with $5\sigma$ EW limit reaching $5$~\AA\ for the N~{\small IV}] doublet or the N~{\small III}] quintuplet. 
This yields a sample of 18 LRDs out of the total 122. 
These 18 LRDs are brighter in the UV continuum than the full LRD population, with $16-50-84$th percentiles of M$_{\rm UV}$ of $-20.4$, $-19.4$, $-18.2$, respectively. 
The 18 LRDs include all the 12 objects with N~{\small IV}] or N~{\small III}] detections described above. 
Ten of these twelve LRDs present N~{\small IV}] or N~{\small III}] emission with EWs $>5$~\AA. 
This results in a fraction of $56^{+14}_{-14}\%$ LRDs presenting relatively strong N~{\small IV}] or N~{\small III}] emission lines. 
This suggests that UV nitrogen emission which links to nitrogen enhanced gas is likely very common among the LRD population. 

We compare the incidence of UV nitrogen emission in LRDs against that of the full galaxy population at $z>4$. 
Recent studies have reported that the N~{\small IV}] and N~{\small III}] emission lines are generally weak in a large population of star-forming galaxies. 
\citet{Isobe2025} have created a composite prism spectrum of 665 non-AGN galaxies at $z=4-7$ and did not detect N~{\small IV}] or N~{\small III}] emission lines. 
\citet{SinghRai2026} have stacked grating spectra of 135 galaxies at $z=6-10$, revealing very weak average N~{\small IV}] (EW $=1.1$~\AA) and N~{\small III}] emission (EW $=0.4$~\AA). 
Strong UV nitrogen emission is also rare among the galaxy population. 
Using a sample of 747 galaxies at $z>4$ constructed from publicly available NIRSpec observations, \citet{Topping2025a} have shown that only $7\%$ of galaxies with sufficiently deep spectra exhibit N~{\small IV}] with EW $>5$~\AA. 
We compare the N~{\small IV}] emitter fraction in LRDs with that in \citet{Topping2025a}.
To make a fair comparison, we select LRDs following the same criteria in \citet{Topping2025a}: identifying objects with deep enough spectra with $5\sigma$ limiting N~{\small IV}] EW reaching $5$~\AA. 
This leaves 11 LRDs from our sample of 122 LRDs. 
We find 6 of these 11 LRDs presenting N~{\small IV}] emission with EW $>5$~\AA, indicating a N~{\small IV}] emitter fraction of $55^{+18}_{-19}\%$. 
Though the sample size is small, the N~{\small IV}] emitter fraction of LRDs is 8 times that of the galaxy population ($7\%$; Figure~\ref{fig:N_emitter}). 


\begin{figure*}
\includegraphics[width=\linewidth]{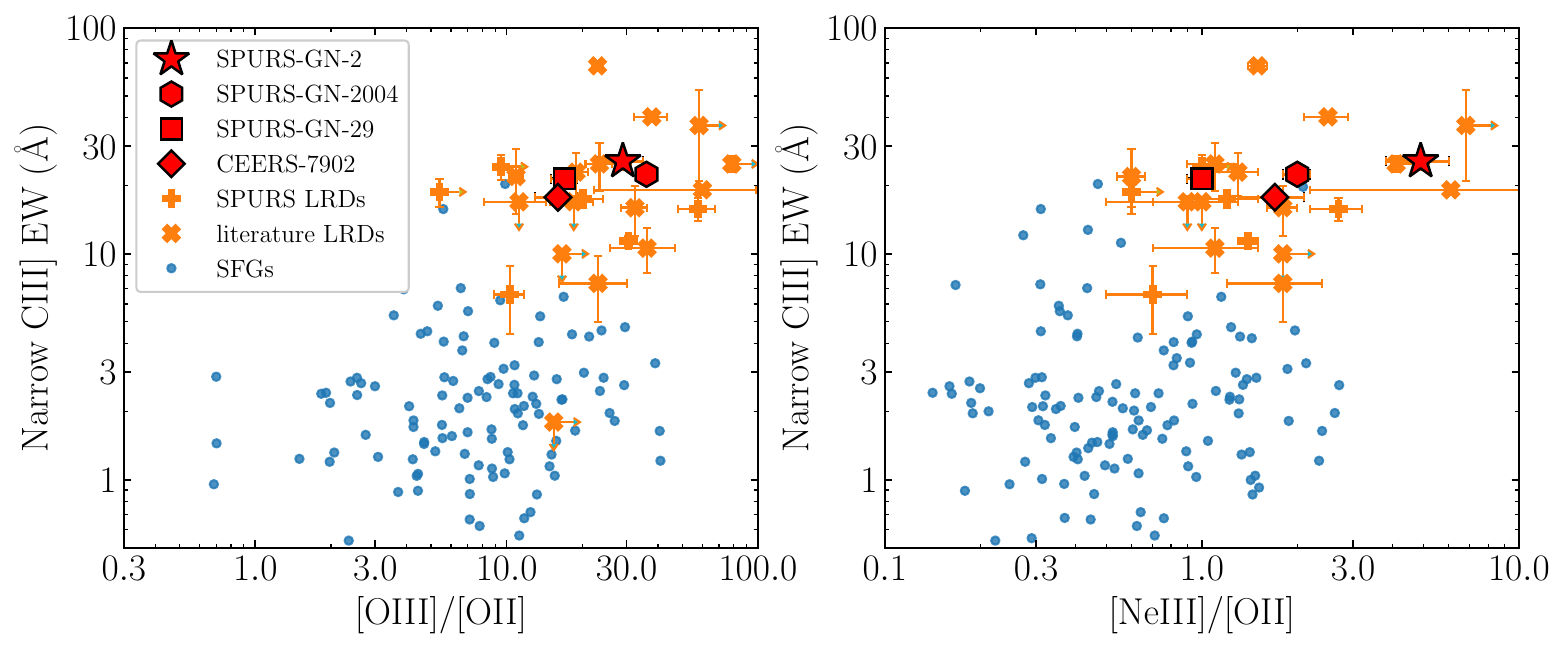}
\caption{C~{\scriptsize III}] EW versus O32 (left) and C~{\scriptsize III}] EW versus Ne3O2 (right) of LRDs. We plot GN-2, GN-2004, GN-29, and CEERS-7902 as red star, red hexagon, red square, and red diamond, respectively. We also show other LRDs with SPURS spectra (orange plus symbols) or in literature (orange cross symbols; \citealt{Harikane2023,Maiolino2024b,deGraaff2025c,Kocevski2025,Tang2025,Juodzbalis2026b,Mascia2026,Morishita2026,Papovich2026}). For comparison, we overplot star-forming galaxies \citep{Curtis-Lake2026,Scholtz2026a,Tang2026a} as blue circles.}
\label{fig:c3ew}
\end{figure*}

Overall, the analysis of the full archive suggests that N~{\small IV}] and N~{\small III}] are fairly common in LRDs. We find that 12 of 18 LRDs with deep spectra show either N~{\small IV}] and N~{\small III}], including 10 with EW $>10$~\AA, implying a nitrogen emitter fraction of $56^{+14}_{-14}\%$. Considering just the subset with deep N~{\small IV}] constraints, this fraction is similar ($55^{+18}_{-19}\%$). While larger grating datasets will be required to reduce the statistical uncertainties, the current samples already suggest that LRDs are more likely to have nitrogen line detections than the full galaxy population. While the physics responsible for nitrogen enhancements has been discussed extensively in the literature \citep[e.g.,][]{Charbonnel2023,Kobayashi2024,Schaerer2024,Senchyna2024,Topping2025a,McClymont2026}, there has not been as much focus on why LRDs also show similar abundance patterns. We will return to the topic of nitrogen in LRDs in Section~\ref{sec:discussion}. 

\subsection{C~{\small III}] EW Distribution in LRDs} \label{sec:ciii_ew}

The C~{\small III}] EW provides a diagnostic of the physical conditions of the narrow-line gas. We have shown that the four LRDs presented in this paper have very large narrow C~{\small III}] EWs, ranging from $18$~\AA\ to $26$~\AA\ (Section~\ref{sec:c3si3o3}). We have reported that these values are well above what is seen in typical star-forming galaxies at similar redshifts \citep[e.g.,][]{Roberts-Borsani2024,SinghRai2026,Tang2026a}. If such large C~{\small III}] EWs are the norm in LRDs, it would hold important clues as to the nature of the ionizing sources powering the narrow-line gas. Using the NIRSpec archive, we characterize the range of C~{\small III}] EWs seen in LRDs.

We again utilize the sample of 122 LRDs at $z>4$ constructed in Section~\ref{sec:N_line}, searching for C~{\small III}] emission lines. 
We identify 20 LRDs with C~{\small III}] line detections. 
We detect C~{\small III}] emission from the archival spectra of 7 LRDs \citep{Labbe2024,Maiolino2024b,deGraaff2025c}, and we also recover the C~{\small III}] detections of the 4 LRDs reported in \citet{Mascia2026,Morishita2026,Papovich2026}. 
The remaining 9 LRDs with C~{\small III}] are from the SPURS program (Table~\ref{tab:new_lrd}), including the 4 presented in this paper. 
The new C~{\small III}] detections of the other SPURS LRDs are shown in Figure~\ref{fig:new_lrd_uv} in the appendix. 
We measure the C~{\small III}] EWs following the same method described in Section~\ref{sec:spurs}. 
The C~{\small III} EWs of these 20 LRDs are generally large, with $16-50-84$ percentiles of $16$~\AA, $22$~\AA, and $26$~\AA\ (Figure~\ref{fig:c3ew}). 
These are above the average C~{\small III} EWs of star forming galaxies, as measured from composite spectra at similar redshifts ($6-8$~\AA; \citealt{Roberts-Borsani2024,SinghRai2026,Tang2026a}). 
We do not find C~{\small III}] emission in the remaining 102 LRDs. 
The $16-50-84$ percentiles of $5\sigma$ limiting C~{\small III}] EWs of these 102 LRDs are $25$~\AA, $55$~\AA, and $112$~\AA.

We plot the C~{\small III}] EWs of LRDs as a function of the O32 and Ne3O2 ratios in Figure~\ref{fig:c3ew}, and we overlay the distribution of galaxies at similar redshifts. It is clear that LRDs lie in a distinct region of the diagram relative to star forming galaxies, with both larger C~{\small III}] EW and larger O32 and Ne3O2. The O32 and Ne3O2 ratios are sensitive to both ionization parameter and electron density. As we will show below, the offset between LRDs and galaxies can be explained by a combination of effects, with larger ionization parameters, larger electron densities, younger stellar population ages, or additional ionizing sources present in the LRDs.  

We now quantify the C~{\small III}] EW distribution of the LRDs.
Here we consider sources with C~{\small III}] EW $>20$~\AA. 
Such strong C~{\small III}] is thought to trace extremely young stellar populations ($<10$~Myr assuming constant star formation history; e.g., \citealt{Jaskot2016,Mainali2020,Tang2021}) or AGN activity \citep{Nakajima2018}. 
Among the 122 LRDs, 31 have deep spectra with $5\sigma$ limiting C~{\small III}] EWs reaching $20$~\AA. 
We identify 10 LRDs with C~{\small III}] detections with EWs $>20$~\AA. 
This results in a strong C~{\small III}] fraction of $32^{+11}_{-9}\%$ for LRDs. 
We compare this fraction against that of the full galaxy population at similar redshift. 
In \citet{Tang2026a}, we have assembled a sample of 401 galaxies at $6<z<9$ from the archival NIRSpec observations, and we found that $11^{+3}_{-3}\%$ of the galaxies show C~{\small III}] with EWs $>20$~\AA. Based on this comparison, it appears that strong C~{\small III}] emission is roughly three times more common in LRDs than in the full galaxy population at $z\simeq7$ (Figure~\ref{fig:c3ew}). 

To understand what the strong C~{\small III}] emission may be revealing about LRDs, we first utilize the \texttt{BEAGLE} tool \citep{Chevallard2016} to explore the stellar populations required to reproduce C~{\small III}] with EW $>20$~\AA. 
Assuming a constant star formation history, it requires young stellar populations with $<10$~Myr with moderate metallicities ($\simeq0.1\ Z_{\odot}$) to produce C~{\small III}] EW $>20$~\AA. 
The two largest C~{\small III}] EWs seen in LRDs are $68$~\AA\ (UNCOVER-45924; \citealt{Labbe2024}) and $40$~\AA\ (C3PO-46403; \citealt{Papovich2026}). 
It requires extremely young ($\simeq2$~Myr) stellar populations to power C~{\small III}] emission with such large EWs.
The UV line diagnostic ratios are consistent with a stellar origin of the strong C~{\small III}] emission in LRDs. 
The large O~{\small III}]/He~{\small II} ($1.2-5.0$) and C~{\small III}]/He~{\small II} ($4.3-11$) flux ratios seen in the four LRDs are consistent with photoionization models driven by star formation \citep[e.g.,][]{Feltre2016,Gutkin2016,Mignoli2019}. 

Alternatively, it is possible that the strong C~{\small III}] EWs of LRDs may reflect an additional ionizing source, perhaps associated with a central AGN. It has been shown that very large C~{\small III}] EWs are often associated with AGN activity at lower redshifts \citep{Nakajima2018}. While LRDs are likely mostly enshrouded in dense gas, it is possible that the covering fraction is not uniform, allowing some ionizing radiation to escape to the narrow line region \citep{Tang2026b}. Indeed some theoretical models for LRDs suggest that they may have polar cavities where ionizing radiation may be able to escape to the narrow line region \citep{Madau2026}.
While the line ratios are consistent with stellar photoionization, it is possible that the ionizing spectrum associated with LRDs may be softer than most AGNs at lower redshifts.
 
In summary, we find that the C~{\small III}] EWs of LRDs are not typical of the galaxy population. They may imply that the host galaxies of LRDs are in the midst of extremely strong bursts of star formation. Among the galaxy population, such bursts are commonly associated with large ionization parameters and high ionized gas densities, both of which would explain the elevated O32 and Ne3O2 ratios we observe in the LRDs. In this picture, the host galaxies of LRDs would preferentially trace the highest specific-star-formation-rate subset of the galaxy population. However, we cannot rule out the possibility that the strong narrow lines instead reflect a significant ionizing contribution from AGN activity. In the following section, we explore whether the SPURS observations reveal massive star signatures that can help distinguish between these two scenarios.

\section{Discussion} \label{sec:discussion}

The deep rest-frame UV spectra of these UV-bright LRDs provide our first opportunity to understand the nature and origin of the UV emission from LRDs, both on scales close to the central engine and within the interstellar medium.
In this section we discuss the implications of our discovery of broad C~{\small IV} emission and dense narrow-line gas for the LRD inner structure (Section~\ref{sec:broad_CIV}), the implications of the interstellar absorption for the nature of the UV continuum (Section~\ref{sec:UV_origin}) and constraints on stellar populations (Section~\ref{sec:vms}) in the LRDs. Finally we discuss the origins of nitrogen enhancement in the LRDs and their potential links to the broader population of nitrogen-emitting galaxies (Section~\ref{sec:nitrogen}).

\subsection{Broad C~{\small IV} and Ultra Dense Narrow-Line Gas} \label{sec:broad_CIV}

The elevated C~{\small III}] EWs we measure (Section~\ref{sec:ciii_ew}) can be explained by either a very young stellar population, or a significant ionizing contribution from the central engine. Distinguishing between these requires evidence that UV continuum photons from the disk or broad-line region reach the narrow-line-emitting gas. Such leakage is a prediction of the orientation-based model of \citet{Madau2026}, in which the gas which makes up the broad-line region occupies an equatorially-concentrated solid angle. This geometry is found in simulations of super-Eddington accretion flows, which indicate that an evacuated, low-density polar structure should be common \citep[e.g.,][]{Jiang2014,Jiang2019,Sadowski2016,Zhang2025}. In this picture, the polar angles will see less of the extremely dense gas clouds, so a substantial fraction of the UV and ionizing continuum should reach the narrow line region. Porosity in the dense gas coverage may also transmit UV radiation beyond the broad-line region \citep[e.g.,][]{Ji2026b,Sok2026,Tang2026b}. Such covering fraction effects have been argued to explain a subset of narrow high ionization lines \citep[e.g.,][]{Papovich2026}.

The broad permitted UV lines offer a test of such leakage, provided the chosen transition is both readily produced by the central engine and minimally attenuated by the surrounding gas on its way out. Bound-free absorption from the $n=2$ level of hydrogen, the process responsible for the Balmer break, extends to all wavelengths shortward of the Balmer limit ($3646$~\AA), but its cross section falls steeply, roughly as $\nu^{-3}$, for photons of increasingly higher energy. Mg~{\small II}~$\lambda2796$ sits close enough to this threshold that it remains strongly attenuated.
C~{\small IV}~$\lambda1548$, at nearly twice the threshold photon energy, will face less attenuation. While He~{\small II}~$\lambda1640$ is also far from the Balmer limit, it is very sensitive to the ionizing spectrum, so absence of 
He~{\small II} is therefore ambiguous between absorption and an intrinsically soft ionizing spectrum. Among the available transitions, C~{\small IV} is therefore best suited to this test.

We have shown in this paper that a subset of LRDs have broad C~{\small IV} emission. GN-2, the bluest source in our sample, shows the strongest broad C~{\small IV} emission. CEERS-7902, the reddest of the three with C~{\small IV} coverage, shows broad C~{\small IV} suppressed by roughly an order of magnitude relative to GN-2. GN-29 is nearly as red as CEERS-7902. It is a weaker C~{\small IV} emitter overall, with its narrow component fainter than the narrow components of GN-2 and CEERS-7902. Its non-detection of a broad component is therefore inconclusive. Scaling its narrow C~{\small IV} flux by the broad-to-narrow ratio in GN2 predicts an expected broad flux below our detection limit. Finally the fourth LRD presented in this paper, GN-2004, does not have spectral coverage of the C~{\small IV} profile.

The presence of broad C~{\small IV} emission indicates that far-UV radiation from the broad-line region or cocoon escapes the surrounding dense gas in a subset of LRDs. If ionizing radiation also reaches the narrow-line emitting gas, it would imply non-stellar contributions to the narrow emission lines in LRDs. This may help explain some of the extreme narrow line properties seen in LRDs, including the very large C~{\small III}] EWs we introduced in Section~\ref{sec:ciii_ew}. We note that (similarly) broad He~{\small II} remains undetected even in GN-2 and CEERS-7902, where broad C~{\small IV} demonstrably escapes. We have argued that this favors an intrinsically soft ionizing spectrum, as has been argued previously \citep[e.g.,][]{Lambrides2026,Wang2026} and not obscuration as the explanation for the absence of broad He~{\small II}.

GN-2 is the clearest example of broad C~{\small IV} transmission in our sample. This LRD has the bluest optical and UV colors, and it is the only source with no detected Balmer line absorption. Its Balmer emission lines have among the largest narrow-to-broad ratios in our sample. Its narrow UV lines imply electron densities of $\gtrsim 1.0\times10^6$~cm$^{-3}$ from C~{\small III}] and $\gtrsim 2.5\times10^6$~cm$^{-3}$ from N~{\small IV}] (Section~\ref{sec:c3si3o3} and \ref{sec:n4n3}), roughly two orders of magnitude above the densities ($\sim1\times10^4$~cm$^{-3}$) typical of the galaxy population at similar redshift \citep{Topping2025a,SinghRai2026,Umeda2026}. These properties can be explained within an orientation-based picture. If our sightline to GN-2 intersects a low-density polar funnel, the blue optical and UV colors may be partially related to leaking continuum emission from the central engine. It is possible that this less obscured sightline allows a deeper view into the AGN, revealing narrow-line gas that is photoionized along the cavity walls. This may boost the narrow-line luminosity and allow visibility of extremely dense material that would not be viewed along more obscured sightlines. 
Super-Eddington accretion is also predicted to drive fast outflows along the polar direction \citep[e.g.,][]{Jiang2014,Jiang2019,Sadowski2016,Zhang2025}, meaning we should expect to see the fastest outflow signatures in the bluest LRDs. This may be consistent with the broad \ion{He}{2} P-Cygni profiles we detect in GN-2 and GN-2004 (Section~\ref{sec:he2}), and the broad emission components in \ion{N}{4}], \ion{N}{3}], and \ion{C}{3}] in GN-2 (Section~\ref{sec:n4n3}).

We emphasize that the optical color of GN-2 classifies it as an LRD, but it is close to the boundary of the little blue dots. In the orientation framework of \citet{Madau2026}, the little blue dots are predicted to be viewed close to pole-on  (see below for an alternative interpretation from \citealt{Sneppen2026a}). In the context of \citet{Madau2026}, this suggests GN-2 may represent an intermediate viewing angle, more polar than the other three LRDs in our sample, but not the fully pole-on view associated with little blue dots. In contrast, the redder colors of GN-29 and CEERS-7902 may reflect a more edge-on viewing angle, resulting in more obscuration of the broad C~{\small IV} photons. Larger spectroscopic samples are required to establish correlations beyond the simple trends we can see with our small SPURS database.

The strength of the broad C~{\small IV} of GN-2 also can be explained within the electron-scattering cocoon picture. It has the weakest Balmer break and the narrowest broad H$\alpha$ width. 
In the context of the column-density sequence linking little blue dots to LRDs \citep{Sneppen2026a}, these suggest GN-2 has a lower column density than the other LRDs in the sample. We note, this could also be explained by orientation effects in these models if the column density is reduced along polar sightlines \citep{Mascia2026}.
This reduced column density (and the associated reduction in $n=2$ bound-free absorption) may explain its transmitted broad C~{\small IV}. 
If the dense neutral gas surrounding the cocoon is clumpy or porous, or has a reduced covering fraction, the transmission of broad C~{\small IV} may be further boosted. 

As C~{\small IV} is a resonant emission line, it can be a sensitive tracer of ionized gas density and kinematics. Future detailed simulations incorporating C~{\small IV} radiative transfer into LRD models should prove promising for providing new constraints on the gas close to the central engine.
As discussed above, we emphasize that GN-2 is at the boundary of the LRD and little blue dot populations. Even in the context of the dense gas cocoon, this is potentially a regime where there is still significant transmission of UV light from the central nucleus, either in terms of nebular continuum or hot blackbody emission from the disk. We will discuss this possibility in more detail in the next subsection.

\subsection{Deeply Embedded UV Emission in SPURS LRDs} \label{sec:UV_origin}

The origin of the UV emission has been one of the central challenges in understanding LRDs.
The dense neutral gas invoked to explain their optical properties should strongly suppress the UV continuum \citep[e.g.,][]{deGraaff2025a,Liu2025,Sneppen2026a}.
The majority of the UV continuum is generally attributed to the host galaxy, though emission from the central source may contribute \citep[e.g.,][]{deGraaff2025c,Rinaldi2025,Cloonan2026,Inayoshi2026,Sun2026,Zhang2026}.

The SPURS spectra open a new window into LRDs' UV continua: revealing a plethora of absorption and fluorescent emission features with no analog in typical star-forming galaxies.
Interpreting these features requires first establishing where in the four LRDs this UV continuum originates.
NIRCam imaging has shown LRDs are often more extended and clumpy in the UV than the optical \citep[e.g.,][]{Rinaldi2025,Cloonan2026,Zhang2026}, motivating the picture that their UV emission is dominated by the host galaxy. 
However, LRDs with compact UV emission make up the majority of the spectroscopically-confirmed population: \citet{Ando2026} find that the UV sizes in the spectroscopic LRD sample of \citet{deGraaff2025c} are significantly smaller than those of star-forming galaxies at fixed UV magnitude and redshift, with many unresolved.
Recent prism integral field unit (IFU) spectroscopy of several LRDs has also shown the dominant UV emission component to be spatially compact, and coincident with the optical emission \citep{Ishikawa2026}.
The four SPURS LRDs belong to this UV-compact population.
Our PSF decomposition demonstrates the majority of their UV emission arises from an unresolved region co-spatial with the optical emission (Section~\ref{sec:psf}).
Thus, SPURS spectroscopy of these four LRDs probes the UV continuum emerging from or extremely close to the optical-emitting region.

The SPURS spectra reveal the majority of the UV continuum in these LRDs is heavily enshrouded in gas.
The detections of strong Ly$\alpha$ damping wings, saturated resonant absorption lines, high EW low oscillator-strength transitions and strong fine-structure absorption lines all indicate a large covering fraction of neutral gas with an extremely high column density ($N_{\rm HI}>10^{22}$~cm$^{-2}$) surrounding the UV continuum in all four LRDs (Section~\ref{sec:abs}).
We also detect multiple Ly$\alpha$- and Ly$\beta$-pumped fluorescent emission lines and a forest of narrow Fe~{\small II} emission lines in the NUV (Section~\ref{sec:low_ion}) that imply comparable neutral gas columns, and that the UV continuum and Lyman series emitting regions must be close to the absorbing gas.
These features are essentially negligible in star-forming galaxies at lower and similar redshifts to the LRDs \citep{Shapley2003,Jones2013,Nakane2026,KeerthiVasan2026}, and imply significantly higher column density gas conditions than we would expect if typical star-forming regions dominate the UV emission.
Together, these results indicate the UV continuum in the SPURS LRDs is deeply embedded behind a very high column density of gas.

The extreme column densities and strong NUV \ion{Fe}{2} forest we detect also offer a potential explanation for the rarity of broad Ly$\alpha$ emission across the LRD population. 
Broad Ly$\alpha$ has, to-date, only been detected in two LRDs \citep{Morishita2026,Tang2026b,Ji2026b,Geris2026,Kageura2026}, including Abell2744-QSO1, the lowest-metallicity LRD identified so far \citep{Tang2026b,Ji2026b}, despite being expected to be produced along with the broad Balmer emission.
Because a large number of \ion{Fe}{2} transitions fall within $\lesssim1000$~km~s$^{-1}$ of the Ly$\alpha$ resonant wavelength \citep{Johansson1984,Sigut2003}, a large \ion{Fe}{2} column density will provide significant opacity to broad Ly$\alpha$ photons emerging from the central source. 
This is consistent with the strong \ion{Fe}{2}~$\lambda2508$ emission we detect in GN-2, GN-2004, and CEERS-7902, which requires Ly$\alpha$ photons at $>600$~km~s$^{-1}$ to reach the \ion{Fe}{2} gas to pump this transition.
Of these, only CEERS-7902 has coverage of Ly$\alpha$, and we do not detect emission. This implies broad Ly$\alpha$ does reach the \ion{Fe}{2} gas, but negligible flux can escape. 
Conversely, Abell2744-QSO1, which does show broad Ly$\alpha$ \citep{Tang2026b,Ji2026b} has a very low metallicity \citep{Maiolino2025} and no clear \ion{Fe}{2} forest, suggesting a lower \ion{Fe}{2} column density. 
Future spectroscopy of the lowest-metallicity LRDs with weak \ion{Fe}{2} emission may reveal more broad Ly$\alpha$ detections.

Given these results, a key question is what powers the UV continuum. 
One possibility is that the UV continuum originates from the central engine of the LRD, either from the accretion disk itself or nebular emission produced by the dense gas invoked to explain the Balmer break.
As the high H~{\small I} column densities required to produce a Balmer break ($N_{\rm HI}>10^{23}$~cm$^{-2}$) would produce a stronger damped Ly$\alpha$ wing than we infer, this implies the UV continuum would need to escape the region where the Balmer break forms along sightlines with lower column densities, as has been proposed in some LRD models \citep{Madau2026,Mascia2026}. 
As discussed above, this emission could escape along low-density polar sightlines, or through a porous or clumpy gas envelope.

A second possibility is that the UV continuum comes from young massive stars embedded within an extended gas envelope around the central engine, exterior to the region where the Balmer break forms \citep[see also, e.g.,][]{Ando2026,Inayoshi2026}. 
To test whether the neutral gas column densities we infer are consistent with this geometry, we adopt the fiducial dense gas cocoon model of \citet{Sneppen2026a}.
While the ionized cocoon in that model is compact ($<0.05$~pc), \citet{Sneppen2026a} note it would be depleted rapidly by accretion and outflows, requiring a larger neutral gas reservoir to sustain the LRD.
Thus we extend the outer radius to $100$~pc, sufficient for a gas depletion timescale $>10$~Myr assuming a radiative efficiency $\eta \sim 0.1$. 
In this configuration, stars within $\lesssim 8$~pc of the nucleus would sit behind a neutral gas column of $N_{\rm HI} \gtrsim 10^{22.6}$~cm$^{-2}$, comparable to the columns we infer from the Ly$\alpha$ damping wings and absorption lines.
This is a conservative estimate: the fiducial model's core column density is extremely high ($N_H \approx 10^{26}$\,cm$^{-2}$), so a lower core density would require stars even closer to the nucleus to reach the same column.

Thus, massive stars embedded deep within a dense gas envelope surrounding the optical emitting region remain a plausible origin for the UV continuum.
Indeed, gravitational instabilities in dense gas disks have been proposed to trigger star formation, with the resulting stars and clusters able to migrate inward via disk interactions to build up nuclear star clusters \citep{Mayer2025,Inayoshi2026}.
If these stars dominate the UV continuum, we may expect to detect their wind and photospheric signatures, which we discuss in more detail below.

Together, these results show that the UV continuum in these LRDs emerges from a compact region deeply embedded in dense gas, possibly probing stars within $\simeq 8$~pc of the central engine. This may indicate that these LRD hosts are in the process of building-up of nuclear star clusters. Since stellar densities are likely to be very high, we may expect dynamical interactions to shape the resulting stellar populations: runaway collisions could build up very massive stars \citep[e.g.,][]{PortegiesZwart2002,Fujii2024,Rantala2025}, close binary interactions could eject nitrogen-rich material via mass transfer or mergers \citep{Senchyna2024}, and tidal disruption of stars by a growing central black hole could expose nitrogen-rich core material as debris \citep[e.g.,][]{Cameron2023}. 
Whether these results are representative of the LRD population will ultimately require larger samples with deep rest-frame UV spectroscopy.

\subsection{Characterizing Massive Star Features in LRDs} \label{sec:vms}

\begin{figure*}
\includegraphics[width=\linewidth]{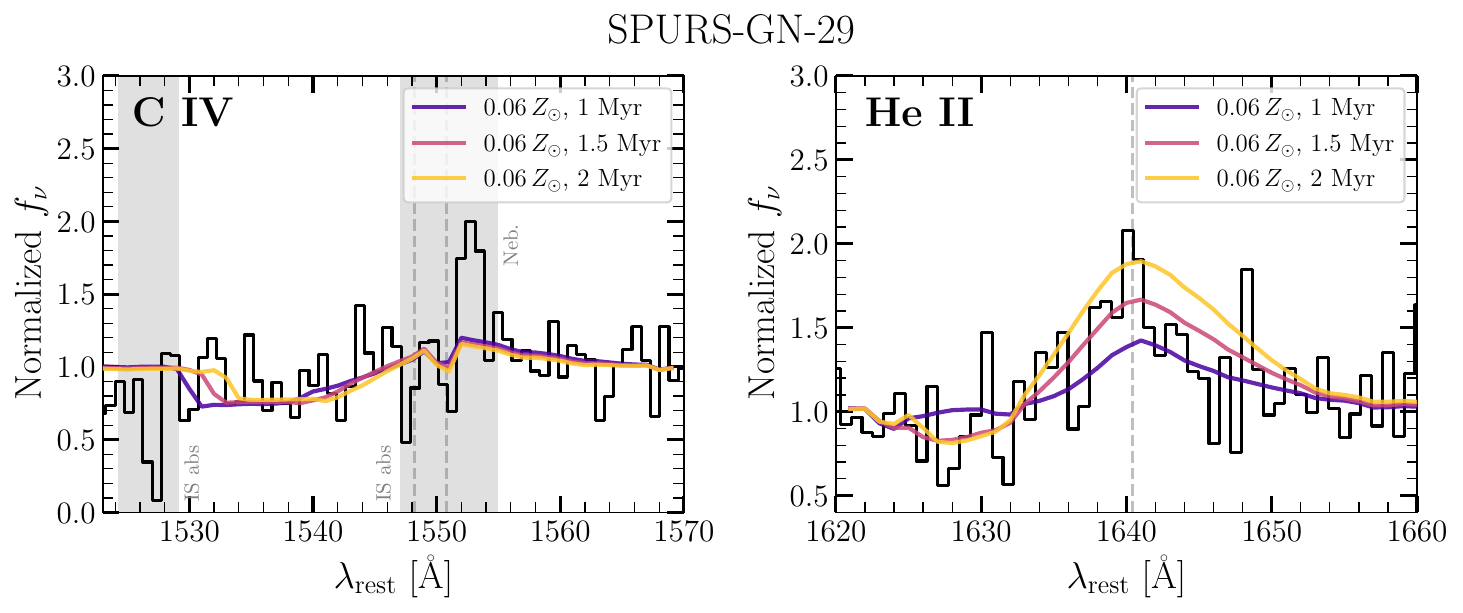}
\caption{Stellar wind features in GN-29: the observed \ion{C}{4} profile (left) and the detected broad \ion{He}{2}~$\lambda1640$ emission (right). We compare the continuum-normalized spectrum (black) to predicted stellar wind profiles from the \citet{Martins2025} SSP models including VMSs, computed at the low metallicity expected for this source ($Z_{\star}=0.06\,Z_{\odot}$). Colored lines show models in which the optically thick wind mass-loss rate is not scaled down with $Z_{\star}$ but is instead fixed to the LMC value. Both the strong broad \ion{He}{2} emission and the weak \ion{C}{4} P-Cygni absorption can be well reproduced by these VMS models at young SSP ages ($1-2$~Myr), despite this low metallicity, provided that the wind mass loss is enhanced. Grey bands mask the interstellar \ion{Si}{2}~$\lambda1526$ and \ion{C}{4} absorption and the narrow nebular \ion{C}{4} emission.}
\label{fig:vms_gn29}
\end{figure*}

We have shown in Section~\ref{sec:uv_emi} that the four LRDs all host highly nitrogen-enhanced ionized gas and high-EW UV emission lines, all of which have now also been observed in a small number of compact, low metallicity galaxies at $z>6$ \citep[e.g.][]{Bunker2023,Cameron2023,Marques-Chaves2024,Schaerer2024,Senchyna2024,Topping2024}. Recently ultra-deep moderate resolution UV spectroscopy has been obtained through SPURS for a subset of the nitrogen emitters. The spectra reveal strong stellar-wind P-Cygni profiles in \ion{C}{4} and \ion{N}{5} and broad \ion{He}{2} emission, that can be explained via the presence of a population of VMSs \citep[$M_{\star}\gtrsim100\,M_{\odot}$;][]{Chen2026b,Marques-Chaves2026}.
It is possible that the winds of VMS drive out the CNO-processed, nitrogen-rich material exposed at their surfaces into the ISM \citep[e.g.,][]{Vink2023,Shi2026}. The deep spectra have also revealed a broad (FWHM $=1078-1670$~km~s$^{-1}$) component to N~{\small IV}]~$\lambda1486$, likely tracing dense WN-like winds or luminous blue variable (LBV)-like outbursts from a population of very massive stars, though an AGN-driven wind cannot be excluded \citep{Chen2026b,Nakane2026}. In either scenario, this broad component may trace the gas delivering the nitrogen enrichment. If the nitrogen enrichment follows similar channels in LRDs, we may expect to see similar wind features in SPURS observations presented in this paper.

We briefly summarize potential stellar features in the four SPURS LRDs. As discussed in Section~\ref{sec:he2}, we detect a broad He~{\small II} emission in GN-29, with an EW of $4.1\pm1.2$\,\AA.
The He~{\small II} profile can be well described with a single Gaussian with a significantly larger (by a factor of 3) FWHM ($929\pm211$~km~s$^{-1}$) compared to narrow forbidden lines, yet narrower than the broad Balmer lines. This feature may be most likely explained through optically thick winds of the most massive O stars near the Eddington limit or from WR stars \citep{Schaerer1996,Crowther2007,Brinchmann2008,Senchyna2021}.
The EW of the broad \ion{He}{2} emission we measure is slightly larger than what is measured in GN-z11 \citep{Chen2026b,Nakane2026} while also comparable to other $z>6$ galaxies suggested to host VMSs \citep{Yang2026}. 
As shown in Figure~\ref{fig:heii}, the \ion{He}{2} emission in both GN-2 and GN-2004 shows a likely P-Cygni profile, with broad blueshifted absorption extending to $\approx2900$ and $2200$~km~s$^{-1}$ respectively. These broad \ion{He}{2} absorption profiles may probe fast AGN-driven winds or VMS populations \citep[e.g.,][]{Martins2025}.
A broad \ion{He}{2} with EW $=4.0$~\AA\ is not ruled out in CEERS-7902, where the value is calculated as $3\sigma$ upper limit assuming the same FWHM as measured in GN-29.
If a recently formed massive star population is present, we also expect to see a strong \ion{N}{5} P-Cygni profile, which is covered by the SPURS observations in three of the four LRDs.
However, as we have shown in Section~\ref{sec:abs_dens}, the spectra near the \ion{N}{5} resonance are strongly attenuated by the neutral \ion{H}{1} at extremely large column densities ($N_{\rm HI} \approx10^{22.0-22.7}$~cm$^{-2}$), leaving an effective transmission of only $0.01-0.04$ at $1239$~\AA\ and washing out their P-Cygni features.
We also look for a \ion{C}{4} P-Cygni profile, but both GN-2 and CEERS-7902 show strong absorption on top of broad \ion{C}{4} emission, making it challenging to separate the stellar wind component, while the \ion{C}{4} in GN-2004 falls into the detector gap. However, a weak \ion{C}{4} P-Cygni feature is seen in GN-29, allowing us to characterize the stellar population using both \ion{C}{4} and \ion{He}{2} wind lines, which we focus on below.

To determine whether the features may be consistent with an origin in massive stars, we compare the observed stellar wind profile in GN-29 with stellar population synthesis models including very massive stars up to $300~M_\odot$.
We consider the models presented in \citet{Martins2025}, which include two variants that differ in their mass-loss prescriptions of optically thick winds ($\dot{M}_{\rm thick}$), one in which the VMS mass-loss rate is calibrated on the LMC \citep[following][]{Grafener2021} and is consequently near-independent of metallicity, and one in which this rate decreases at a lower metallicity.
We consider models at a stellar metallicity that is informed by the gas-phase oxygen abundance, $12+\log({\rm O/H}) = 8.13\pm0.10$ ($0.28\,Z_{\odot}$), derived in Section~\ref{sec:opt}.
As the stellar metallicity is set by iron production dominated by Type~Ia supernovae that is delayed relative to oxygen (dominated by Type~II supernovae), we expect high-redshift galaxies to be $\alpha$/Fe-enhanced.
To approximate this effect, we consider models at $Z_{\star}=0.06\,Z_{\odot}$, consistent with the factor of five ratio found between gas-phase and stellar metallicities found among $z\simeq2-3$ galaxies \citep[e.g.][]{Steidel2016,Sanders2020,Topping2020,Runco2021} and expected from theoretical limit for Type~II supernova yields \citep[$\simeq0.2\,({\rm O/Fe})_{\odot}$;][]{Nomoto2006}.
We also consider models at higher metallicities where these winds are easier to launch, but we consider them unlikely for this source.

As shown in the right panel of Figure~\ref{fig:vms_gn29}, the \cite{Martins2025} models at $0.06\,Z_{\odot}$ with $\dot{M}_{\rm thick}$ fixed to LMC metallicity reproduce the observed profile at ages of $1-1.5$~Myr, predicting equivalent widths of $4.1$ and $7.0$~\AA\ that bracket the observed $4.1\pm1.2$~\AA.
We note that the models with $\dot{M}_{\rm thick}$ decreasing as a function of metallicity produce at most $1.9$~\AA\ of \ion{He}{2} emission over the $1-2$~Myr ages at which VMS dominate, and no more than $3.7$~\AA\ even at an implausibly high $0.2\,Z_{\odot}$, suggesting that enhanced mass-loss rates in the VMS are required to match the observed profile.
As the \ion{C}{4} P-Cygni absorption is known to trace metallicity for integrated light clusters at fixed other properties, we expect it to weaken as metallicity decreases \citep[e.g.][]{Rix2004,Leitherer2010,Leitherer2011,VidalGarcia2017,Chisholm2019}.
As expected, at the low metallicity relevant for GN-29, the same models also produce a weak \ion{C}{4} P-Cygni profile (EW $=-1.9$ to $-2.1$~\AA\ at $1-1.5$~Myr), in agreement with our data ($-1.5\pm0.6$~\AA).

We also consider other possibilities that may explain the broad \ion{He}{2} without requiring very strong stellar winds, but find them to be unlikely.
Given the much smaller (by $3\times$) FWHM of \ion{He}{2} in GN-29 compared to the Balmer lines, they are unlikely to share the same kinematics or originate from the BLR \citep[e.g.][]{Clavel1991}.
In addition, in classical type I AGN, broad \ion{He}{2} is accompanied by broad \ion{C}{4} emission line of comparable or greater strength \citep{VandenBerk2001,Telfer2002}, whereas the \ion{C}{4} emission of GN-29 is narrow (FWHM $=318\pm70$~km~s$^{-1}$). 
The electron scattering that has been proposed for the broad Balmer lines of LRDs \citep{Rusakov2026,Sneppen2026a}  may also produce a broad \ion{He}{2} line.
However, since the Thomson cross-section is independent of wavelength, its width should also be roughly comparable to those of the Balmer lines, making it unlikely to explain the very different widths we measure for these lines. Winds from an AGN disk may also produce a broad or P-Cygni \ion{He}{2}. For the broad-absorption-line and accretion-disk winds that dominate the quasar population, broad \ion{He}{2} is generally found with much larger FWHM ($5000-20\,000$~km~s$^{-1}$; e.g., \citealt{Weymann1991,Murray1995,Gibson2009}).
We cannot rule out the case that the outflows associated with the LRDs may be considerably slower, and indeed in the case of GN-2 and GN-2004 we have suggested that the \ion{He}{2} P-Cygni profiles may probe AGN-driven winds seen through the polar axis. 

\begin{figure}
\includegraphics[width=\linewidth]{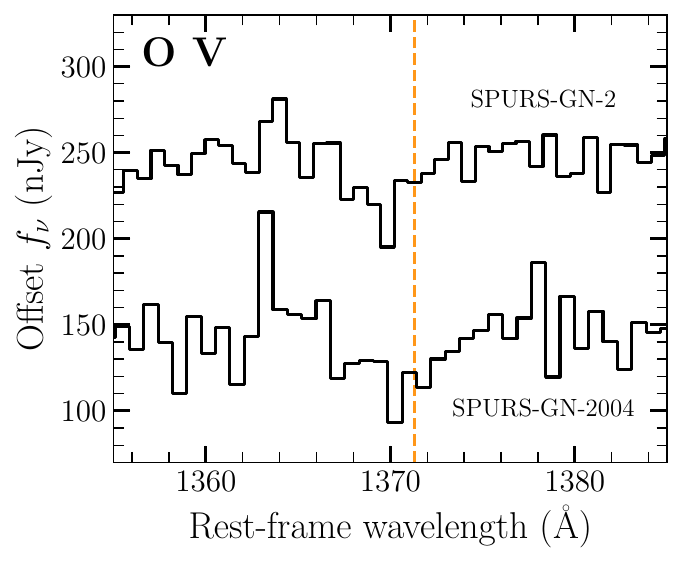}
\caption{Zoom-in on the \ion{O}{5}~$\lambda1371$ region in the observed spectra of GN-2 (top) and GN-2004 (bottom).
Both sources show a broad absorption feature spanning $\approx1367$--$1372$~\AA, consistent with being from stellar winds of VMS or hot WR populations.}
\label{fig:vms_ov}
\end{figure}

We also search for other features that might trace the outflows suggested by the \ion{He}{2} profiles.
In addition to the broad \ion{N}{4}] discussed further below, we detect broad \ion{C}{3}] in GN-2, which can also be powered in WR atmospheres \citep[e.g.][]{Hillier1989,Hamann1992,Niedzielski1994}.
We search for other lines which typically accompany these features; and find that the data for both GN-2 and GN-2004 is potentially consistent with blueshifted absorption in \ion{O}{5}~$\lambda 1371$
and emission in \ion{C}{3}~$\lambda 2297$, both permitted transitions with a significantly elevated lower level (the direct analogues of \ion{N}{4}~$\lambda 1719$ for these ions) which are common in hot stellar atmospheres and rare to entirely absent in other environments.
Indeed, \ion{O}{5}~$\lambda 1371$ in particular is rare enough in lower-mass stars but strong enough at high temperatures to be employed as a tracer of VMS and hot WR populations in integrated light spectra, emerging in many young clusters including R136 \citep[e.g.][]{woffordRareEncounterVery2014,smithVeryMassiveStar2016,crowtherR136StarCluster2016,martinsInferringPresenceVery2023}.
If confirmed, this \ion{O}{5} absorption (see Figure~\ref{fig:vms_ov}) would suggest that a large fraction of the UV continuum is likely stellar in origin (or otherwise cloaked in exceptionally hot and dense gas).
We emphasize that these features and their associations are tentative at the current SNR and will require dedicated modeling; but this suggests that deeper spectroscopy, which could more clearly constrain these fainter features, might shed significant light on the origin of these other broad UV lines.

To summarize, we detect wind signatures in the SPURS LRD spectra that may indicate the presence of VMSs. In the case of  GN-29, we have demonstrated that population synthesis models with VMS are able to reproduce the observed strong, broad \ion{He}{2} and weak \ion{C}{4}. In GN-2 and GN-2004, the \ion{He}{2} profiles appear P-Cygni in shape, with very fast winds ($\gtrsim 2200$~km~s$^{-1}$). Whether these trace winds are powered by massive stars or an AGN is not clear, but in either case the winds may be important for delivering the nitrogen-enhanced abundance pattern to the ISM, as suggested in \citet{Chen2026b}. We consider the prevalence of nitrogen emission in the LRDs in the next subsection.

\subsection{Nitrogen Emission in LRDs} \label{sec:nitrogen}

\begin{figure}
\includegraphics[width=\columnwidth]{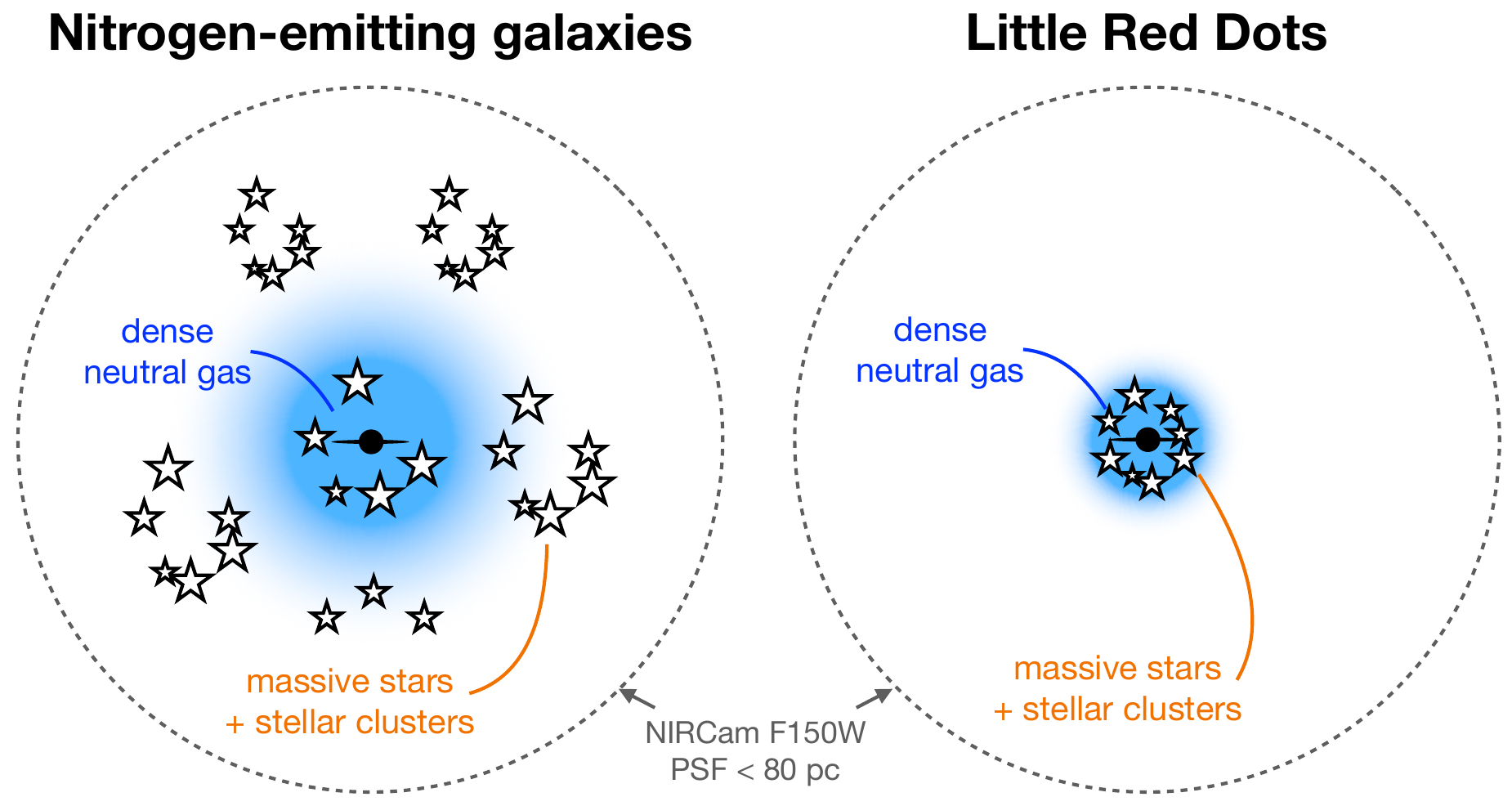}
\caption{Schematic illustrating similarities and differences between LRDs and nitrogen emitting galaxies and the sub-PSF dense stellar systems they potentially trace.}
\label{fig:N_cartoon}
\end{figure}

We have shown that all four LRDs in our paper host highly nitrogen-enhanced gas (Section~\ref{sec:n4n3}). 
Our archival analysis demonstrates this is not unique to our small sample: nitrogen emission appears roughly $8\times$ more common among LRDs than in the broader galaxy population at these redshifts (Section~\ref{sec:N_line}). 
The origin of nitrogen enhancement in early galaxies has been debated in the literature \citep[e.g.,][]{Cameron2023,Charbonnel2023,Kobayashi2024,Senchyna2024,McClymont2026}, and its prevalence in LRDs may offer new clues to the physical conditions that produce it.

JWST has now obtained spectra for over a thousand galaxies at $z>6$ \citep[e.g.,][]{Heintz2025,Valentino2025}, spanning a wide diversity of ISM conditions and stellar populations. Yet within this range, LRDs and nitrogen-emitting galaxies\footnote{We will refer to \ion{N}{4}] and \ion{N}{3}] emitters that are not LRDs as nitrogen-emitting galaxies.} do not appear to be broadly distributed; instead, their properties overlap, occupying only a small region of the parameter space populated by the full sample (see Figure~\ref{fig:N_cartoon} for a schematic).
One of the hallmarks of both populations is extremely compact UV emission ($\lesssim 80$~pc), suggesting a very dense complex of star clusters \citep[e.g.,][]{Tacchella2023,Topping2024}. In addition, both show high ionized gas densities (Section~\ref{sec:n4n3}, \citealt{Topping2024,Topping2025a,Maiolino2024a,Chen2026b,Papovich2026}), fluorescent O~{\small I}* emission, requiring dense neutral gas close to the ionizing sources \citep{Choe2025,Tripodi2025a,Chen2026b,Kokorev2026}, and in some cases, potential signatures of very massive stars \citep[see Section~\ref{sec:vms};][]{Chen2026b}. 

These similarities point to a common origin of nitrogen enhancement within high density environments. In the LRDs studied in this paper, the gas columns we infer are consistent with a UV-emitting source confined to within as little as $\sim8$~pc of the central engine, pointing to an environment where nuclear clusters are being assembled.
Dynamical interactions are likely to be common in dense star clusters on such small scales, with runaway stellar collisions potentially building-up populations of VMS \citep[e.g.,][]{Rantala2024,Rantala2026}. It is even plausible that supermassive stars (SMS; $M_\star >10^3\,M_\odot$) might be effectively formed in these dense metal poor conditions \citep{Chon2025,Gieles2025}. There are many papers emphasizing why such high density environments may spark the nitrogen enhancement. The winds of VMS or SMS populations can efficiently transport CNO-processed, nitrogen-rich material into the surrounding gas, offering a natural explanation for the elevated N/O and N/C ratios seen in both LRDs and nitrogen emitters \citep[e.g.,][]{Charbonnel2023,Vink2023,Nandal2025,Shi2026}.
Close binary interactions among massive stars in these same dense environments can eject comparably nitrogen-rich material through non-conservative mass transfer or stellar mergers \citep{Senchyna2024}.

These same dense nuclear environments also raise the possibility that the central black hole itself contributes directly to the nitrogen budget, through tidal disruption events (TDEs). In sufficiently dense clusters, close stellar encounters can readily scatter stars onto orbits that bring them within the tidal disruption radius of the central black hole. Since intermediate- and high-mass stars carry substantial CNO-processed material in their cores, the resulting debris can be strongly nitrogen-enhanced, offering an additional enrichment channel \citep{Cameron2023,Mockler2024,Watanabe2026}. The high frequency of nitrogen enhancement among LRDs suggests this possibility deserves serious consideration. Whether the nitrogen-emitting galaxies also harbor supermassive black holes in their nuclei remains debated \citep[e.g.,][]{Larson2023,Maiolino2024a}. Deep observations from SPURS have revealed broad N~{\small IV}] emission lines in nitrogen-emitting galaxies \citep{Chen2026b,Nakane2026}, suggesting the presence of dense, nitrogen-rich outflows. This has been interpreted as evidence for WN-like winds or LBV-like outbursts, though an AGN-driven wind cannot be ruled out. In the latter case, rather than producing the nitrogen-rich material directly, the AGN outflow may instead be driving out a pre-existing reservoir of TDE or dense cluster debris, polluting the surrounding interstellar medium with nitrogen-rich material. The presence of broad N~{\small IV}] in GN-2 (Section~\ref{sec:n4n3}) suggests such winds may be present in a subset of LRDs as well. 

Despite their similarities, the nitrogen-emitting galaxies' spectra also differ from LRDs in several ways. They do not show the prominent red optical colors of LRDs, and their spectra exhibit neither Balmer breaks nor Balmer line absorption. While samples remain small, deep UV spectra from SPURS have revealed another key difference: low-ionization interstellar absorption lines are much weaker in nitrogen-emitting galaxies than in LRDs, implying a lower covering fraction of neutral gas and a UV continuum that is less deeply embedded than in LRDs. It may be that these two populations represent a continuum along the density sequence of high-redshift galaxies, with LRDs corresponding to the most concentrated and densest systems. 
The nitrogen-emitting galaxies may share a similar, albeit slightly less dense, nuclear structure to LRDs ($\lesssim10$~pc scales). But while the UV light of LRDs we present in this paper may be very highly concentrated (resulting in the strong absorption lines)\footnote{We note that LRDs are often surrounded by blue clumps, but these are often found on larger physical scales ($>100$~pc) than we are discussing here \citep{Baggen2026,Pacucci2026}.}, the UV output of the nitrogen-emitting galaxies is instead dominated by surrounding young star clusters ($\lesssim100$~pc scales) that are not deeply embedded within neutral gas\footnote{It is worth emphasizing that the young star clusters that are proposed to surround the nucleus in this picture remain unresolved at NIRCam resolution.}. This framework was recently proposed by \citet{Chen2026b} to simultaneously explain the presence of broad Ly$\alpha$ and fluorescent emission lines alongside weak low-ionization absorption lines in the nitrogen-emitting galaxy GN-z11. This picture remains speculative owing to the small samples with deep spectroscopy. 
Larger spectroscopic databases are required to test whether the LRDs and nitrogen-emitting galaxies studied to-date are representative.

\section{Summary} \label{sec:summary}

We present ultra-deep (29 hours) JWST/NIRSpec G140M (rest-frame UV) spectroscopy of four UV-bright (M$_{\rm UV}<-20$) LRDs in the GOODS-N and EGS fields at $z\simeq7$: GN-2, GN-2004, GN-29, and CEERS-7902. 
The data were obtained as part of the SPURS Cycle 4 Large Program.
These observations provide the deepest rest-frame UV view of LRDs yet obtained at medium resolution ($R\simeq1000$) 
and provide new insights into the origin of LRD's UV emission and the conditions responsible for extreme nitrogen enhancements in the early universe.
The ultra-deep spectra reveal the first detections of broad high ionization UV emission lines in LRDs and provide the first view of the gas surrounding the UV continuum-emitting sources in LRDs.
We summarize the key results below.

1. With the SPURS spectra, we characterize the rest-frame UV to optical spectroscopic properties of the four LRDs. 
These systems span wide ranges of continuum slopes among the LRD population. 
GN-2 and GN-2004 are blue from rest-frame FUV ($\beta_{\rm FUV}=-2.39$ to $-1.95$) to optical ($\beta_{\rm opt}=0.06-0.09$), while GN-29 and CEERS-7902 are redder ($\beta_{\rm FUV}=-1.41$ to $-0.76$, $\beta_{\rm opt}=0.40-0.48$). 
The Balmer lines show both narrow and broad emission, with larger narrow-to-broad H$\alpha$ flux ratios in GN-2 and GN-2004 than GN-29 and CEERS-7902. 
These may reflect different relative contributions of host galaxy or AGN versus dense gas cocoon emission among the four LRDs. 
We find large ionization-sensitive line ratios (O32 $=16-36$, Ne3O2 $=1.0-4.9$) in all four LRDs, likely indicating both large ionization parameters and high electron densities in the narrow-line emitting gas. 

2. The ultra-deep spectra reveal the first detections of broad high ionization lines in the UV. 
We detect a strong broad (FWHM $=2830$~km~s$^{-1}$) C~{\small IV} emission line in the G140M spectrum of GN-2 and a weaker broad C~{\small IV} in CEERS-7902. 
The broad C~{\small IV} to H$\beta$ flux ratios of both LRDs are well below those of typical type I AGN, consistent with previous results suggesting LRDs have soft ionizing spectra. 
We also detect many other narrow UV emission lines (N~{\small IV}], He~{\small II}, O~{\small III}], Si~{\small III}], C~{\small III}]) in the four LRDs. 
We detect nitrogen emission lines in all four LRDs, which indicate nitrogen-enhanced abundance patterns with super-solar N/O and N/C ratios.
In GN-2 we detect broad emission components in N~{\small IV}] (similar to that detected in GN-z11; \citealt{Chen2026b}), N~{\small III}], and C~{\small III}], indicating fast moving ionized gas.
Using the density-sensitive N~{\small IV}], Si~{\small III}], and C~{\small III}] doublet ratios, we derive large electron densities for the four LRDs, well above that of typical galaxies at $z\simeq7$. 
In particular, GN-2 has an extremely high electron density ($\gtrsim10^6$~cm$^{-3}$) that is two orders of magnitude above the typical $z\simeq7$ value.

3. The continuum spectra of all four LRDs show strong downturns near the Ly$\alpha$ wavelength, extending to $\gtrsim1300$~\AA\ in the rest frame, and extremely strong absorption lines, including detections of usually weak fine-structure absorption.
Reproducing these downturns requires exceptionally large \ion{H}{1} column densities ($N_{\rm HI}\approx10^{22.0-22.7}$~cm$^{-2}$). 
The low-ionization absorption lines are strong and saturated, with EWs (median ranging $-2.2$~\AA\ to $-1.2$~\AA\ in each source) well above those measured in star-forming galaxies at comparable redshifts, and implying near-unity covering fractions ($C_f=0.7-1.0$). 
Together, these results indicate the UV-emitting regions are deeply embedded in neutral gas that covers nearly the entire source. 
We detect strong \ion{Al}{2}~$\lambda1670$ and \ion{Al}{3}~$\lambda\lambda1854,1862$ absorption, motivating investigation of aluminum abundances that potentially offer a signpost of supermassive and very massive star enrichment, but find that line saturation and ionization effects prohibit conclusive results.

4. Consistent with the picture from the absorption lines, we detect a suite of low-ionization fluorescent emission lines, indicating pumping by the UV continuum and Lyman series photons in dense neutral gas. 
Fluorescent \ion{O}{1}*~$\lambda1306$ in GN-2 and GN-2004 requires Ly$\beta$ pumping in dense \ion{H}{1} gas close to the ionizing sources, while the rare semi-forbidden \ion{O}{1}]~$\lambda1641$ line detected in GN-2 implies extreme column densities $N_{\rm HI}>10^{22.5}$~cm$^{-2}$. 
The NUV spectra of all four LRDs also show a forest of strong, narrow, permitted \ion{Fe}{2} emission, with the total NUV \ion{Fe}{2} EWs ranging from 72 to 356 \AA. (integrated over $2200-3090$~\AA).
Three LRDs show extremely strong narrow \ion{Fe}{2}~$\lambda2508$ emission (EW $=6-10$~\AA, deconvolved FWHM $\lesssim464$~km~s$^{-1}$), requiring pumping by broad Ly$\alpha$ in a dense, iron-rich region. We suggest this region may regulate the visibility of broad Ly$\alpha$ from the central engine in the LRD population.

5. We explore the incidence of N~{\small IV}], N~{\small III}] emission and strong C~{\small III}] emission in a sample of 122 $z>4$ LRDs with NIRSpec rest-frame UV spectroscopy. 
Strong C~{\small III}] (EW $>20$~\AA) in LRDs is about three times more common than in the galaxy population.
This suggests that the host galaxies of LRDs are more likely to be in the midst of intense bursts of star formation or with a significant ionizing contribution from an AGN. 
We find that $56^{+14}_{-14}\%$ of $z>4$ LRDs present N~{\small IV}] or N~{\small III}] detections with EW $>5$~\AA. 
This fraction is $\sim8\times$ larger than that of the full galaxy population \citep{Topping2025a}. 

6. We investigate the dense gas properties that can explain the strong broad \ion{C}{4} emission in GN-2. 
GN-2 has one of the bluest UV and optical colors among LRDs, no clear Balmer line absorption, and signatures of extremely dense and fast-moving ionized gas. 
The presence of broad \ion{C}{4} requires FUV radiation from the central engine to escape through the surrounding dense gas, suggesting GN-2 is viewed along a less obscured sightline than other LRDs in our sample. 
This could be explained by orientation-based models, in which sightlines closer to the polar direction intercept a smaller column of broad-line-region clouds and may be more aligned with ionized outflows from the accretion disk. 
Alternatively, in the dense gas cocoon picture, GN-2 may have a lower covering fraction or column density of dense neutral gas surrounding its nucleus than the rest of our sample.

7. The absorption lines and Ly$\alpha$ damping wing demonstrate the UV continuum in these four LRDs is deeply embedded in neutral gas. In our sample we find the UV continuum emerges from a compact, unresolved ($<80$~pc) region co-spatial with the optical emission.
The UV continuum could be consistent with escaping emission from the central engine along low column density sightlines, or young massive stars embedded within an extended dense gas envelope surrounding the central engine.
If massive stars produce the UV emission, we find they would need to be confined to within just $\lesssim8$~pc of the nucleus to be consistent with the column densities inferred from the UV absorption features, implying extremely concentrated stellar systems.

8. The spectra reveal \ion{He}{2} wind signatures in three LRDs that may indicate the presence of very massive stars (VMS), but can also be consistent with AGN-driven winds. In either case, we argue these winds could deliver the nitrogen-enhanced material into the ISM.
In GN-29, we detect a strong, broad \ion{He}{2} emission line ($\mathrm{EW}=4.1\pm1.2$~\AA, FWHM $=929\pm211$~km~s$^{-1}$) together with weak \ion{C}{4} P-Cygni absorption, which are well reproduced by stellar population synthesis models including VMS.
In GN-2 and GN-2004, the \ion{He}{2} profiles instead show blueshifted P-Cygni absorption, indicating fast winds out to $\gtrsim2200$~km~s$^{-1}$ that could be driven either by massive stars or by an AGN.

9. Among the broader population of $z>4$ galaxies observed with JWST, LRDs and nitrogen-emitting galaxies stand out as a small, tightly overlapping subset, sharing nitrogen-enhanced abundance patterns, elevated narrow-line electron densities, and compact UV emission not seen in typical star-forming systems at these redshifts. We argue that this points to a common origin in dense nuclear environments, where massive stars are concentrated in compact clusters surrounded by extremely dense gas, and where dynamical interactions, including runaway stellar collisions, binary mass transfer, and tidal disruption of stars by the central black hole, can efficiently enrich the surrounding gas with nitrogen. Despite these similarities, nitrogen-emitting galaxies show markedly weaker low-ionization absorption than LRDs, implying a lower covering fraction of neutral gas and a less deeply embedded UV continuum. We propose that LRDs and nitrogen emitters may represent a sequence in gas column density around a similar underlying nuclear structure, with LRDs at the densest, most deeply embedded extreme.

\begin{acknowledgments}

The authors thank Liang Dai, Yuzo Ishikawa, and Harley Katz for helpful conversations, St\'ephane Charlot and Jacopo Chevallard for providing access to the \texttt{BEAGLE} tool, and Fabrice Martins for sharing their population synthesis models.
MT is supported by the National Natural Science Foundation of China (grant No. 12673014).
DPS acknowledges support by the National Science Foundation under Grant No. AST-2109066. 
CAM acknowledges support by the European Union ERC grant RISES (101163035), Carlsberg Foundation (CF22-1322), and VILLUM FONDEN (37459).
ZC acknowledges support by VILLUM FONDEN under grant 37459.
SSG and VG acknowledge support by the Carlsberg Foundation (CF22-1322).
LW acknowledges support from the Gavin Boyle Fellowship at the Kavli Institute for Cosmology, Cambridge and from the Kavli Foundation. 
KVGC is supported by NASA through the STScI grants JWST-GO-03777, JWST-GO-04265, and JWST-GO-05974.
The Cosmic Dawn Center (DAWN) is funded by the Danish National Research Foundation under grant DNRF140.

This work is based in part on observations made with the NASA/ESA/CSA JWST. The data were obtained from the Mikulski Archive for Space Telescopes at the Space Telescope Science Institute, which is operated by the Association of Universities for Research in Astronomy, Inc., under NASA contract NAS 5-03127 for JWST. 
These observations that form the basis of this work are associated with program GO 9214 (SPURS).
The authors thank the program coordinator, Christian Soto, and NIRSpec reviewer, Diane Karakla.
This work also uses observations associated with the following publicly available programs: GTO 1180, 1181, GTO 1210, GTO 1286, GTO 1287, and GO 3215 (JADES, doi:10.17909/8tdj-8n28), ERS 1345 and DDT 2750 (CEERS, doi:10.17909/z7p0-8481), GO 2561 (UNCOVER), GO 4233 (RUBIES), GO 6368 (CAPERS), and GO 8018 (DIVER). 
The authors acknowledge the above teams led by Daniel Eisenstein \& Nora L\"uetzgendorf, K. Isaak, Steven L. Finkelstein, Pablo Arrabal Haro, Ivo Labb\'e \& Rachel Bezanson, Anna de Graaff \& Gabriel Brammer, Mark Dickinson, and Xiaojing Lin for developing their observing programs.
Some of the data products presented herein were retrieved from the Dawn JWST Archive (DJA). DJA is an initiative of the Cosmic Dawn Center (DAWN), which is funded by the Danish National Research Foundation under grant DNRF140.
The Tycho supercomputer hosted at the SCIENCE HPC center at the University of Copenhagen was used for supporting this work.

\end{acknowledgments}





%
\facilities{JWST(NIRSpec)}

\software{\texttt{msaexp} \citep{Brammer2023}, \texttt{emcee} \citep{Foreman-Mackey2013}, \texttt{pysersic} \citep{Pasha2023}, \texttt{fantasy} \citep{Ilic2023}, \texttt{NumPy} \citep{Harris2020}, \texttt{SciPy} \citep{Virtanen2020}, \texttt{Astropy} \citep{AstropyCollaboration2013,AstropyCollaboration2018,AstropyCollaboration2022}, \texttt{Matplotlib} \citep{Hunter2007}, \texttt{BEAGLE} \citep{Chevallard2016}.}


\appendix

\counterwithin{figure}{section}
\counterwithin{table}{section}

\section{BIC Analysis of Emission Line Fitting} \label{sec:line_fit}

The N~{\small IV}], N~{\small III}], C~{\small III}], He~{\small II}, and [O~{\small III}] emission lines of the four LRDs present complex profiles with multiple components. 
To interpret each of these lines, we fit the line profile with different models. 
We evaluate the goodness of fit of each model using the Bayesian Information Criterion (BIC; \citealt{Schwarz1978,Liddle2007}). 
The BIC of each model fit is defined as: BIC $\equiv -2\ln{\mathcal{L}}+k\ln{N}$, where $\mathcal{L}$ is the maximum likelihood achievable by the model, $k$ is the number of parameters of the model, and $N$ is the number of data points used in the fit.
When comparing two models, we take $\Delta$BIC $>10$ as very strong evidence that the model with the lower BIC provides a better fit to the data than the other. 
The best-fit results are presented in Table~\ref{tab:uv_lines} (for N~{\small IV}], N~{\small III}], C~{\small III}], He~{\small II} emission lines) and Table~\ref{tab:opt_lines} (for [O~{\small III}] lines). 

We first fit the C~{\small III}] profile of GN-2 with two narrow Gaussians, fixing the two centroids to the rest-frame wavelengths of [C~{\small III}]~$\lambda1907$, C~{\small III}]~$\lambda1909$. 
We allow the peak fluxes of the two Gaussians and the line width to vary as free parameters, but require the two line widths to be the same. 
Here we mask out the region with Fe~{\small III}~$\lambda1914$. 
We find excess emission on both blue and red sides of the C~{\small III}] profile, likely suggesting the presence of a broad C~{\small III}] component. 
With this model, we derive a large reduced chi-square value ($\chi^2_{\nu}=2.7$) and a large BIC value ($135$). 
We then simultaneously fit the C~{\small III}] of GN-2 with two narrow Gaussians and two broad Gaussians. 
Again we fix the centroids of the two broad Gaussians to the rest-frame wavelengths of the doublet.
The fitting results are presented in Section~\ref{sec:c3si3o3}. 
This model well reproduces the C~{\small III}] profile of GN-2 (Figure~\ref{fig:fuv_lines}) with broad C~{\small III}] with FWHM of $2271\pm371$~km~s$^{-1}$. 
We find much lower $\chi^2_{\nu}$ ($1.3$) and also lower BIC ($83$). 
The difference between BIC values of the two fits ($\Delta$BIC $=52$) indicates that the model with a broad component provides a much better fit to the C~{\small III}] profile. 

We also fit the C~{\small III}] profiles of GN-2004, GN-29, and CEERS-7902 with the two models described above: two narrow Gaussians only and a combination of two narrow and two broad Gaussians. 
For each of these three LRDs, we perform the BIC test for C~{\small III}]. 
We do not find a significant difference between fitting C~{\small III}] with both models ($\Delta$BIC $\leq1$). 
When fitting C~{\small III}] with two narrow and two broad Gaussians, the inferred broad C~{\small III}] components have S/N below 3. 
These indicate that we do not detect clear broad C~{\small III}] in these three LRDs (Figure~\ref{fig:fuv_lines}).

The SPURS spectrum of GN-2 reveals the detection of a broad N~{\small IV}] component. 
Motivated by this, we simultaneously fit the N~{\small IV}] profile of GN-2 with two narrow Gaussians and two broad Gaussians, with centroids fixed to the rest-frame wavelengths of [N~{\small IV}]~$\lambda1483$, N~{\small IV}]~$\lambda1486$. 
We present the fitting results in Section~\ref{sec:n4n3}. 
This model well reproduce the N~{\small IV}] profile (Figure~\ref{fig:fuv_lines}), with a $\chi^2_{\nu}$ of $1.1$ and a BIC of $66$. 
The best-fit broad N~{\small IV}] component has a FWHM of $1514\pm137$~km~s$^{-1}$. 
We also examine whether the N~{\small IV}] of GN-2 can be fitted by two narrow Gaussians only. 
With this fit, we derive considerably larger $\chi^2_{\nu}$ ($4.1$) and BIC ($159$).
The difference in BIC ($\Delta$BIC $=60$) suggests that the model including a broad component is strongly preferred. 

GN-29 and CEERS-7902 also show N~{\small IV}] detections. 
We fit each of the N~{\small IV}] profiles with models with and without a broad component. 
We then perform the BIC test for these fits.
For both CEERS-7902 and GN-29, we do not find a significant difference between fitting N~{\small IV}] with the two models ($\Delta$BIC $\leq1$). 
The broad components inferred from the model with broad Gaussians also have S/N below 3. 
These results demonstrate that there is no clear broad component underlying the N~{\small IV}] of these two LRDs (Figure~\ref{fig:fuv_lines}). 

We model the individual components of the N~{\small III}]~$\lambda1746-1754$ quintuplets of GN-2 and GN-2004, which show high S/N ($=17-27$) N~{\small III}] detections. 
We first fit the N~{\small III}] profile of GN-2 with five narrow Gaussian functions following the methodology described in \citet{Maiolino2024a} and \citet{Chen2026b}. 
The centroids of the five narrow Gaussians are fixed to the rest-frame wavelengths of individual quintuplet lines. 
We also require the line widths of the five components to be the same, and fix the following flux ratios according to the theoretical values: $f_{\rm NIII]\lambda1746}/f_{\rm NIII]\lambda1752}=0.14$, $f_{\rm NIII]\lambda1748}/f_{\rm NIII]\lambda1754}=0.95$. 
With this model, we derive a large $\chi^2_{\nu}$ value ($3.6$) and a large BIC ($151$). 
We find excess emission in the blue side of the N~{\small III}] profile, likely suggesting the presence of a broad N~{\small III}] component. 
We then model the N~{\small III}] profile of GN-2 with five narrow Gaussians and a broad Gaussian, allowing the centroid, peak flux, and line width of the broad component to vary freely.
The fitting results are presented in Section~\ref{sec:n4n3}. 
By adding the broad component (FWHM $=1797\pm219$~km~s$^{-1}$), we find a significantly lower $\chi^2_{\nu}$ ($1.6$) and lower BIC ($81$). 
The difference in BIC ($\Delta$BIC $=70$) indicates that the model with broad component indeed fits the N~{\small III}] profile of GN-2 better (Figure~\ref{fig:fuv_lines}). 
We perform the same fits to the N~{\small III}] emission of GN-2004. 
Again, we find that the model including a broad component is strongly preferred to reproduce the N~{\small III}] profile of GN-2004 with a much lower BIC ($27$) than the model with five narrow components only. 
For GN-29 and CEERS-7902, we do not resolve the N~{\small III}] quintuplet lines owing to the relatively low S/N ($=4-6$). 

The He~{\small II}~$\lambda1640$ emission line is coupled with nearby O~{\small I}]~$\lambda1641$ emission. 
To resolve both lines for the four LRDs, we fit each He~{\small II} profile with two Gaussians. 
We fix the two centroids to the rest-frame wavelengths of He~{\small II}~$\lambda1640$ and O~{\small I}]~$\lambda1641$, respectively. 
We allow the two peak fluxes and the line width of He~{\small II} to vary as free parameters, while we fix the line width of O~{\small I}] to be the same as O~{\small III}]~$\lambda1666$. 
This model well reproduces the He~{\small II} profile of GN-2 ($\chi^2_{\nu}=1.2$, BIC $=21$), and we identify the O~{\small I}]~$\lambda1641$ emission with S/N $=3$ (Section~\ref{sec:low_ion}). 
We also fit the He~{\small II} of GN-2 with a single Gaussian with centroid fixed at the rest-frame wavelength of He~{\small II}~$\lambda1640$. 
With this model, we find excess emission on the red side of the line profile with a large $\chi^2_{\nu}$ ($2.2$) and a large BIC ($35$). 
The difference in BIC ($\Delta$BIC $=14$) indicates that the model including both He~{\small II} and O~{\small I}] is significantly preferred over the model with He~{\small II} only. 
For GN-2004, GN-29, and CEERS-7902, we do not find O~{\small I}] (S/N $<1.5$) when modeling the line profile with two Gaussians. 
We then fit their He~{\small II} profiles with single Gaussians. 
With this model, we find no difference compared to fitting with two Gaussians ($\Delta$BIC $=0-2$), suggesting that there is no clear O~{\small I}]~$\lambda1641$ detection in these three LRDs (Figure~\ref{fig:heii}). 

Finally, we test whether the [O~{\small III}]~$\lambda4959$, $\lambda5007$ emission lines of the four LRDs present broad components, as have been revealed in a few LRDs in the literature \citep[e.g.,][]{Juodzbalis2024}. 
For each LRD, we consider two models: one fitting [O~{\small III}]~$\lambda4959$, $\lambda5007$ with two narrow Gaussians only, and one with two narrow and two broad Gaussians. 
We mask out the region including H$\beta$ when performing the fits.
For two narrow Gaussian model, we fix the centroids to the rest-frame wavelengths of [O~{\small III}]. 
We allow the peak flux and line width of [O~{\small III}]~$\lambda5007$ to vary as free parameters, fix the flux of [O~{\small III}]~$\lambda4959$ according to the theoretical line ratio ($f_{\rm [OIII]\lambda4959}:f_{\rm [OIII]\lambda5007}=1:2.98$), and require the line width of $\lambda4959$ to be the same as $\lambda5007$. 
With this model, we find excess emission in both blue and red sides of [O~{\small III}] profiles of each LRD, potentially indicating the presence of broad components. 
We then simultaneously fit the [O~{\small III}] doublet of each LRD with two narrow and two broad Gaussian functions, setting the parameters of broad Gaussians in the same way as narrow Gaussians described above. 
This model can well reproduce the [O~{\small III}] doublet profiles of all the four LRDs ($\chi^2_{\nu}=1.2-1.5$), revealing the presence of broad [O~{\small III}] components (FWHM $=1130-1627$~km~s$^{-1}$). 
The fitting results are presented in Section~\ref{sec:opt}. 
Comparing to models including narrow Gaussians only, the differences in BIC ($\Delta$BIC $=39-621$) strongly prefer models with broad components (Figure~\ref{fig:oiii}).

\section{H$\alpha$ Emission Line Fitting} \label{sec:balmer_fit}

The H$\alpha$ emission lines of the four LRDs show both narrow and broad components, and we also find H$\alpha$ absorption features in three LRDs (GN-2004, GN-29, CEERS-7902). 
In order to interpret the H$\alpha$ complex, we fit each H$\alpha$ line with multiple different models. 
Following the framework presented in Appendix~\ref{sec:line_fit}, we assess the goodness of fit using the Bayesian Information Criterion and reduced chi-square values.

We first assume the Doppler velocity broadening of H$\alpha$, and consider whether the line profile can be fitted by a combination of a narrow (FWHM $\lesssim500$~km~s$^{-1}$) and a broad (FWHM $\gtrsim1000$~km~s$^{-1}$) Gaussian. 
When an absorption feature is present, we fit the absorption line with an additional Gaussian. 
We allow the centroid, peak flux, and line width of each Gaussian to vary as free parameters. 
We additionally fit the [N~{\small II}]~$\lambda\lambda6548,6584$ emission lines with two Gaussians, fixing the centroids to the rest-frame wavelengths of [N~{\small II}] and fixing the $\lambda6548:\lambda6584$ flux ratio to the theoretical value ($3.05$). 
We do not find significant [N~{\small II}] emission (S/N $<3$) in any of the four LRDs, and the results do not change with and without fitting [N~{\small II}]. 
Therefore, we only consider fitting H$\alpha$ for the four LRDs.
With this model, we find excess flux in both blue and red wings of the H$\alpha$ profiles of all four LRDs. 
The $\chi^2_{\nu}$ values are large ($2.2-5.7$), indicating that the H$\alpha$ profiles cannot be well reproduced with a single broad and a narrow emission. 


\begin{figure*}
\includegraphics[width=\linewidth]{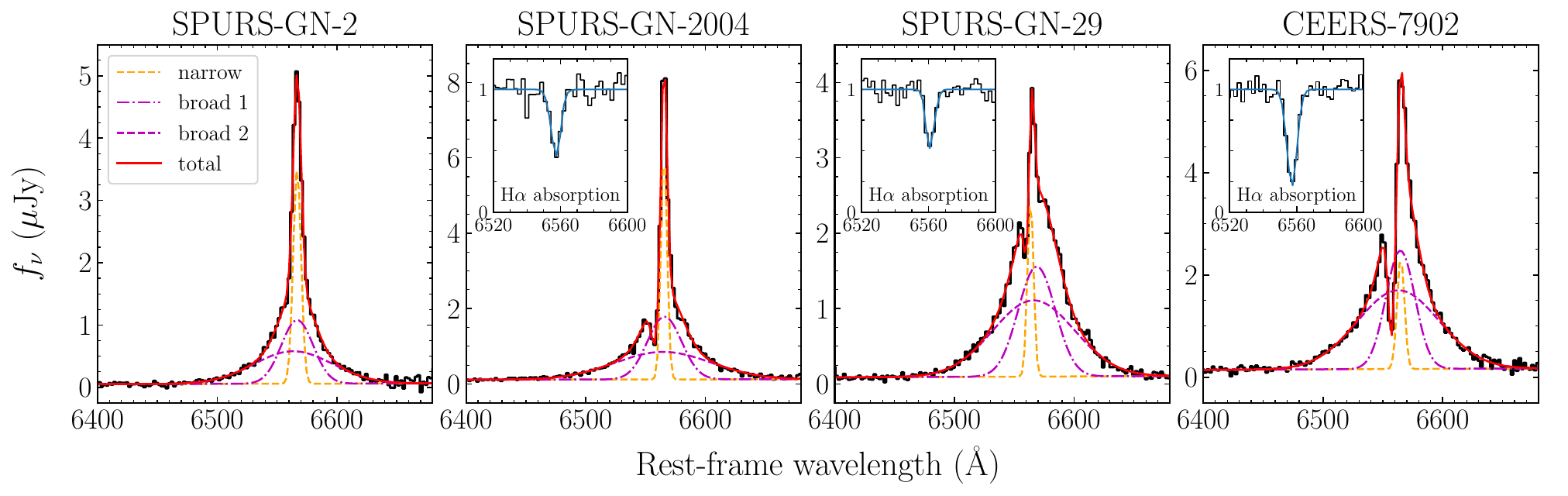}
\caption{Three-Gaussian fits to the H$\alpha$ emission line profiles of the four LRDs. For each object, the narrow H$\alpha$ emission component is shown as the orange dashed line. The broad H$\alpha$ is described by the sum of two components (magenta dashed and dotted lines). In the upper left corner of GN-2004, GN-29, and CEERS-7902, we show the absorption components. The total best-fit H$\alpha$ line profiles are shown as red solid lines.}
\label{fig:ha_gauss}
\end{figure*}

\begin{figure*}
\includegraphics[width=\linewidth]{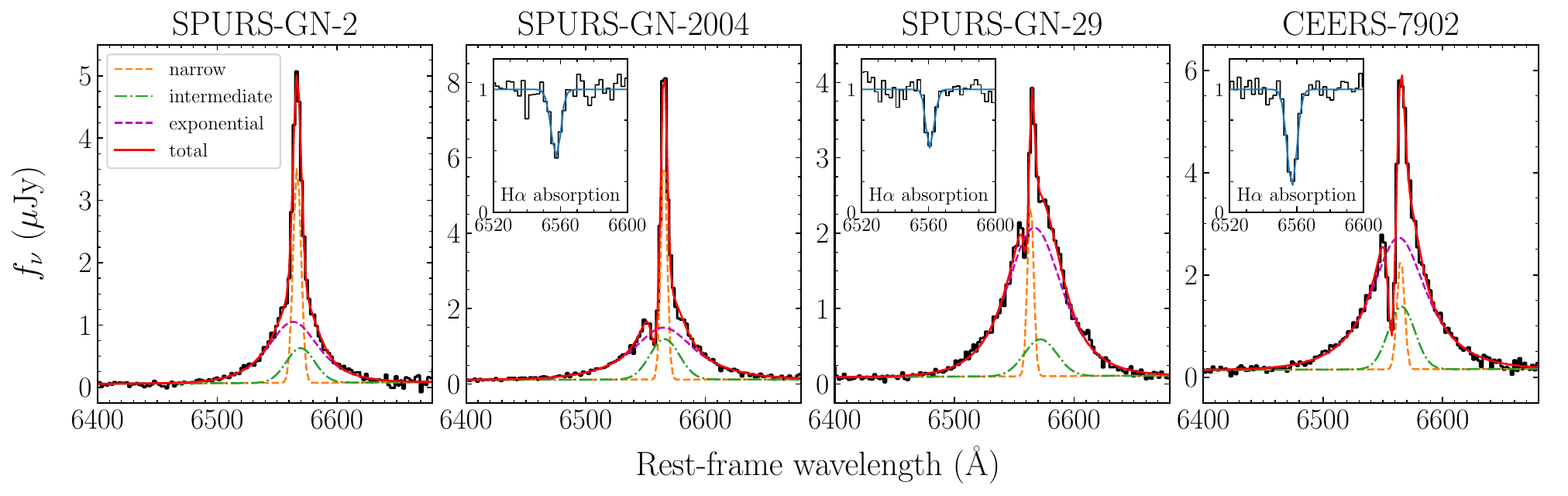}
\caption{Exponential model fits to the H$\alpha$ emission line profiles of the four LRDs. The narrow emission component is shown as the orange dashed line. The intermediate component, which describes the unscattered BLR emission, is shown as green dash-dotted line. The broad exponential wing (convolution of intrinsic BLR emission with an exponential kernel) is shown as magenta dashed line. We also show the absorption components in the upper left corner of GN-2004, GN-29, and CEERS-7902. The total best-fit H$\alpha$ line profiles are shown as red solid lines.}
\label{fig:ha_exp}
\end{figure*}


\begin{deluxetable}{l|cccc}
\tablecaption{H$\alpha$ properties inferred from exponential model fits to the four LRDs.}
\tablehead{
Properties & SPURS-GN-2 & SPURS-GN-2004 & SPURS-GN-29 & CEERS-7902 \\
}
\startdata
FWHM$_{\rm narrow}$ (km s$^{-1}$) & $356\pm10$ & $275\pm7$ & $276\pm8$ & $329\pm7$ \\
EW$_{\rm narrow}$ (\AA) & $510\pm19$ & $335\pm12$ & $149\pm32$ & $109\pm7$ \\
$f$ & $0.78\pm0.12$ & $0.76\pm0.04$ & $0.88\pm0.15$ & $0.83\pm0.04$ \\
FWHM$_{\rm int}$ (km s$^{-1}$) & $1243\pm211$ & $1291\pm87$ & $1436\pm330$ & $1159\pm95$ \\
FWHM$_{\rm exp}$ (km s$^{-1}$) & $1180\pm95$ & $1690\pm74$ & $1233\pm70$ & $1485\pm59$ \\
EW$_{\rm broad}$ (\AA) & $1323\pm185$ & $1181\pm67$ & $1395\pm277$ & $1341\pm89$ \\
FWHM$_{\rm abs}$ (km s$^{-1}$) & -- & $355\pm41$ & $255\pm15$ & $333\pm14$ \\
EW$_{\rm abs}$ (\AA) & -- & $-93\pm12$ & $-106\pm8$ & $-146\pm8$ \\
$\Delta v_{\rm abs}$ (km s$^{-1}$) & -- & $-296\pm50$ & $-142\pm48$ & $-319\pm48$ \\
\enddata
\tablecomments{$f$ is the fraction of intrinsic BLR emission being scattered. FWHM$_{\rm int}$ is the FWHM of the intermediate Gaussian which describes the intrinsic (unscattered) BLR emission. FWHM$_{\rm exp}$ is the FWHM of the broad exponential wing. EW$_{\rm broad}$ is the sum of the EWs of the intermediate component and the exponential wing, both contributed from the BLR emission.}
\label{tab:ha_exp}
\end{deluxetable}

Recent studies have found that the broad H$\alpha$ emission of LRDs can be better fitted by double-Gaussian profiles when S/N is high enough \citep[e.g.,][]{DEugenio2025,DEugenio2026}. 
Motivated by this, we simultaneously fit each H$\alpha$ line with a narrow and two broad Gaussian functions. 
Again, for the three LRDs presenting absorption features, we include another Gaussian. 
The fitting results are described in Section~\ref{sec:opt}, and we show the best-fit models in Figure~\ref{fig:ha_gauss}. 
The model can well reproduce the H$\alpha$ profile, with $\chi^2_{\nu}$ values of $1.1-1.4$. 
Comparing to model including one narrow and one broad Gaussian, the differences in BIC ($\Delta$BIC $=108-469$) indicate that the three-Gaussian model is strongly preferred in fitting H$\alpha$ of the four LRDs. 

It has also been argued that the Balmer emission lines of LRDs can be broadened by electron scattering, and thus the broad Balmer lines show exponential wings \citep[e.g.,][]{Matthee2024,Rusakov2026}. 
We then consider this scenario by fitting the H$\alpha$ profile with an exponential model following the methodology described in \citet{Matthee2026}. 
This model can be described by the following equation:
\begin{eqnarray*}
I(\lambda) = N(\lambda) + (1-f) \cdot G(\lambda) + f \cdot (G * E)(\lambda),
\end{eqnarray*}
where $N(\lambda)$ is a Gaussian function describing the narrow emission, $G(\lambda)$ is a Gaussian with intermediate width describing the intrinsic (unscattered) emission from BLR. 
The exponential wing produced by electron scattering is described by $f \cdot (G * E)(\lambda)$, that a fraction ($f$) of intrinsic BLR emission is scattered, where $E$ is an exponential convolution kernel:
\begin{eqnarray*}
E(\lambda) \equiv (1/2W) \cdot \exp{(-|\lambda-\lambda_0|/W)}.
\end{eqnarray*}
Where $\lambda_0$ is the centroid of H$\alpha$, $W$ is the $e$-folding scale. 
The FWHM of the broad exponential wing can be expressed by FWHM $\approx2\ln{2}\cdot W$. 
Here we allow the centroids, peak fluxes, line widths of Gaussians, as well as the scattering fraction $f$ and $e$-folding scale $W$ to vary as free parameters. 

We show the exponential model fitting results in Table~\ref{tab:ha_exp}, and we plot the best-fit models in Figure~\ref{fig:ha_exp}. 
The exponential model can also well reproduce the H$\alpha$ line profiles of the four LRDs ($\chi^2_{\nu}=1.1-1.5$). 
With this model, the broad H$\alpha$ lines have narrower FWHMs ($1180-1690$~km~s$^{-1}$) than Gaussian fits. 
We derive that a large fraction ($76-88\%$) of the intrinsic BLR H$\alpha$ emission experiences electron scattering. 
The total broad H$\alpha$ EWs ($1181-1395$~\AA) are similar to those derived from Gaussian fits. 
For narrow H$\alpha$ emission lines, we find FWHMs ($275-356$~km~s$^{-1}$) and EWs ($109-510$~\AA) that are consistent with Gaussian fits (Table~\ref{tab:opt_info}). 
The differences in BIC between exponential model fits and three-Gaussian model fits are very small ($\Delta$BIC $=1-3$), indicating that both models fit the H$\alpha$ profiles equally well for the four LRDs.

\section{Methodology of Deriving Ionized Gas Properties} \label{sec:gas_method}

Using the narrow (FWHM $\simeq330$~km~s$^{-1}$) emission line detections in SPURS spectra, we derive the electron densities, temperatures, and chemical abundance ratios (O/H, N/O, N/C) of the ionized gas for the four LRDs. 
The derived narrow-line gas properties are presented in Table~\ref{tab:properties}. 
Here we use the \texttt{Python} package \texttt{PyNeb} \citep{Luridiana2015}. 
The electron densities and temperatures are derived by jointly fitting the density-sensitive and temperature-sensitive line ratios. 
In each LRD, we derive the densities of the carbon, nitrogen, and silicon emitting gas with the available C~{\small III}]~$\lambda1909$/[C~{\small III}]~$\lambda1907$, N~{\small IV}]~$\lambda1486$/[N~{\small IV}]~$\lambda1483$, and Si~{\small III}]~$\lambda1892$/Si~{\small III}]~$\lambda1883$ ratios, respectively. 
For temperature-sensitive line ratios we primarily focus on [O~{\small III}]~$\lambda4363$/[O~{\small III}]~$\lambda5007$, since both lines are close in wavelength and the ratio is less impacted by dust attenuation. 
We also derive temperatures using the O~{\small III}]~$\lambda1666$/[O~{\small III}]~$\lambda5007$ ratios. 

We characterize the gas-phase oxygen abundances of the four LRDs following the methodology commonly used in recent LRD studies \citep[e.g.,][]{Isobe2025,Nikopoulos2026}.
The oxygen abundance O/H is estimated as the sum of O$^{2+}$/H and O$^{+}$/H. 
Based on the derived densities (inferred from C~{\small III}] doublet ratios) and temperatures, we infer O$^{2+}$/H from the narrow [O~{\small III}]~$\lambda5007$/H$\beta$ flux ratios. 
For the O$^+$ zone, the temperature and density are likely different from those of the O$^{2+}$ zone. 
Since we do not detect [O~{\small II}]~$\lambda\lambda7320,7330$ auroral lines, we derive the O$^+$ zone temperature following \citet{Campbell1986}: $T({\rm O}^+)=0.7\times T({\rm O}^{2+})+3000$~K. 
For the O$^+$ zone density, since we are not able to deblend the [O~{\small II}]~$\lambda\lambda3727,3729$ doublet due to the resolution of the spectra and directly infer the density, we assume a typical value of the low-ionization zone ($n_{\rm e}\simeq300$~cm$^{-3}$) derived in the literature \citep[e.g.,][]{Topping2025b,Sanders2026}. 
The singly ionized oxygen abundances (O$^+$/H) are inferred from the narrow [O~{\small II}]/H$\beta$ flux ratios based on the above O$^+$ zone densities and temperatures. 

We derive the nitrogen-to-oxygen abundance ratio following the similar procedures described in \citet{Martinez2025}. 
For the LRDs with both N~{\small IV}] and N~{\small III}] detections (GN-2, GN-29, CEERS-7902), we calculate the N/O ratios using (N$^{3+}$+N$^{2+}$)/O$^{2+}$ derived from (N~{\small IV}]+N~{\small III}])/O~{\small III}] flux ratios. 
For GN-2004 whose N~{\small IV}] is not available, we calculate N$^{2+}$/O$^{2+}$ from the N~{\small III}]/O~{\small III}] ratio. 
To convert (N$^{3+}$+N$^{2+}$)/O$^{2+}$ or N$^{2+}$/O$^{2+}$ to N/O abundance ratio, we apply an ionization correction factor (ICF) for each LRD.
This step accounts for the contribution of ionic species that are not detected in the spectra (e.g., N$^+$, N$^{2+}$ or N$^{3+}$). 
Following the equations presented in \citet{Martinez2025}, we compute the ICF for N/O as a function of metallicity and the N~{\small IV}]/N~{\small III}] (N43) ratio (or O32 ratio when N43 is not available). 
We also compute the nitrogen-to-carbon abundance ratios using the N~{\small IV}] or N~{\small III}] and C~{\small III}] detections.

\section{Emission Lines in SPURS LRDs} \label{sec:more_lines}

\subsection{NUV Fe~{\small II} Emission Line Measurements} \label{sec:nuv_feii}

In Table~\ref{tab:nuv_feii}, we present the EWs of NUV Fe~{\small II} emission lines of the four LRDs.

\begin{deluxetable}{lcccc}
\tablecaption{EW (\AA) measurements of the NUV Fe~{\scriptsize II} complexes.}
\tablehead{
 & GN-2 & GN-2004 & GN-29 & CEERS-7902
}
\startdata
Fe~{\scriptsize II}~2200--2660      & $202\pm13$  & $76\pm8$    & $40\pm12$   & $101\pm14$ \\
Fe~{\scriptsize II}~UV62,63         & $70\pm6$    & $22\pm2$    & $13\pm3$    & $28\pm5$ \\
Fe~{\scriptsize II}~UV61            & $44\pm6$    & $17\pm2$    & $5\pm3$     & $38\pm6$ \\
Fe~{\scriptsize II}~UV60,78         & $28\pm6$    & $10\pm2$    & $7\pm3$     & $10\pm6$ \\
Fe~{\scriptsize II}~(UV)~2200--3090 & $356\pm24$  & $141\pm12$  & $72\pm16$   & $194\pm26$ \\
Fe~{\scriptsize II}~$\lambda$2508   & $9.4\pm1.0$ & $6.1\pm1.0$ & $2.9\pm1.5$ & $10.4\pm1.5$
\enddata
\tablecomments{We report the rest-frame EWs of individual bands and the total value from $2200-3090$~\AA, followed by the narrow Fe~{\scriptsize II}~$\lambda2508$ line.}
\label{tab:nuv_feii}
\end{deluxetable}

\subsection{Rest-frame optical spectra} \label{sec:full_opt}

\begin{figure*}
\includegraphics[width=\linewidth]{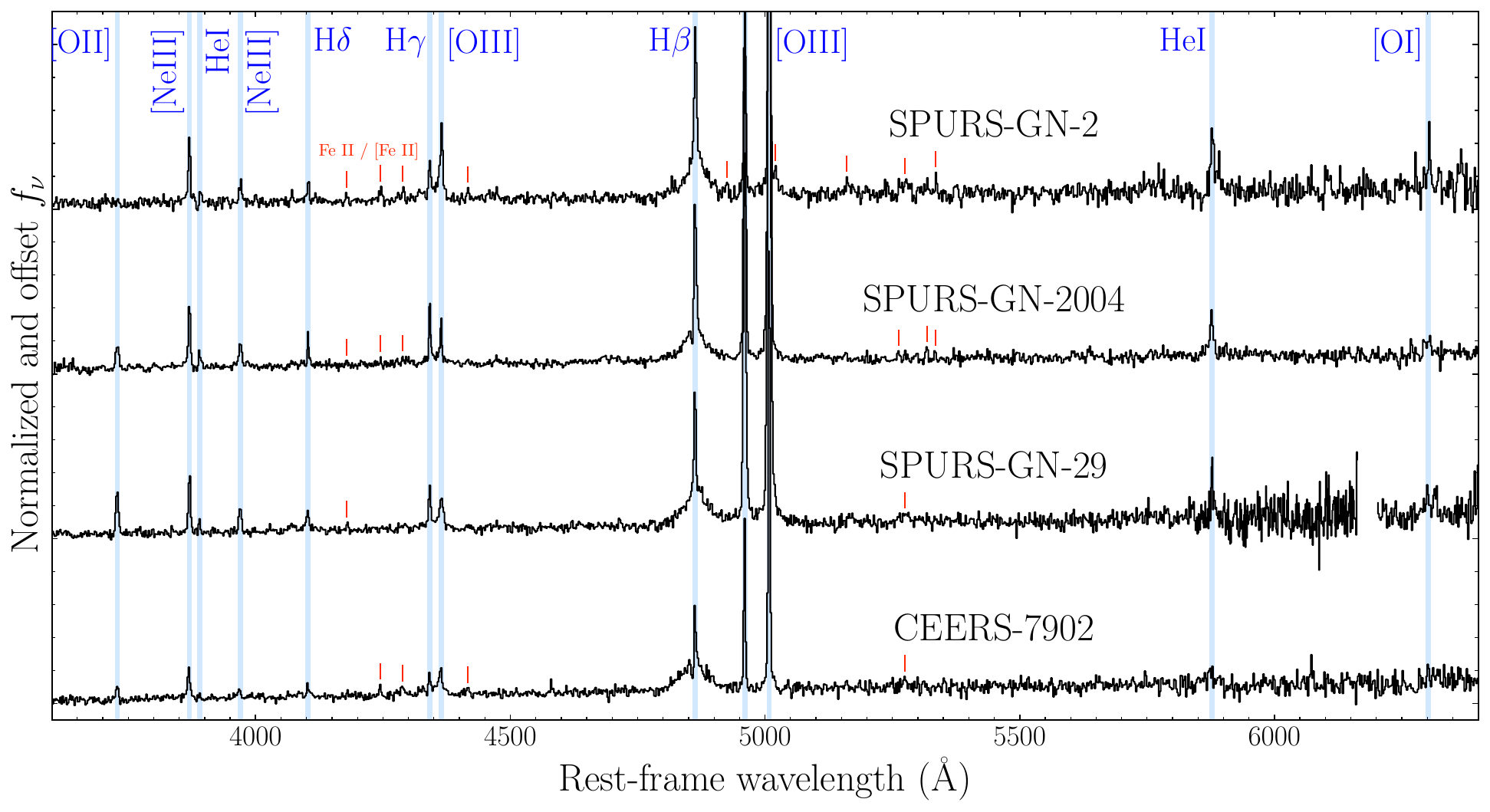}
\caption{SPURS rest-frame optical spectra of SPURS-GN-2, SPURS-GN-2004, SPURS-GN-29, and CEERS-7902. The figure is shown in the same way as Figure~\ref{fig:fuv_spec}.
We also mark the faint emission feature corresponding to \ion{Fe}{2} or [\ion{Fe}{2}] lines with red bars.}
\label{fig:opt_spec}
\end{figure*}

\begin{figure*}
\centering
\includegraphics[width=0.95\linewidth]{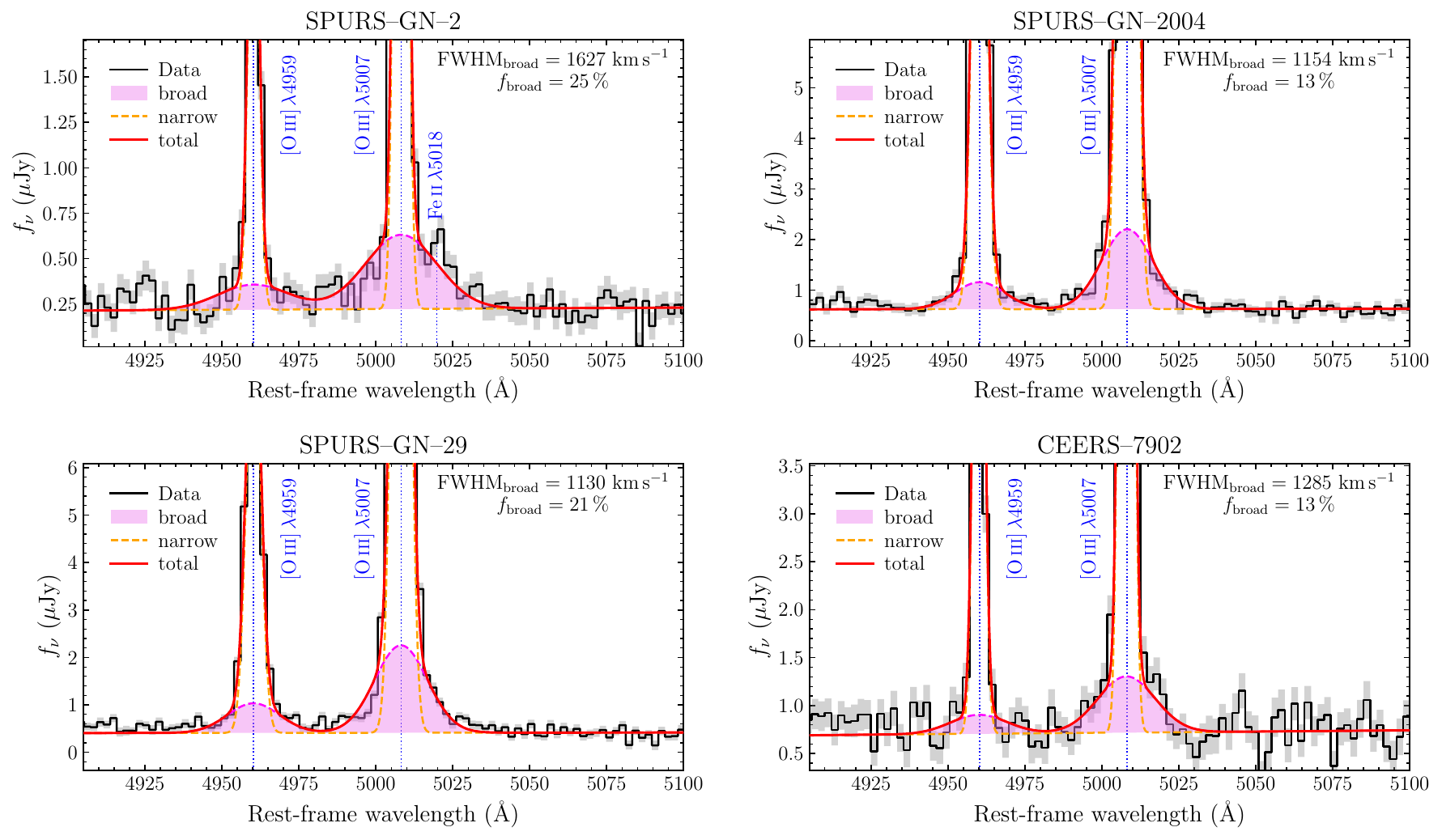}
\caption{[O~{\scriptsize III}]~$\lambda\lambda4959,5007$ emission lines of the four LRDs. We overplot the best-fit line profile to each spectrum (red), with the narrow components shown as orange dashed lines and the broad components as magenta shaded regions. The FWHM of the broad component (corrected for the instrumental resolution) and its contribution to the total [O~{\scriptsize III}] line flux are quoted in each panel. A narrow \ion{Fe}{2}~$\lambda5018$ line is marginally detected ($2.8\sigma$) on the red wing of [\ion{O}{3}]~$\lambda5007$ in SPURS-GN-2, and adding it to the fit leaves the broad width unchanged and only slightly lowers the broad line fraction $f_{\rm broad}$ from $25\%$ to $22\%$.}
\label{fig:oiii}
\end{figure*}

The full rest-frame optical spectra of the four LRDs are shown in Figure~\ref{fig:opt_spec}. 
We present the optical emission line measurements (fluxes, EWs, FWHMs) in Table~\ref{tab:opt_lines}. 
We also show the fits to the [O~{\small III}]~$\lambda4959$, $\lambda5007$ emission lines of the four LRDs in Figure~\ref{fig:oiii}. 


\begin{figure*}
\centering
\includegraphics[width=0.95\linewidth]{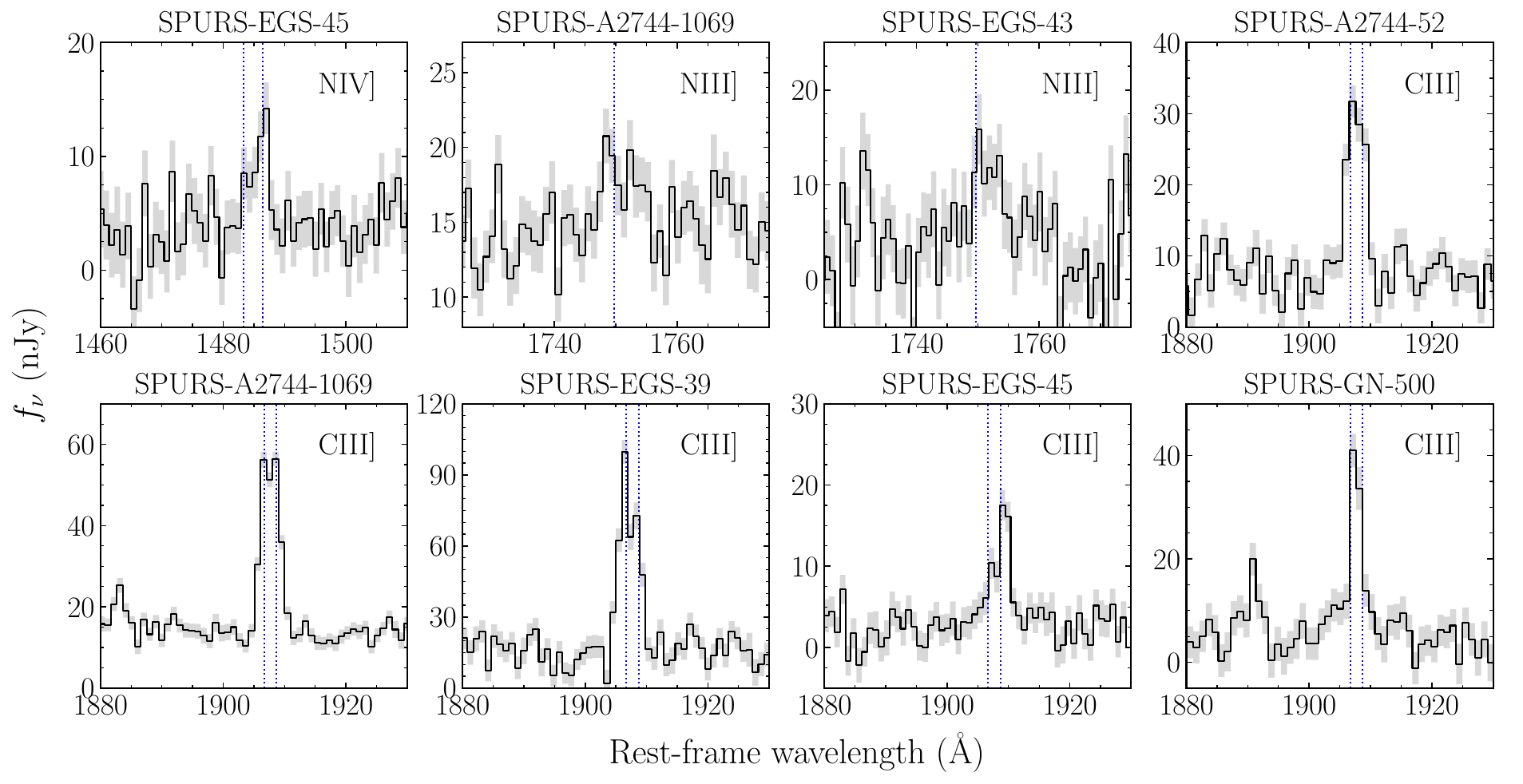}
\caption{N~{\scriptsize IV}], N~{\scriptsize III}], and C~{\scriptsize III}] emission lines identified in SPURS G140M spectra of LRDs listed in Table~\ref{tab:new_lrd}.}
\label{fig:new_lrd_uv}
\end{figure*}

\subsection{Rest-frame UV Emission Lines of New SPURS LRDs} \label{sec:new_lrd_uv}

In Figure~\ref{fig:new_lrd_uv}, we show the UV emission lines (N~{\small IV}], N~{\small III}], C~{\small III}]) identified in the SPURS G140M spectra of LRDs listed in Table~\ref{tab:new_lrd}. 


\begin{deluxetable*}{l|ccc|ccc|ccc|ccc}
\setlength{\tabcolsep}{3pt}
\tablecaption{Rest-frame optical emission line fluxes ($10^{-19}$~erg~s$^{-1}$~cm$^{-2}$), EWs (\AA), and FWHMs (km~s$^{-1}$) of the four LRDs.}
\tablehead{
 & \multicolumn{3}{c|}{SPURS-GN-2} & \multicolumn{3}{c|}{SPURS-GN-2004} & \multicolumn{3}{c|}{SPURS-GN-29} & \multicolumn{3}{c}{CEERS-7902} \\
Line & Flux & EW & FWHM & Flux & EW & FWHM & Flux & EW & FWHM & Flux & EW & FWHM
}
\startdata
{[}O~{\scriptsize II}]~$\lambda3728$ & $2.5\pm0.5$ & $10.8\pm2.2$ & $363\pm145$ & $13.1\pm1.2$ & $22.8\pm2.1$ & $360\pm24$ & $18.6\pm1.6$ & $53.3\pm4.6$ & $380\pm31$ & $12.6\pm2.7$ & $34.3\pm7.3$ & $389\pm80$ \\
{[}Ne~{\scriptsize III}]~$\lambda3869$ & $12.0\pm0.8$ & $43.7\pm2.9$ & $275\pm13$ & $26.6\pm1.1$ & $39.5\pm1.7$ & $273\pm8$ & $19.2\pm1.4$ & $48.4\pm3.5$ & $367\pm19$ & $21.5\pm2.3$ & $28.7\pm3.1$ & $457\pm36$ \\
He~{\scriptsize I}~$\lambda3890$+H8 & $2.0\pm0.9$ & $7.7\pm3.5$ & $353\pm123$ & $6.8\pm1.1$ & $9.9\pm1.6$ & $266\pm31$ & $3.1\pm1.0$ & $7.8\pm2.4$ & $247\pm59$ & $1.4\pm0.4$ & $2.0\pm0.5$ & $244\pm84$ \\
{[}Ne~{\scriptsize III}]~$\lambda3968$+H$\epsilon$ & $4.7\pm1.1$ & $17.4\pm3.9$ & $467\pm78$ & $13.4\pm1.4$ & $18.1\pm1.9$ & $424\pm33$ & $9.5\pm1.3$ & $21.8\pm3.1$ & $448\pm46$ & $9.4\pm0.9$ & $13.3\pm1.2$ & $420\pm65$ \\
H$\delta_{\rm total}$ & $3.1\pm0.8$ & $11.0\pm2.9$ & -- & $9.2\pm1.0$ & $11.2\pm1.2$ & -- & $7.9\pm1.4$ & $16.5\pm2.9$ & -- & $21.3\pm4.7$ & $28.6\pm6.4$ & -- \\
\hspace{1em}H$\delta_{\rm narrow}$ & $3.1\pm0.8$ & $11.0\pm2.9$ & $296\pm57$ & $9.2\pm1.0$ & $11.2\pm1.2$ & $306\pm17$ & $7.9\pm1.4$ & $16.5\pm2.9$ & $469\pm73$ & $2.5\pm0.7$ & $3.4\pm0.9$ & $244\pm31$ \\
\hspace{1em}H$\delta_{\rm broad}$ & -- & -- & -- & -- & -- & -- & -- & -- & -- & $18.8\pm4.7$ & $25.2\pm6.3$ & $1974\pm319$ \\
H$\gamma_{\rm total}$ & $17.9\pm3.0$ & $58.9\pm9.8$ & -- & $44.6\pm3.9$ & $52.5\pm4.6$ & -- & $24.0\pm3.6$ & $47.4\pm7.0$ & -- & $38.9\pm7.5$ & $45.6\pm8.8$ & -- \\
\hspace{1em}H$\gamma_{\rm narrow}$ & $3.7\pm0.8$ & $12.2\pm2.8$ & $279\pm47$ & $16.6\pm1.2$ & $19.6\pm1.4$ & $306\pm12$ & $8.2\pm0.9$ & $16.2\pm1.9$ & $285\pm24$ & $5.6\pm1.7$ & $6.6\pm2.0$ & $244\pm48$ \\
\hspace{1em}H$\gamma_{\rm broad}$ & $14.2\pm2.9$ & $46.7\pm9.4$ & $2745\pm390$ & $27.9\pm3.7$ & $32.9\pm4.4$ & $2375\pm231$ & $15.8\pm3.4$ & $31.2\pm6.8$ & $2781\pm430$ & $33.3\pm7.3$ & $39.0\pm8.5$ & $3083\pm458$ \\
{[}O~{\scriptsize III}]~$\lambda4363$ & $11.5\pm1.2$ & $37.9\pm4.0$ & $427\pm33$ & $10.0\pm1.0$ & $11.8\pm1.2$ & $306\pm16$ & $10.8\pm1.5$ & $21.4\pm3.0$ & $476\pm60$ & $13.3\pm2.7$ & $15.5\pm3.1$ & $437\pm67$ \\
H$\beta_{\rm total}$ & $61.3\pm4.5$ & $181\pm13$ & -- & $139.4\pm9.5$ & $133\pm9$ & -- & $91.0\pm8.3$ & $146\pm13$ & -- & $146.2\pm9.4$ & $140\pm9$ & -- \\
\hspace{1em}H$\beta_{\rm narrow}$ & $17.6\pm1.5$ & $51.9\pm4.4$ & $337\pm21$ & $46.3\pm3.3$ & $44.3\pm3.1$ & $306\pm16$ & $20.3\pm2.3$ & $32.5\pm3.8$ & $274\pm23$ & $16.2\pm1.9$ & $15.5\pm1.8$ & $244\pm3$ \\
\hspace{1em}H$\beta_{\rm broad}$ & $43.8\pm4.2$ & $129\pm13$ & $2955\pm197$ & $93.1\pm8.9$ & $89.1\pm8.6$ & $3125\pm207$ & $70.8\pm7.9$ & $113\pm13$ & $3357\pm263$ & $130.0\pm9.2$ & $125\pm9$ & $3982\pm206$ \\
{[}O~{\scriptsize III}]~$\lambda4959_{\rm total}$ & $24.1\pm0.4$ & $74.1\pm3.3$ & -- & $137.0\pm1.1$ & $139\pm3$ & -- & $98.5\pm1.8$ & $159\pm9$ & -- & $56.9\pm0.9$ & $52.9\pm1.6$ & -- \\
\hspace{1em}{[}O~{\scriptsize III}]~$\lambda4959_{\rm narrow}$ & $18.2\pm0.4$ & $55.7\pm2.3$ & $241\pm5$ & $119.6\pm1.6$ & $122\pm3$ & $315\pm3$ & $78.4\pm6.3$ & $126\pm15$ & $330\pm16$ & $49.9\pm0.6$ & $46.4\pm1.2$ & $236\pm3$ \\
\hspace{1em}{[}O~{\scriptsize III}]~$\lambda4959_{\rm broad}$ & $6.0\pm0.5$ & $18.3\pm1.8$ & $1627\pm181$ & $17.3\pm1.2$ & $17.6\pm1.2$ & $1130\pm117$ & $20.1\pm4.5$ & $32.4\pm6.2$ & $1183\pm306$ & $7.0\pm1.0$ & $6.5\pm1.0$ & $1285\pm135$ \\
{[}O~{\scriptsize III}]~$\lambda5007_{\rm total}$ & $70.5\pm1.3$ & $217\pm9$ & -- & $400.4\pm3.1$ & $413\pm9$ & -- & $287.9\pm5.2$ & $470\pm24$ & -- & $166.4\pm2.6$ & $155\pm4$ & -- \\
\hspace{1em}{[}O~{\scriptsize III}]~$\lambda5007_{\rm narrow}$ & $53.1\pm1.0$ & $163\pm6$ & $241\pm5$ & $349.7\pm4.6$ & $361\pm9$ & $315\pm3$ & $229.2\pm18.4$ & $374\pm42$ & $330\pm16$ & $146.0\pm1.8$ & $136\pm3$ & $236\pm3$ \\
\hspace{1em}{[}O~{\scriptsize III}]~$\lambda5007_{\rm broad}$ & $17.5\pm1.3$ & $53.6\pm5.1$ & $1627\pm181$ & $50.7\pm3.6$ & $52.3\pm3.7$ & $1130\pm117$ & $58.7\pm13.3$ & $95.9\pm18.8$ & $1183\pm306$ & $20.4\pm2.9$ & $19.0\pm2.8$ & $1285\pm135$ \\
He~{\scriptsize I}~$\lambda5876$ & $10.6\pm0.8$ & $34.6\pm2.8$ & $432\pm51$ & $18.8\pm1.4$ & $22.9\pm1.7$ & $392\pm45$ & $8.4\pm1.4$ & $15.2\pm2.5$ & $360\pm90$ & $19.1\pm3.4$ & $19.0\pm3.4$ & $1531\pm469$ \\
{[}O~{\scriptsize I}]~$\lambda6302$ & $3.9\pm0.8$ & $12.6\pm2.6$ & $281\pm37$ & $8.5\pm1.4$ & $10.9\pm1.7$ & $665\pm181$ & $5.2\pm1.2$ & $9.4\pm2.1$ & $343\pm99$ & $<4.9$ & $<5.1$ & -- \\
\enddata
\tablecomments{For non-detections, $3\sigma$ upper limits of line fluxes and EWs are provided.}
\label{tab:opt_lines}
\end{deluxetable*}


\bibliography{SPURS_LRD_2}{}
\bibliographystyle{aasjournalv7}



\end{document}